# Origin and evolution of the Galactic inventories of interstellar dust and its composition

Anuj Gupta*, Sandeep Sahijpal
*Department of Physics, Panjab University, Chandigarh 160014, India*

**ABSTRACT**

Interstellar dust is a significant component of matter in the galaxies. The dust owns its origin and reprocessing in a wide range of astrophysical environments. In order to understand the origin and evolution of the distinct types of interstellar dust grains, we have attempted a comprehensive correlated study of the thermodynamics condensation of dust grains in distinct stellar environments with the Galactic chemical evolution of the Milky Way Galaxy. The Galaxy is evolved in terms of elemental evolution resulting from stellar nucleosynthetic contributions of several generations of stars. Based on the elemental composition of the evolving Galaxy, the relative abundances of the major constituents of interstellar dust are assessed. The major aim is to redistribute the various condensable elements at any epoch during the evolution of the Galaxy into various grain constituents and understand their abundance evolution based on a mass-balance formalism. We also performed thermodynamical equilibrium condensation calculations to understand the stellar origin of various grain constituents that could carry the isotopic signatures of the various stellar nucleosynthetic sources. This is perhaps a novel attempt to estimate the bulk dust mass budget in the evolving Galaxy. The normalized mass of the Galactic dust is predicted to decrease with the increase in distance from the Galactic centre. It increases over time. The supernovae SN Ia are predicted as the most prominent sources of Fe-dust mass, the supernovae SN II+Ib/c produces oxides and silicate dust mass, and the AGB stars contributes to carbonaceous dust mass.

## 1 INTRODUCTION

Interstellar dust is a major constituent of the baryonic matter distribution in the interstellar medium (ISM) as the bulk abundance of refractory and moderately volatile elements are trapped in dust grains. These grains can originate only in the stellar environments such as circumstellar envelopes associated with the evolving stars, planetary nebulae, supernovae ejecta and the outflows of massive stars. Subsequently, the grains are injected into interstellar space where further reprocessing and evolution of dust takes place (Evans 1993). The dust plays an important role in numerous physicochemical processes in interstellar space. For instance, the most plausible process associated with the formation of hydrogen molecule in the interstellar space is through the adsorption of hydrogen atoms on the surface of dust grains, followed by molecule formation on the surface (Gould & Salpeter 1963; Islam 2010). Thus, even the most predominant molecular species cannot form in interstellar space without the presence of dust. Further, the distinct type of circumstellar and interstellar dust grains condensed in diverse environments carry distinct isotopic and chemical signatures representing a wide-range of stellar evolutionary and stellar nucleosynthetic processes. The condensation of dust grains plays a vital role in the commencement of the formation of planetary systems. These dust grains coagulate and coalesce through a wide range of physicochemical processes operating within the accretion disk to form planetesimals and protoplanets (Weidenschilling & Cuzzi 1993; Weidenschilling 2000).

The commencement of the condensation of interstellar dust could have initiated in the protogalaxies within the circumstellar environments associated with the earliest generations of stars formed in the galaxies. The formation and evolution of the Milky Way Galaxy is believed to have initiated well within the initial ~1 Gyr (Giga years) from the time of the Big-Bang origin of the Universe (Mo, van den Bosch & White 2010). The Galaxy formation commenced with the merging of protogalaxies that were formed by the amalgamation of diffuse neutral hydrogen gas clouds. While the primordial nucleosynthesis after the Big-Bang resulted in hydrogen and helium enriched primordial gas, the heavier elements came into existence through the stellar nucleosynthesis operating within several generations of stars (Clayton 1968; Pagel 1997). The stellar nucleosynthesis within these numerous generations of stars of different masses and metallicities resulted in the gradual abundance evolution of the elemental and isotopic inventories of the Galaxy. An evolving ensemble of several generations of stars chemically enriched the interstellar medium by ejecting their freshly synthesized yields at the final stages in their evolution. The galactic chemical evolution (GCE) models incorporate the accretional history of the Galaxy, the star formation rate and the stellar nucleosynthetic inventories from stars formed over the galactic time-scale (Matteucci & François 1989; Chiappini, Matteucci & Romano 2001; Kobayashi et al. 2006; Minchev et al. 2012,2015; Sahijpal & Gupta 2013; Grisoni, Spitoni & Matteucci 2018; Sahijpal & Kaur 2018). The GCE models deal with the understanding of the origin and temporal evolution of elemental and isotopic abundances over the entire Galaxy. Based on the evolution of the elemental abundance distribution across the Galaxy, in the present work, we assess the evolving trends in the interstellar dust abundance and composition across the Galaxy over its temporal evolution. The analysis is based on a comprehensive mass-balance calculation along with the thermodynamical condensation trends associated with the distinct evolved stars. Based on a detailed GCE model (Sahijpal & Kaur 2018), we present a case study of the spatial and temporal abundance distribution evolution of various condensable elements into distinct grain constituents. We estimate the dust mass gradients across the Milky Way Galaxy with distinct chemical composition. The distributions include 22 distinct grain components such as silicates, oxides, graphite, SiC, and Fe-Ni metal, in addition to the possibility of condensation of ice as a mantle on these grains. The mass-balance calculations performed here are relevant in understanding the abundance evolution of these grains across the entire Galaxy.

---

* E-mail: mr.anuj@pu.ac.in

The Galactic dust constitutes only ~1 per cent of the gas mass in the ISM (McKinnon, Torrey & Vogelsberger 2016; Giannetti et al. 2017). The presence of dust is inferred primarily by the light extinction curves and meteorites analysis. On the Galactic scale, the spectroscopic techniques are used to analyse the presence of dust grains. The dust in the ISM absorbs the starlight and re-radiates it in the infra-red wavelength resulting in the reddening of a star. Dust particles also polarize the starlight. However, the dust is mostly inferred in the form of extinction of starlight in terms of a correlated behaviour between the light extinction curves and the size of dust particles. The silicates and oxides rich grains of several hundred Angstrom are generally considered to be responsible for the light extinction trends in the ultraviolet region. An almost identical sized silicates and oxides constituents as cores of dust grains with a mantle of ices ($H_2O$, $CO_2$, and $NH_4$), totally amounting to several thousand Angstrom, are generally considered to be responsible for the extinction features in the visible region (Li 2005). The spectroscopic observations give us direct information about the composition of interstellar dust (Draine 2003; Gibb et al. 2004). For instance, the prominent absorption feature at 2200 $Å$ in the light extinction curve supports graphite in ISM (Stecher & Donn 1965; Wickramasinghe & Guillaume 1965). At the solar system scale, the interstellar dust grains separated from primitive meteorites provide a unique opportunity to understand the variety of stellar nucleosynthetic processes operating in various astrophysical environments (Nguyen, Keller & Messenger 2016). For instance, the presolar grains separated from primitive meteorites infer a widespread presence of carbides apart from the silicates and oxide grains in the interstellar medium (Amari et al. 1993; Gyngard et al. 2018).

The condensation of dust requires high-density regions at moderately low temperatures (Lord 1965). In general, such conditions are achieved in the circumstellar envelopes of low- to intermediate-mass stars during the late stages of their evolution, and in the ejecta of supernovae (Zinner 1997; Ebel 2000). Diverse stellar environments have distinct chemical compositions and physicochemical conditions, which further vary significantly during their evolutionary stages. The stars of distinct masses are mostly formed in clusters with the stellar mass following an initial mass function, IMF (Salpeter 1955; Lada & Lada 2003). The stars in the mass range, 0.1-100 $M_\odot$ are formed in the Galaxy according to the IMF and the star formation rate (SFR).

The diverse astrophysical environments and the associated distinct physicochemical processes provide a wide range of possible dust condensation scenarios with grains of different chemical and isotopic compositions. In the atmosphere of red giant stars and planetary nebula (PN) envelopes, the number densities are high, and the gas kinetic temperature is close to the condensation temperature of many elements. During the red giant phase and later during the asymptotic giant branch (AGB) phase, the star undergoes substantial mass-loss in the form of stellar wind (Höfner & Olofsson 2018). The circumstellar dust grains can condense in this expanding and cooling gas. Several late-type stars are observed to be surrounded by dust shells of grains whose mineral compositions imitate the major chemistry of the gas (Gail & Sedlmayr 1999; Cherchneff et al. 2000; Williams 2014). Distinct from the red-giant and AGB stellar environments, the ejecta from the supernovae vary in chemical and isotopic compositions. The cause of this variability is the variations in the distinct nuclear burning experienced during stellar evolution and explosive nucleosynthesis of various shells inside the stars (Lattimer, Schramm & Grossman 1978; Clayton, Deneault & Meyer 2001). After the explosion, the hot ejecta of a supernova cools down to temperatures at which grains can condense. The interstellar dust grains mostly start as silicate- or carbon-rich grains depending upon their astrophysical environment (Lodders & Fegley 1995). At later stages, volatile elements are accumulated to form icy mantles composed of water ice, methane, carbon monoxide, and ammonia (Saslaw & Gaustad 1969; Fraser, Collings & McCoustra 2002). Subsequent to the condensation in the stellar atmosphere, the dust grains are ejected into interstellar space by the stellar radiation pressure or in the supernova ejecta. These grains are further reprocessed in the interstellar medium. These processes involve sputtering, heating, evaporation, and re-condensation (see e.g. Evans, 1993). In the present work, we make an attempt to understand the stellar origin of the various grain constituents that could carry the chemical and isotopic signatures of the various stellar nucleosynthetic sources. We could specifically decipher the stellar contribution of dust grains from low to intermediate-mass stars and supernovae type II, Ib/c, and Ia during the galactic chemical evolution.

As mentioned earlier, the condensation of dust depends upon the prevailing pressure, temperature and the elemental abundance of the environment. These factors collectively determine the partial pressure of various elements in a gas (Ebel 2006). The difference in the relative abundance of the elements in that environment leads to the condensation of distinct types of grains. For instance, carbon and oxygen form CO molecule. The less abundant element out of two gets exhausted in the CO form, whereas, the remaining amount of more abundant element reacts to form other gaseous or solid species. Therefore, in the case of a C-rich environment, condensates like carbides are more stable. On the other hand, if oxygen is more abundant than carbon in a stellar environment, oxides and silicates are condensed. The equilibrium calculations provide an insight into the chemical nature of the dust that can condense in a gaseous system cooling over time. The equilibrium condensation calculations assume that dust gets sufficient time to condense as gas cools down slowly and interact with the remaining gas in terms of the further chemical reaction. The temperatures and pressures are very high in the inner regions of the stellar envelope, and the cooling time is relatively more than the time taken by a typical thermochemical reaction. Although there is a possibility that thermochemical equilibrium prevails in such a region, the astrophysical environments can, in general, be in a non-equilibrium state (Donn & Nuth 1985; Amari, Zinner & Lewis 1996; Cherchneff 2009). The dynamic pressure that represents the non-equilibrium state in a mechanical sense changes the condensation temperatures of the condensates as shown by Gupta and Sahijpal (2020) (See also, Tanaka, Tanaka & Nakazawa 2002). Therefore, we have performed the simulations at static pressure to understand thermodynamics associated with the condensation of dust grains in the vicinity of AGB stars and supernovae ejecta.

The theoretical technique involved in the modelling of the Galactic evolution and interstellar dust abundance evolution is discussed in Section 2. GCE model for estimating the elemental abundance evolution of the Milky Way Galaxy is discussed in Subsection 2.1 with the mass-balance calculations discussed in Subsection 2.2. The thermodynamical calculations to compute the chemical equilibria for various evolved stars such as AGB stars, SN II, and SN Ia are explained in Subsection 2.3. Subsection 2.4 contains the details of the mapping of these two approaches in order to make an attempt to deconvolute the contributions of various evolved stellar sources to the ISM dust inventories. The results of our calculations are presented in Section 3. Section 4 includes the discussion of the estimates of the dust mass in the Milky Way Galaxy along with the condensation scenarios considered in this study. Finally, the main conclusions drawn from the present work are mentioned in section 5.

## 2 METHODOLOGY

In the present work, we have followed two distinct approaches to make an assessment regarding the abundance evolution of interstellar dust grains of varied compositions across the Galaxy over the Galactic time-scales. This includes a detailed mass-balance calculation in estimating the relative abundance of the dust grains of distinct compositions based on the elemental abundance evolution of the Galaxy. In order to achieve this goal, we deduced the elemental abundance evolution using one of the recent GCE model developed by Sahijpal & Kaur (2018). Subsequently, we performed detailed mass-balance calculations at different epochs of the evolution of the Galaxy across its radial extent. The relative dust abundances thus estimated are correlated with the thermodynamical calculations of dust grain condensation in diverse stellar environments ranging from the evolved AGB stars, supernovae II, Ia of varied metallicities to the accretion disk associated protostars where planet formation occurs, thereby, enabling us to make an assessment regarding the abundance evolution of dust in the Galaxy. In the following, firstly, we present our GCE model for estimating the elemental abundance evolution of the Milky Way Galaxy. This is followed by our mass-balance calculations and the thermodynamical condensation models for a wide range of stellar evolutionary phases. Finally, we present the mapping of these two approaches in order to make an attempt to deconvolute the contributions of various evolved stellar sources to the ISM dust inventories.

### 2.1 The elemental abundance evolution across the Galaxy

Stars in different mass ranges evolve distinctly. Low to intermediate-mass stars (< 8 $M_\odot$) evolve as red giants (RG) and asymptotic giant branch (AGB) stars. The stars in the mass range, 11-100 $M_\odot$ evolve through core-collapse supernovae (SN II). The massive stars (30-100 $M_\odot$) evolve through Wolf-Rayet (WR) phase prior to their explosion as a supernova (SN Ib/c). The massive stars evolve faster than the low mass stars, thereby, quickly recycling the galactic matter. The supernovae SN Ia are considered to be associated with the explosion of a white dwarf star subsequent to the accretion from a binary companion. The nucleosynthetic yield from various stellar sources is an important input component in both the mass-balance calculations and the thermodynamical equilibrium condensation calculations. In order to perform the mass-balance calculations for the entire Milky Way Galaxy, we have used the recently developed GCE model (Sahijpal & Gupta 2013; Sahijpal & Kaur 2018). Based on this model, we radially partitioned the Milky Way into eight concentric annular rings, initializing from 2 kpc to 18 kpc distance from the Galactic centre. The width of each annular ring was taken to be 2 kpc. The solar neighbourhood is represented by the 8-10 kpc annular ring. We adopted the homogeneous GCE model, whereby each annular ring evolved independently as a single unit due to instant assumed homogenization. We adopted the essential features of the Model-A of the GCE model developed by Sahijpal & Kaur (2018). The Galaxy was assumed to accrete by two distinct episodes involving the Halo-thick disk phase, followed by the thin disk phase. While the former phase lasted over the initial one billion years, the latter phase led to the gradual accretion of the Galaxy over several billion years. The Galactic thin disc was formed by assuming an inside-out accretion criterion, whereby, the inner regions rapidly accreted matter compared to the outer regions (Sahijpal & Kaur 2018). The metallicity of the accreting matter was assumed to be 0.1 times the assumed solar metallicity (Sahijpal & Gupta 2013). We modelled the successive formation and evolution of stars in the mass range 0.1-100 $M_\odot$ in the simulation according to the prevailing star formation rate that was assumed to be a function of the surface mass densities of gas and stars. The star formation rate was assumed to be a function of the Galactocentric distance of an annular ring and time (see e.g., Figure 1 and Table 1 of Sahijpal & Kaur 2018). The choice of the simulation parameters dealing with the star formation rates was made to reproduce the observed metallicity gradients across the Galaxy. The stars were formed according to the IMF prescribed by Sahijpal & Kaur (2018). Subsequent to the evolution of the stars, the stellar nucleosynthetic yields carried out by stellar ejecta were homogenized over the annular ring. We adopted the stellar nucleosynthetic yields of low and intermediate-mass AGB stars (0.8-8 $M_\odot$) from Karakas and Lattanzio (2007). The nucleosynthetic yields of massive stars (> 11 $M_\odot$) were taken from Woosley & Weaver (1995). The SN Ia yields were taken from Iwamoto et al. (1998). The simulation was run for 13.5 billion years, with a temporal resolution of one million years, and a constraint to reproduce a metallicity ($Z_\odot$) of 0.019 (Anders and Grevesse 1989) and [Fe/H] =0 at the time of the solar system formation around 4.5 Gyr ago in the solar annular ring at an assumed distance of 8-10 kpc from the Galactic centre. This was achieved by appropriately parameterizing the fraction, $f_{SNIa}$, of low- and intermediate-mass stars that evolve as stars in binary systems that eventually explode as supernovae SN Ia (Sahijpal & Kaur 2018). This fraction was assumed to be 0.025. The supernovae SN Ia are the main contributor of iron in the Galaxy. Due to the uncertainty in the revised value (Asplund et al. 2009) of the solar metallicity (Sahijpal & Gupta 2013; von Steiger & Zurbuchen 2015; Vagnozzi 2019), we assumed the pre-revised value (Anders & Grevesse 1989) in the present work. This is distinct from the recent work by Sahijpal & Kaur (2018). However, it should be noted that the choice of the value of solar metallicity will not significantly influence the significant conclusions drawn from the present work.

## 2.2 Mass-balance calculations and relative abundance of dust grains

Based on the elemental abundance evolution trends estimated for the Galaxy, we performed mass-balance calculations to approximate the most probable estimates of the composition of dust grains that could condense. The estimates were obtained at 15 different epochs for the Galaxy across the 8 annular independent rings of width 2 kpc each from 2-18 kpc distance from the Galactic centre. A formalism has been developed which evaluates the approximate relative dust mass with the evolution of the Galaxy in an internally consistent manner. In order to consider the entire wide range of condensable elements, we performed the mass-balance calculations for 12 elements (H, C, O, Na, Mg, Al, Si, S, K, Ca, Ti, and Fe). These are the major constituents of grains generally observed to condense in ISM. In addition, we assumed 22 distinct compositions of grains along with the possibilities of icy mantles. This would represent an ensemble of dust grains that are likely to condense in astrophysical environments such as the accretion disks associated with protostars. The condensation of dust in these environments eventually leads to the formation of planets. In the considered annular ring at a specific epoch, we started the mass-balance calculations by considering the mass fractions of all the stable isotopes till iron with the total gas mass normalized to unity. We arranged all the considered elements in increasing order of abundance and volatility. Subsequently, the grain compositions were sorted in correspondence to the elements. In order to estimate the general composition of dust, we introduced a *fractional parameter, $f_{Xg}$*, which characterizes the condensed mass fraction of an element $X$ for grain composition '$g$' that will be subsequently removed from the gas phase. This is a free parameter and can be assumed to have any value between 0 and 1. Its value is assumed in a manner so as to take care of the abundance order of elements and the probability of formation of dust grains based on their volatility. The combination of the order of the abundance and refractory nature does not produce a unique value of the fraction, so we imposed another constraint on it. We assumed a value of a *fractional parameter* such that the total dust-to-gas mass ratio remains within the observationally modelled estimates as given by equation (4) of Giannetti et al. (2017). The mass of grain that can condense is always constrained by its least abundant constituent element because that gets exhausted before all the other constituents. Therefore, the fraction value $f_{Xg}$ was chosen only for the least abundant grain constituent corresponding to every grain. If $l$ is such a constituent element of the grain '$g$', then its mass fraction $m_{lg}$ condensed in the considered environment is given in equation (1):

$$m_{lg} = f_{lg} M_{li} \qquad (1)$$

Here, $M_{li}$ is the initial mass fraction of element $l$ in the given chemical composition at the iteration $i$, and $f_{lg}$ represents a corresponding fraction of the element that is condensed corresponding to grain '$g$'. Using this condensed mass of the grain constituent, we calculated the mass of the grain composition using equation (2):

$$M_g = \frac{m_{lg}}{W_{lg}} \qquad (2)$$

Here, $M_g$ represents the normalized mass of the condensed grain, and $W_{lg}$ represents the mass fraction of the least abundant element '$l$' in the corresponding grain '$g$'. It is calculated using equation (3):

$$W_{lg} = v_{lg} \frac{A_l}{W_g} \qquad (3)$$

Here, $v_{lg}$ is the valency of an element in the grain condensate, $A_l$ is the atomic weight of an element and $W_g$ represents the molecular weight of the grain condensate. With the condensation of grain composition $g$, a fix fraction of all the constituent elements gets locked in that grain. The condensed mass of constituent element '$k$' other than the least abundant constituent element '$l$' in any grain '$g$' can be calculated using equations (2) and (3). The formula is given in equation (4):

$$m_{kg} = M_g W_{jg} \qquad (4)$$

The condensed mass $m_{kg}$ of element '$k$' is subtracted from the total mass fraction of that element available in the system, and the remaining mass of the element is calculated at every iteration step. Thus, the remaining mass fraction $M_{k(i+1)}$ of the element '$k$' in the gas phase at $i+1$ iteration is calculated as:

$$M_{k(i+1)} = M_{ki} - m_{kg} \qquad (5)$$

Here, $M_{ki}$ represents the remaining mass fraction of the element '$k$' in the gas phase at $i^{th}$ iteration. The remaining mass is utilized in the formation of all other consecutive grains and the redistribution proceeds. The grains are formed following the increasing order of their volatility, and the elements are exhausted following the increasing order of their abundance in the system. The process is iterated until the formation of all the grain compositions. The grain masses are calculated at every step in an internally consistent manner. Based on the composition and the thermodynamical condensation formulation, the grains are categorized as silicates, oxides, carbonaceous, and iron-dust. Finally, the separately calculated grain masses are added up in different groups to obtain the condensed dust mass of the corresponding group. The present work aims to provide an overall assessment of the composition of the dust. It should be noted that various grain growth and destruction processes, such as accumulation or thermal sputtering that occur in ISM, have a significant impact on the size and the composition of the dust. Although the processes are important in accurately modelling dust composition and abundance evolution, these processes are not considered in the present work. However, the dust mass that finally survives in ISM after reprocessing is ~1 per cent of the gas mass.

### 2.3 Thermodynamic equilibrium condensation calculations of dust grains

Gupta & Sahijpal (2020) recently developed a condensation model based on the thermodynamical approach, which includes equilibrium as well as non-equilibrium scenarios of dust condensation. The model provides the condensation sequences, condensation reactions, condensation temperatures, and the normalized masses of the condensed dust grains in any astrophysical environment. In the present work, we have performed the thermodynamical condensation calculations to understand the stellar origin of the various grain constituents that could carry the isotopic signatures of distinct nucleosynthetic sources. Here, we avoid a detailed discussion of the condensation model. However, numerous features of the numerical simulations are summarized in brief.

The modelling of equilibrium condensation calculations involves simultaneous solution of the non-linear mass-balance equations of the elements and the mass-action equations of the condensed species (Grossman 1972; Lattimer et al. 1978; Ebel et al. 2000). The program involves a mass-balance equation corresponding to each element with their partial pressures as the variables. A set of 20 abundant elements (H, He, C, N, O, F, Na, Mg, Al, Si, P, S, Cl, K, Ca, Ti, Cr, Fe, Co, and Ni) was considered for thermodynamical calculations. By simultaneously converging the mass-balance equations of the elements, we computed their partial pressures. The modified Powell method was used to achieve the point of convergence (Chen & Stadtherr 1981), which like other root-finding techniques requires an initial guess to start with. In order to fulfil the requirement of initial guess in such a vast parametric space, we exploited the Monte Carlo approach. At an initial temperature step, the random numbers were used as an initial guess with a strict chi-square test to ensure the global convergence. The random numbers were generated by using a library named 'random' in python (version 3.5) that uses the Mersenne Twister core generator, one of the most extensively tested generators. At successive temperature steps, the result of the current iteration was used as an initial guess for the next iteration and the partial pressures of the elements were calculated.

Using the partial pressures of the elements from which a condensate is formed, the Gibbs energy value of the formation of that condensate was calculated. Also, the Gibbs energy value of the condensate was calculated from the available thermodynamical data. The stability of the solid species was assessed by the comparison of these two Gibbs energy values at every temperature step. The mass-action equation of condensate was added to the system when the condensate was found to be stable. This numerical process continued in agreement with the thermodynamical laws that allow the system to evolve with decreasing temperature. The appeared condensate was removed from the system when it was found in a vanishingly small amount. The entire process was repeated for the modified system assemblage in which the matter was redistributed among the available phases. We iterated the calculations to successively lower temperatures for a bulk system composition at an assumed total pressure.

The thermodynamical data library is an important input component of the condensation calculations. The details of the species considered in the gaseous and solid phases and the sources of their thermodynamic data are the same as that of a recently developed model (Gupta & Sahijpal 2020). It should be mentioned that some minor changes were incorporated in the thermodynamical library while running the simulations in the present work. The changes consist of the removal of some complex solid species from the data library which are not observed in meteorites. Apart from this fassaite was not considered as a solid-solution. Further, the thermodynamical code developed in python deals with the equilibrium as well as non-equilibrium condensation scenarios. The thermodynamic non-equilibrium condensation scenario generally results in the systematic isolation of a fraction of the earlier condensed grains from the system. The residual fraction of the dust grain remains available and can interact with the gas in the system. The non-equilibrium scenarios, in general, show the same condensation sequence as is shown by equilibrium scenarios (Tanaka et al. 2002; Gupta & Sahijpal 2020). The continuous isolation of a fraction of dust leads to a lesser number of 'late' condensates in the non-equilibrium scenarios. Because the type of condensates remains almost the same in both the scenarios except for the 'late' condensates, the present work deals only with the equilibrium condensation scenarios.

The dust grain condensation depends upon the pressure, temperature, and chemical composition of the environment. All the simulations were performed at a pressure of $10^{-5}$ bar. This pressure value represents an intermediate pressure between the values leading to liquid stability on one side and larger gaseous stability fields on the other side. Moreover, the effect of pressure is primarily on the condensation temperature only. The condensation sequence just gets shifted downwards or upwards in terms of temperature with the decrease or increase in pressure, respectively. Liquids get stabilized above a pressure of $10^{-2}$ bar (Yoneda & Grossman 1995), which may produce significant changes in the condensation sequence. Below this pressure, the condensation calculations produce the condensation sequence to an appropriate degree. Apart from this, the chemical compositions obtained using the recently developed GCE model (Sahijpal & Kaur 2018) were fed to the condensation model as an input. The thermodynamical code (Gupta & Sahijpal 2020) simulated the diverse stellar environments and produced the distinct condensation sequence of the grains. Distinct nucleosynthetic yields produced various types of grains with distinct abundances. In order to understand the composition and stellar sources of various dust grains, a detailed thermodynamical analysis is required for each annular ring and temporal epoch of the Galaxy. Rather than performing a detailed analysis of each star during its various stages, this study aims to provide the overall production of dust grains and the most prominent stellar sources where the dust could condense.

### 2.4 Mapping of the two distinct approaches for dust condensation & the Galactic dust inventories

As already mentioned, we have estimated the dust formation and dust inventory evolution using two distinct approaches. The code developed in this study consists of internally consistent mass-balance calculations that are further supported by the thermodynamical equilibrium condensation calculations. In the mass-balance calculations, we considered a basis matrix of four groups categorized as oxides-, silicates-, carbonaceous-, and iron-dust consisting of 22 distinct compositions of dust grains. The 22 distinct compositions were extracted out by running the thermodynamical models for the accretion disks associated with protostars and thus, represent an ensemble of dust grains that are plausible to be formed in the considered environments. Some of the grains, particularly iron-type, are part of more than one basis matrix. For instance, $FeAl_2O_4$ falls in oxide-type as well as in iron-type dust. It should be noted that the oxides- and silicates-dust are chosen as disjoint sets. We have not considered the condensation of Polycyclic Aromatic Hydrocarbons (PAHs) in our models. We performed mass-balance calculations for fifteen temporal epochs at eight distinct annular rings of the Galaxy. A value of the fractional parameter was chosen corresponding to each annular ring-temporal point over the entire Galaxy. In addition, we also represent three basis matrices whose components are given by thermodynamical models simulated for SN Ia, SN II+Ib/c and AGB stars. In the present work, we have not considered various evolutionary stages of different stellar environments in order to avoid further complications. Thus, the chemical compositions of the supernovae ejecta and the AGB stellar winds were deduced using the homogenized approach. This homogenization also influences the formation of the basis vectors. Therefore, the complete injective mapping of mass-balance and thermodynamical basis vectors is not possible due to the variation in the formation of basis vectors. A detailed analysis based on the thermodynamical approach adopted for various stages of Wolf-Rayet stars (Gupta & Sahijpal 2020) as an example could be more useful in future works. Since the present work is first of its kind that quantitively investigates the general composition evolution of dust and its origin, we present an approach based on homogenized stellar ejection.

## 3 RESULTS

We have performed calculations to estimate the abundance evolution of various grain compositions in the Milky Way Galaxy. The simulations involve mass-balance calculations at distinct epochs of the Galaxy evolution over its entire radial extent. We considered 15 temporal epochs in 8 annular rings of a width of 2 kpc each. The mass-balance calculations were performed for each of the rings corresponding to every epoch. The normalized mass fractions for all the considered dust grain compositions within each annular ring and temporal epoch are presented in Table 1, with the full details available in the Supporting Information. The Galactic gas surface mass density (in $M_\odot$ pc$^{-2}$) is also provided in Table 1. The absolute Galactic dust surface mass density can be obtained by multiplying the gas mass density in an annular ring of the Galaxy with its corresponding normalized dust mass. The 22 distinct compositions are categorized into four distinct groups as silicates-, oxides-, iron-, and carbonaceous-dust. The list of dust grains considered in the basis groups is presented in Table 2. The temporal evolution of the normalized mass distribution of distinct dust grain basis components over the extent of the entire Galaxy is shown in Fig. 1. It contains six panels 1a–f corresponding to the normalized masses of the oxides-, carbonaceous-, iron-, silicates-, and total-dust, and the gas surface mass density. The panels labelled 1a–e represent the temporal evolution of the distinct normalized dust masses of the basis-components. Along with this temporal evolution, the pie graphs representing the distribution of the grain in the corresponding basis components at 9 and 13.5 Gyr for 8-10 kpc distance from the Galactic centre is shown in the panels labelled 1a–d. The pie graphs in the panel labelled 1e show the composition of dust in the solar neighbourhood. The smaller pie graphs represent the composition at the time of formation of the solar system, and the larger pie graphs represent the present time in all the panels. The gas surface mass density is also plotted in Fig. 1f. As already mentioned, the absolute dust mass density can be directly calculated by multiplying the gas mass density with the corresponding normalized mass fraction of the dust grain. The spatial mass distribution of dust is plotted in Fig. 2. It contains five panels 2a–e corresponding to dust mass in the form of oxides-, carbonaceous-, iron-, silicates- and total-grains labelled accordingly. The initial gas mass was normalized to one for all the nucleosynthetic yields considered in mass-balance calculations. The results obtained from various calculations are tabulated for all the compositions (see Tables 3–10). Column 1 represents the epoch value and columns 2 to 6 represent the corresponding oxides-, carbonaceous-, iron-, silicates- and total-dust normalized masses.

In addition, the simulations include thermodynamical equilibrium condensation calculations for distinct stellar populations. As already explained in Subsection 2.3, only the equilibrium calculations were performed in this work owing to our experience gained by modelling condensation of grains in Wolf-Rayet stellar winds. The stellar environments modelled thermodynamically include supernovae (SNe II+Ib/c) ejecta with four distinct metallicities, SN Ia ejecta and stellar winds of low- to intermediate-mass stars which evolve as RG and AGB phases. Along with these calculations, we also simulated four distinct extreme combinations of the annular ring and temporal epochs of the Milky Way Galaxy that could represent the accretion disk associated protostars (Figure 3). These include environments associated with protostars at different times and annular rings. These calculations can be performed for any space-time combination in the evolving Galaxy, and it will represent the most likely composition of the dust that would condense in the vicinity of a protostellar environment. The abundances used in all these models were normalized with respect to silicon. All the sets of simulations were performed at an assumed initial total pressure of $10^{-5}$ bar. The relative elemental abundances and the initial total pressure of the system collectively determine the partial pressure of the elements. Thus, distinct minerals stabilize accordingly producing distinct condensation sequences that were studied for all the models. The variations in the appearance and disappearance temperatures and mass distributions of the condensates were studied for the distinct stellar clusters. The normalized mass distribution of the condensed phases and gas phases for the considered stellar environments are shown in Figs 3–8. Figs 3 and 5–8 representing the normalized mass distribution of condensates, each contain four panels with different considered models. Fig. 3 represents the mass distribution of the condensed phases for the Milky Way Galaxy. Fig. 4 represents the mass distribution of the condensates and the gaseous species for the supernovae SN Ia in panels labelled Fig. 4a and Fig. 4b, respectively. In Figs 5–7, the panels (a–d) for AGB stars are labelled as "AGB_stellarMass_Metallicity", and in Fig. 8, panels are labelled as "SNII_Metallicity".

The results obtained from the various simulations are tabulated in the form of the temperatures of the appearance (and disappearance) of various condensates for all the compositions (see Tables 11a–16a). The condensates are written by their names for solid solutions and by their formulas for pure solid phases. The various columns in Tables 11a–16a represent the name of the condensates and the condensation temperatures. The condensation temperature represents the value of the temperature at which the condensed phase stabilizes for the first time for a given composition. The peak values of the normalized mass of stable condensates in the simulated temperature range are also provided (see Tables 11b–16b). The various columns in Tables 11b–16b give the name of the condensates and the maximum value of normalized mass that the condensates attain at a certain temperature. The condensation sequence and peak normalized mass for four distinct compositions in the Galaxy are presented in Tables 11. Column 1 represents the name of the condensates that become stable with decreasing temperature, and columns 2 to 5 represent the corresponding condensation temperatures in Tables (a) and the peak mass value in the corresponding Tables (b) for the models considered in the present work. Table 12 represents the results for supernovae SN Ia. Columns 1–4 in Table 12 represent the name of the condensates, the corresponding appearance temperature, the disappearance temperature, and the peak mass value attained, respectively. The results for the AGB stellar models at 0.019 and 0.0001 metallicities are presented in Tables 13–15 for the considered masses. Tables 16 presents the results for the supernovae SNe II+Ib/c at four distinct metallicities.

We performed the thermodynamical simulations of a system assemblage by initially assuming it to be completely homogenized and vaporized. Before the onset of stability of any solid phase, different elements remained in their stable gaseous forms according to the relative chemical composition of the considered environment. For instance, silicon remained in SiO, $SiO_2$ forms in O-rich environments such as AGB stellar models at solar metallicity and in Si, $SiC_2$ forms in C-rich environments such as AGB stellar models at $10^{-4}$ metallicity. Similarly, titanium remained mostly in a monatomic form in C-rich environments, whereas it remained in TiO and $TiO_2$ forms in the case of O-rich environments. Carbon and oxygen combined to form CO molecules. The more abundant element out of these two is found in other forms. Aluminium mostly remained in monatomic Al form, and a small part of it remained in other forms such as AlCl, $AlO_2$. Calcium, magnesium, and iron mostly remained in their monatomic form. The results are discussed in detail in the following section.

## 4 DISCUSSION

The purpose of this work is to understand the general composition of dust grains and the basic trends in their abundance evolution in the Galaxy. This is understood on the basis of galactic chemical evolution (GCE) of the elements. An internally consistent mass-balance formalism is developed in this study to deduce dust formation and evolution in the Milky Way Galaxy. This newly developed formalism is further supported by the thermodynamical analysis of dust grains condensation at different epochs across the Galaxy. In addition to the radial and temporal evolution of dust, this study also attempts to understand the contributions of dust from various stellar sources at a qualitative level. This is estimated by performing the thermodynamical equilibrium condensation calculations for the winds of evolved stars and the supernovae ejecta. The thermodynamical condensation code is already tested and verified in our previous study (Gupta & Sahijpal 2020). Subsequent to the validation of codes, we ran a set of simulations for the entire Galaxy and distinct stellar clusters. The significant features that can be extracted from the systematic study of various simulations are discussed below.

### 4.1 Temporal evolution of dust in the Galaxy

The mass-balance calculations developed in this work were performed for all the epochs considered over the Galactic time-scales. Fig. 1a–e represents the normalized mass distribution of dust grain basis components as a function of temporal epochs at different radial distances from the Galactic centre. The calculations infer an increase in the mass of the condensed dust in time from the Big-Bang origin of the Universe around 13.7 Gyr ago. With the temporal evolution of the Galaxy, the stellar nucleosynthesis enriches the ISM with the heavier elements. Since, the heavier elements are more refractory, in general, the normalized dust production is enhanced with the evolving chemical composition of the Galaxy. The rapid increase in the dust mass production, specifically during the initial 1 Gyr, is due to the assumed enhanced star formation rates during the initial 1 Gyr that corresponds to the Halo-thick disc stage of the Galaxy (Sahijpal & Kaur 2018). This results in a rapid contribution of stellar nucleosynthetic contributions of condensable elements from SN II+I b/c to the Galaxy. The massive stars (> 11 $M_\odot$) that eventually explode as SN II+I b/c, generally contribute to the Galaxy over short timespans of tens of million years after their formation in stellar clusters. The dust mass fraction acquires a maximum of around ~1 Gyr for the various annular rings and drops subsequently. The maxima are marked by the transition from the Halo-thick disc stage to the thin disc stage of the accreting Galaxy. We have assumed the accretion of the Galaxy in two episodes, the Halo-thick disc stage lasting for the initial ~1 Gyr that is followed by a prolonged accretion of the thin disc. The rapid accretion of low metallicity (0.1×$Z_\odot$) gas around the beginning of the thin disc stage of the Galaxy (~1 Gyr) lowers the average prevailing metallicity of the Galaxy, thereby, resulting in the sharp decline of the dust condensable material. This results in a rapid reduction in dust production (Fig. 1a-e) around 1-1.5 Gyr. The reduction in the normalized dust production correlates with the increase in the gas mass density (Fig. 1f) during the period on account of accretion of low metallicity matter. It should be noted that the maximum drop in the dust production takes place in the inner regions of the Galaxy compared to the outer regions. This is essentially due to the assumed inside-out accretion scenario of the Galaxy (Sahijpal & Kaur 2018), whereby, the inner regions of the Galactic thin disc accrete rapidly compared to the outer regions. Eventually, with the gradual accretion of the thin disc, the metallicity begins to increase on account of stellar nucleosynthetic contributions from SN II+I b/c, and hence, the dust production increases over the Galactic timescales. (Fig. 1; Tables 3–10). Thereafter, the normalized dust mass gradually increases for subsequent epochs.

The Fig. 1 also indicates that the normalized oxides dust mass is the smallest, and the silicate dust mass is largest compared to other types of dust at all epochs. Also, the oxides dust, which is chosen as a disjoint set of silicates dust, is slightly less than the carbonaceous dust. Iron dust mass falls between carbonaceous- and silicate-dust masses. As the GCE model adopted in the present work is homogeneous and all the annular rings are O-enriched, the formation of carbonaceous grains is less probable. This also signifies that if oxygen is more abundant than carbon, the dust remains mainly in silicate form. Oxides dust other than silicates is less probable to form. Thus, all the annular ring compositions primarily produce silicates- and iron- dust masses.

**Table 1.** The normalized masses for all the considered dust grain compositions within each annular ring and temporal epoch.

| Annular ring 1 (2-4 kpc) | | | | |
|---|---|---|---|---|
| Time (Gyr) | 0.2 | 0.4 | 0.6 | 0.8 |
| Gas surface mass density | 2.21 | 2.87 | 3.04 | 3 |
| $CaTiO_3$ | $1.64 \times 10^{-6}$ | $2.71 \times 10^{-6}$ | $3.53 \times 10^{-6}$ | $4.20 \times 10^{-6}$ |
| $Ti_2O_3$ | $9.64 \times 10^{-9}$ | $1.59 \times 10^{-8}$ | $2.07 \times 10^{-8}$ | $2.47 \times 10^{-8}$ |
| TiC | $7.23 \times 10^{-8}$ | $1.20 \times 10^{-7}$ | $1.56 \times 10^{-7}$ | $1.85 \times 10^{-7}$ |
| $Al_2O_3$ | $1.14 \times 10^{-5}$ | $2.24 \times 10^{-5}$ | $3.09 \times 10^{-5}$ | $3.77 \times 10^{-5}$ |
| $CaAl_4O_7$ | $8.71 \times 10^{-6}$ | $1.71 \times 10^{-5}$ | $2.37 \times 10^{-5}$ | $2.88 \times 10^{-5}$ |
| $Ca_2SiO_4$ | $1.08 \times 10^{-5}$ | $1.63 \times 10^{-5}$ | $2.00 \times 10^{-5}$ | $2.27 \times 10^{-5}$ |
| $Ca_2Al_2SiO_7$ | $2.07 \times 10^{-5}$ | $3.12 \times 10^{-5}$ | $3.81 \times 10^{-5}$ | $4.34 \times 10^{-5}$ |
| CaS | $5.09 \times 10^{-6}$ | $7.67 \times 10^{-6}$ | $9.37 \times 10^{-6}$ | $1.07 \times 10^{-5}$ |
| $Mg_2SiO_4$ | $2.08 \times 10^{-4}$ | $3.24 \times 10^{-4}$ | $3.96 \times 10^{-4}$ | $4.48 \times 10^{-4}$ |
| $MgAl_2O_4$ | $9.83 \times 10^{-7}$ | $1.53 \times 10^{-6}$ | $1.87 \times 10^{-6}$ | $2.11 \times 10^{-6}$ |

*Here, only the first few rows and columns are given. The full table is available online in the Supporting Information as Table S1.

**Table 2.** The list of dust grains considered in distinct five basis components.

| | |
|---|---|
| Total dust | $CaTiO_3$, $Ti_2O_3$, TiC, $Al_2O_3$, $CaAl_4O_7$, $Ca_2SiO_4$, $Ca_2Al_2SiO_7$, CaS, $Mg_2SiO_4$, $MgAl_2O_4$, MgS, $Ca_2MgSi_2O_7$, $MgSiO_3$, SiC, $FeSiO_3$, Fe-metal, $FeAl_2O_4$, $Fe_3C$, FeS, $KAlSi_3O_8$, $NaAlSi_3O_8$, Graphite, $H_2O$ |
| Silicate dust | $Ca_2SiO_4$, $Ca_2Al_2SiO_7$, $Mg_2SiO_4$, $Ca_2MgSi_2O_7$, $MgSiO_3$, $FeSiO_3$, $KAlSi_3O_8$, $NaAlSi_3O_8$ |
| Oxides dust | $CaTiO_3$, $Ti_2O_3$, $Al_2O_3$, $CaAl_4O_7$, $MgAl_2O_4$, $FeAl_2O_4$ |
| Iron dust | $FeSiO_3$, Fe-metal, $FeAl_2O_4$, $Fe_3C$, FeS |
| Carbonaceous dust | TiC, SiC, $Fe_3C$, C |

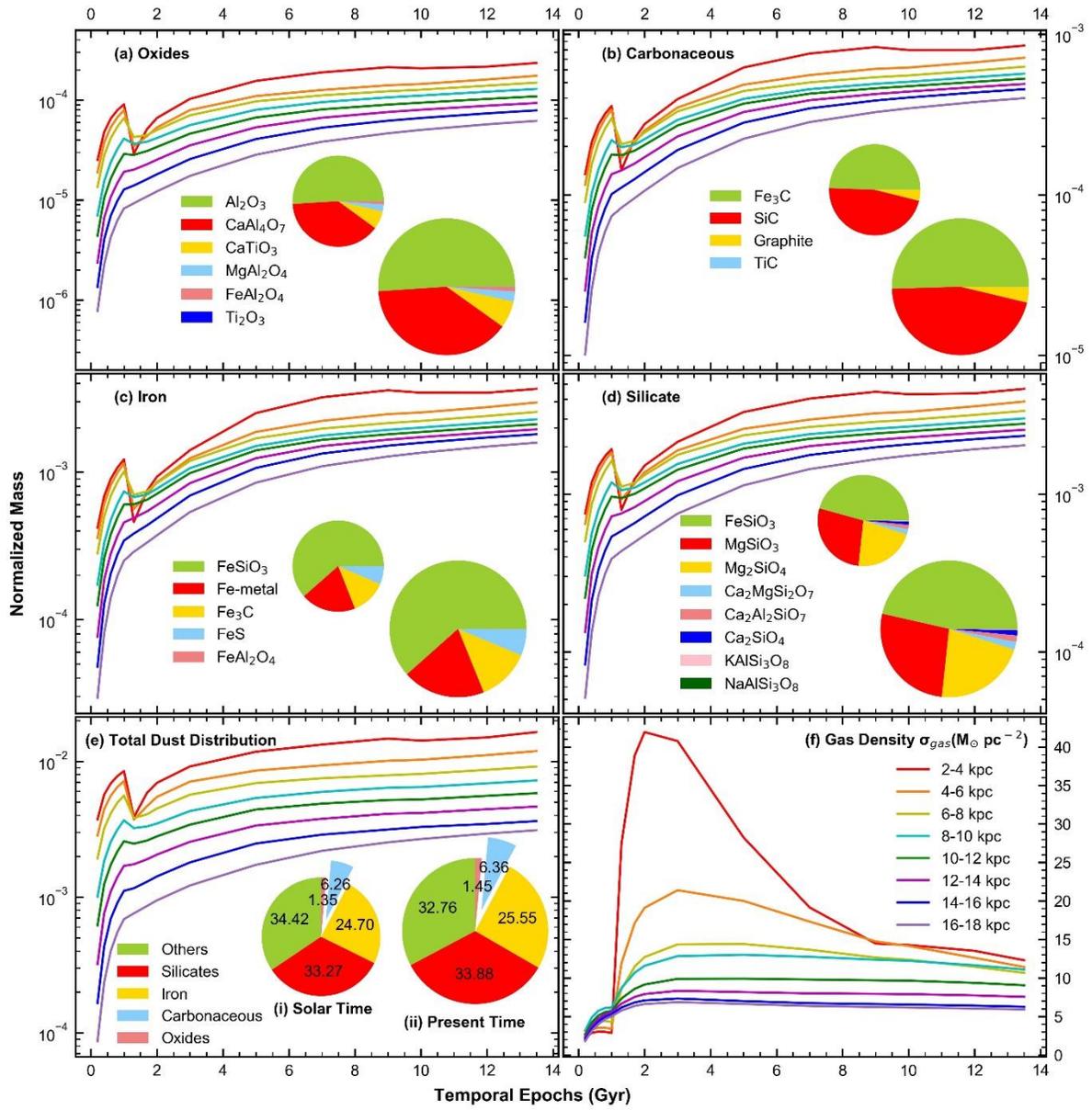

**Figure 1.** Temporal evolution of the normalized mass distribution of distinct dust grain basis components corresponding to the eight annular rings of the Milky Way Galaxy. The pie graphs show the distribution of the grain in the corresponding basis components at 9 and 13.5 Gyr for 8-10 kpc distance from the Galactic centre. The pie graphs in the panel 'e' show the composition of dust in the solar neighbourhood. The smaller pie graphs represent the formation time of the solar system ~4.5 Gyr ago, and the larger pie graphs represent the present time. The panel 'f' represents the temporal evolution of the gas surface mass density corresponding to the eight annular rings of the Galaxy. The absolute dust surface mass density can be obtained by multiplying the gas surface mass density with the normalized dust mass. The data points in the figure are taken only at 15 specific temporal epochs for mass balance calculations.

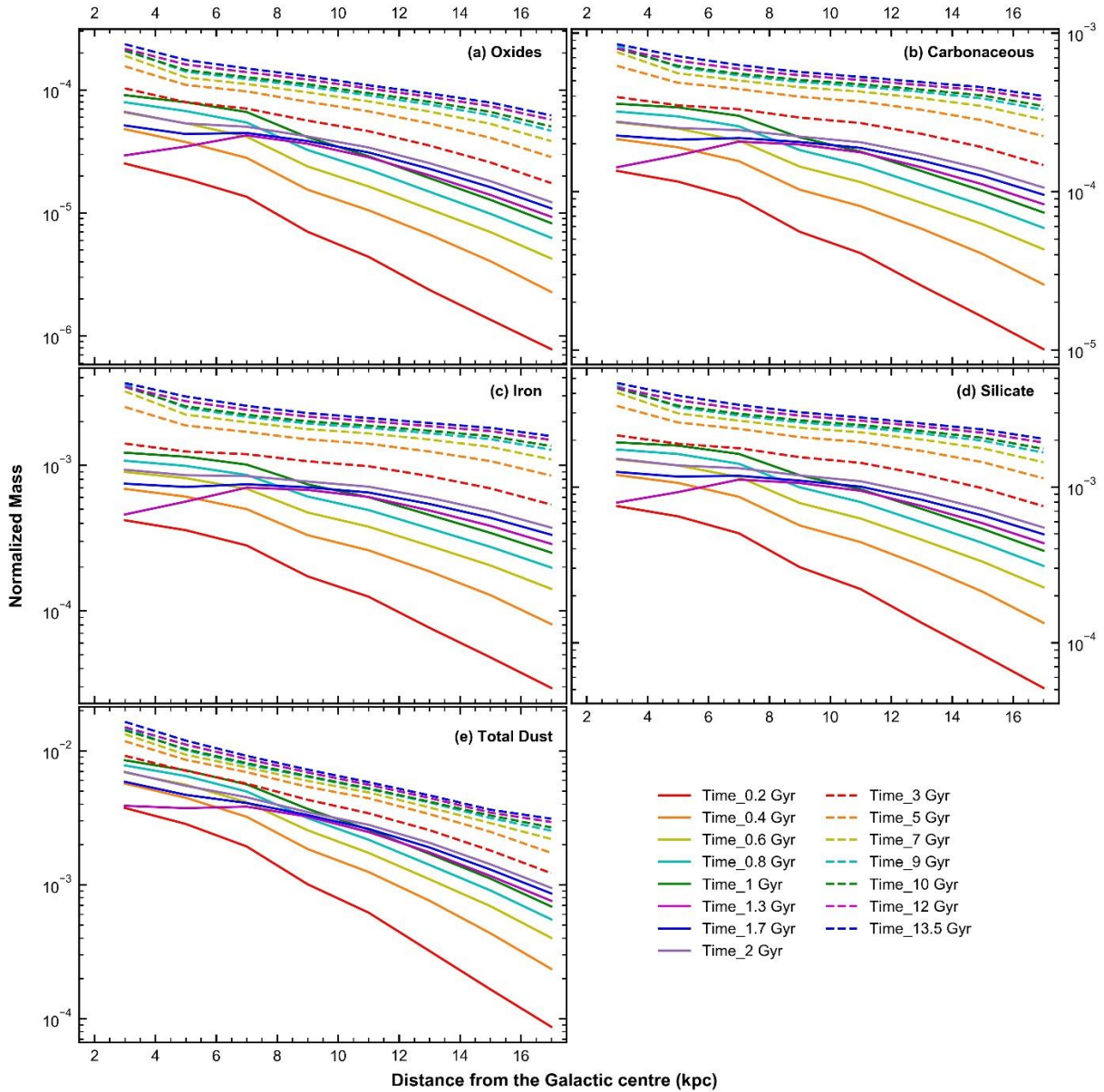

**Figure 2.** Spatial evolution of the Galactic gradients of the normalized mass distribution of distinct dust grains at 15 distinct temporal epochs of the evolution of the Milky Way Galaxy from the Big-Bang origin of the universe.

In order to understand the evolution of distinct groups in the Galaxy, we also performed thermodynamical calculations for distinct epochs. The temporal evolution of dust was explored for three distinct epochs at a distance of 2–4 and 16–18 kpc from the Galactic centre. The mass distributions of condensed phases are presented in Fig. 3. The panels labelled (a) and (c) in Fig. 3 infer the temporal evolution from initial time (~1 Gyr) to the time of solar system formation (~9 Gyr) at the outermost annular ring at 16–18 kpc distance from the Galactic centre. The panels labelled (b) and (d) infer the temporal evolution from solar system formation (~9 Gyr) to the present time (~13.5 Gyr) at the innermost annular ring at 2–4 kpc distance from the Galactic centre. The condensation of dust grains depends on the chemical composition of the system assemblage. Thus, different stellar environments having different chemical compositions produce different condensation sequences. The temperatures of the appearance and disappearance of all the condensates for these simulations are summarized in Table 11a. The corresponding peak values of the normalized mass of condensed phases are summarized in Tables 11b. The shift in condensation temperatures and the variation in dust mass can be seen by comparing the panels (a) and (c) for annular ring 8 and the panels (b) and (d) for annular ring 1 (Fig. 3; Tables 11).

It can be inferred from Fig. 3 that more dust is condensed at later epochs. The condensation sequences are almost the same in all these compositions. But the condensation temperatures, as well as condensed dust masses, increase as we move forward in time from the temporal origin of the Galaxy to the present time (Tables 11). The gaseous species CO utilizes the total carbon in the system because the C/O ratio is less than one in all the models. This results in the stability of silicates and oxides only. In addition, the thermodynamical calculations predict larger dust mass for silicates- and iron-groups, increasing from carbonaceous to silicates, as inferred by mass-balance calculations. In both calculation approaches, silicate mass is approximately one order of magnitude higher than that of oxide mass.

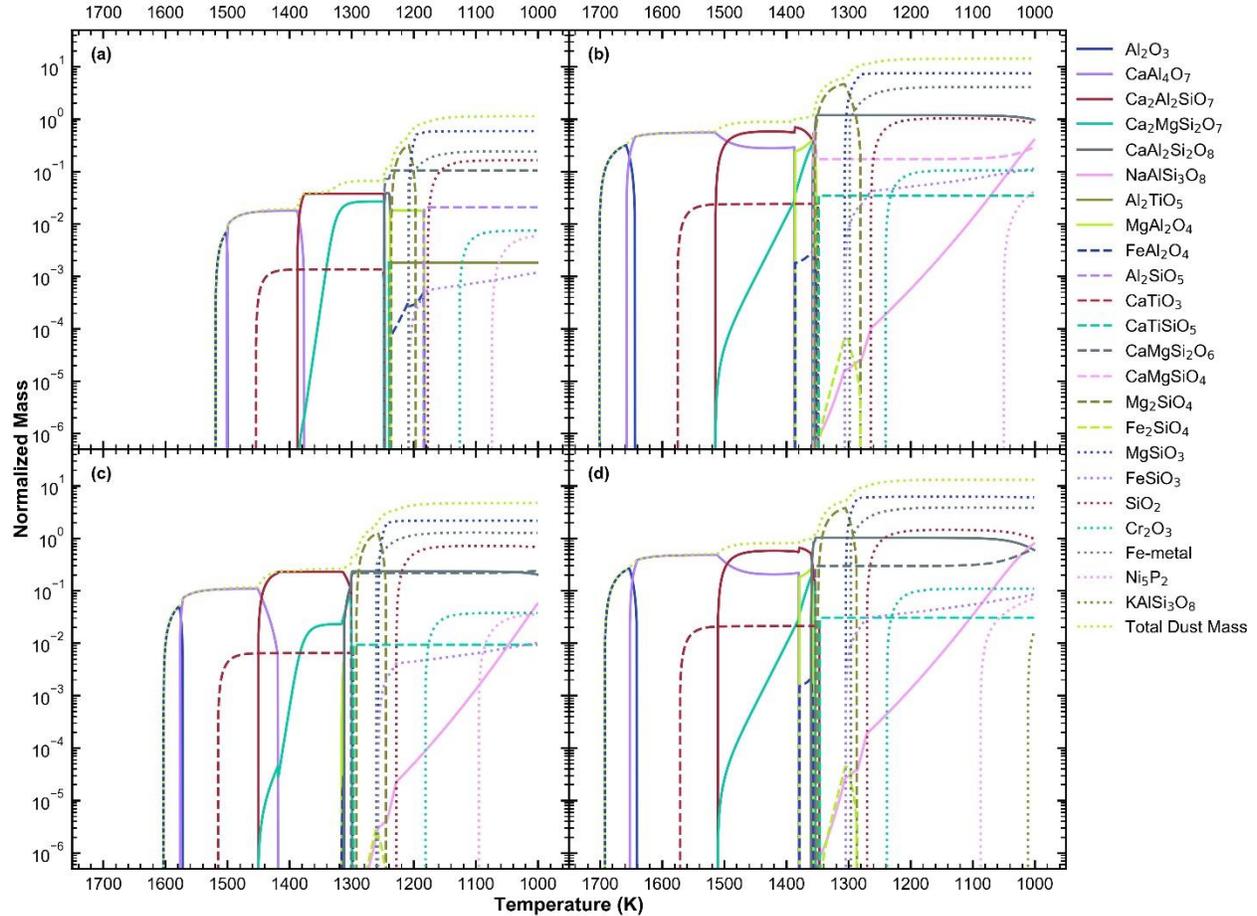

**Figure 3.** The normalized mass distribution of the condensed phases as a function of temperature for the Milky Way Galaxy at, (a) 1 Gyr and 16-18 kpc from the Galactic centre, (b) 13.5 Gyr and 2-4 kpc from the Galactic centre, (c) 9 Gyr and 16-18 kpc from the Galactic centre, (d) 9 Gyr and 2-4 kpc from the Galactic centre at an assumed pressure of $10^{-5}$ bar.

### 4.2 Spatial evolution of dust in the Galaxy

Similar to the temporal evolution of dust mass, the mass-balance calculations were also performed for the entire radial extent of the Galaxy. The dust mass is inferred to decrease with the increase in distance from the Galactic centre. Fig. 2a–e represents the normalized mass distribution of dust grain basis components as a function of annular ring distance from the Galactic centre at different epochs. The panels (a–e) indicate that oxides dust mass is the smallest, and silicate dust mass is the largest compared to other types of dust throughout the Galaxy (Tables 3–10). The carbonaceous dust mass is also very small. It is slightly more than the oxides dust. This is because of the homogeneous model considered for Galaxy evolution. GCE model gives a C/O ratio less than one for all the annular rings that lead to condensation of silicate dust. The order of increasing dust mass from oxides to silicate dust is the same as already discussed in Subsection 4.1.

It should be noted that the mass-balance calculations involve several assumptions and free parameters to deduce the abundance evolution of dust grain. This includes uncertainties related to the elemental abundance evolution of the Galaxy. These uncertainties are elaborately discussed in previous Galactic Chemical Evolution models (see e.g., Sahijpal & Kaur 2018). These uncertainties are mostly related with the assumed scenario associated with the accretion of the Galaxy, the assumed star formation rates across the entire extent of the Galaxy and their time dependence, the assumed stellar initial mass function (IMF), the assumed supernovae SN Ia model prescription, and to an extent the stellar nucleosynthetic yields. The most uncertain parameters specifically dealing with dust condensation include the value of condensed fractions of dust that is removed from the gas phase. Distinct values of these *fractional parameters*, $f_{Xg}$,

$f_{lg}$ can produce a set of distinct solutions. Hence, the solutions are not unique. However, the estimates provided here give the general composition in terms of different groups. The decrease in the fractional value for one component of a group leads to increase in the value for another component. For instance, silicate may be in Ca-, Al- or Mg- forms. Thus, a decrease in the fraction of Ca-silicate may increase the fraction of Al- or Mg-silicates. Similarly, a decrease in the fraction of condensation of Ca in silicate groups may increase its fraction in oxide group. In order to minimize the uncertainties, several logical and observational constraints were imposed on the values of these *fractional parameters* as discussed in Subsection 2.2. For instance, the total dust-to-gas mass ratio were taken according to the available observational data (Giannetti et al. 2017). Although it is difficult to quantify the relative abundance of various dust mass fractions on the basis of observations, a broad composition of the interstellar and circumstellar dust grains can be deciphered on the basis of spectroscopy. For instance, the strong absorption features at 9.7 and 18 µm are correlated with the Si—O stretching, and O—Si—O bending modes in amorphous silicates (Gibb et al. 2004; Mauney & Lazzati 2018). The weak, narrow absorption features at 11, 16, 19, 23 and 28 µm are linked with crystalline silicates (Spoon et al. 2006). The prominent feature at 2200 Å can be approximately reproduced by graphite particles (Weingartner & Draine 2001). SiC exhibits a strong spectroscopic feature at 11.3 µm (Chen et al. 2009).The variation in the mass fractions of different compositions of dust grains in different stars has been observed in several studies (see e.g., Verhoelst et al. 2009 and references therein). For instance, the analysis of observed spectra of the circumstellar dust around two bright Herbig Ae stars, AB Aur and HD 163296, shows a wide variation in the composition with Mg mainly confined to silicates and Fe forming metal, oxides, and sulphides (Bouwman et al. 2000). Narrow to broad silicate features have been observed for O-rich evolved stars (Speck et al. 2000). This study presents a qualitative estimate of the different compositions of dust grains in terms of different groups.

**Table 3.** The normalized mass distribution of distinct dust grains at 15 distinct temporal epochs of the evolution of the Milky Way Galaxy at a distance of 2-4 kpc from the Galactic centre.

| Time (Gyr) | Normalized Dust Mass | | | | |
|---|---|---|---|---|---|
| | Oxide | Carb. | Iron | Silicate | Total |
| 0.2 | $2.5\times10^{-5}$ | $1.3\times10^{-4}$ | $4.1\times10^{-4}$ | $7.5\times10^{-4}$ | $3.7\times10^{-3}$ |
| 0.4 | $4.7\times10^{-5}$ | $2.1\times10^{-4}$ | $6.8\times10^{-4}$ | $1.1\times10^{-3}$ | $5.6\times10^{-3}$ |
| 0.6 | $6.5\times10^{-5}$ | $2.7\times10^{-4}$ | $8.9\times10^{-4}$ | $1.5\times10^{-3}$ | $6.8\times10^{-3}$ |
| 0.8 | $7.9\times10^{-5}$ | $3.1\times10^{-4}$ | $1.0\times10^{-3}$ | $1.7\times10^{-3}$ | $7.7\times10^{-3}$ |
| 1.0 | $9.0\times10^{-5}$ | $3.5\times10^{-4}$ | $1.2\times10^{-3}$ | $1.9\times10^{-3}$ | $8.4\times10^{-3}$ |
| 1.3 | $2.9\times10^{-5}$ | $1.4\times10^{-4}$ | $4.5\times10^{-4}$ | $7.9\times10^{-4}$ | $3.8\times10^{-3}$ |
| 1.7 | $5.1\times10^{-5}$ | $2.2\times10^{-4}$ | $7.4\times10^{-4}$ | $1.2\times10^{-3}$ | $5.8\times10^{-3}$ |
| 2.0 | $6.6\times10^{-5}$ | $2.7\times10^{-4}$ | $9.3\times10^{-4}$ | $1.5\times10^{-3}$ | $6.9\times10^{-3}$ |
| 3.0 | $1.0\times10^{-4}$ | $3.9\times10^{-4}$ | $1.4\times10^{-3}$ | $2.1\times10^{-3}$ | $9.1\times10^{-3}$ |
| 5.0 | $1.5\times10^{-4}$ | $6.1\times10^{-4}$ | $2.5\times10^{-3}$ | $3.3\times10^{-3}$ | $1.1\times10^{-2}$ |
| 7.0 | $1.8\times10^{-4}$ | $7.5\times10^{-4}$ | $3.2\times10^{-3}$ | $4.0\times10^{-3}$ | $1.3\times10^{-2}$ |
| 9.0 | $2.1\times10^{-4}$ | $8.3\times10^{-4}$ | $3.5\times10^{-3}$ | $4.4\times10^{-3}$ | $1.4\times10^{-2}$ |
| 10.0 | $2.0\times10^{-4}$ | $7.9\times10^{-4}$ | $3.4\times10^{-3}$ | $4.3\times10^{-3}$ | $1.4\times10^{-2}$ |
| 12.0 | $2.1\times10^{-4}$ | $7.9\times10^{-4}$ | $3.4\times10^{-3}$ | $4.3\times10^{-3}$ | $1.5\times10^{-2}$ |
| 13.5 | $2.3\times10^{-4}$ | $8.4\times10^{-4}$ | $3.6\times10^{-3}$ | $4.6\times10^{-3}$ | $1.6\times10^{-2}$ |

**Table 4.** As Table 3, but at a distance of 4-6 kpc from the Galactic centre.

| Time (Gyr) | Normalized Dust Mass | | | | |
|---|---|---|---|---|---|
| | Oxide | Carb. | Iron | Silicate | Total |
| 0.2 | $1.8\times10^{-5}$ | $1.1\times10^{-4}$ | $3.5\times10^{-4}$ | $6.4\times10^{-4}$ | $2.8\times10^{-3}$ |
| 0.4 | $3.7\times10^{-5}$ | $1.9\times10^{-4}$ | $6.0\times10^{-4}$ | $1.0\times10^{-3}$ | $4.4\times10^{-3}$ |
| 0.6 | $5.3\times10^{-5}$ | $2.4\times10^{-4}$ | $8.1\times10^{-4}$ | $1.3\times10^{-3}$ | $5.5\times10^{-3}$ |
| 0.8 | $6.7\times10^{-5}$ | $2.9\times10^{-4}$ | $9.8\times10^{-4}$ | $1.6\times10^{-3}$ | $6.4\times10^{-3}$ |
| 1.0 | $7.9\times10^{-5}$ | $3.3\times10^{-4}$ | $1.1\times10^{-3}$ | $1.8\times10^{-3}$ | $7.1\times10^{-3}$ |
| 1.3 | $3.4\times10^{-5}$ | $1.6\times10^{-4}$ | $5.6\times10^{-4}$ | $9.2\times10^{-4}$ | $3.7\times10^{-3}$ |
| 1.7 | $4.3\times10^{-5}$ | $2.1\times10^{-4}$ | $7.0\times10^{-4}$ | $1.1\times10^{-3}$ | $4.6\times10^{-3}$ |
| 2.0 | $5.3\times10^{-5}$ | $2.5\times10^{-4}$ | $8.5\times10^{-4}$ | $1.3\times10^{-3}$ | $5.4\times10^{-3}$ |
| 3.0 | $7.9\times10^{-5}$ | $3.4\times10^{-4}$ | $1.2\times10^{-3}$ | $1.8\times10^{-3}$ | $7.0\times10^{-3}$ |
| 5.0 | $1.1\times10^{-4}$ | $4.8\times10^{-4}$ | $1.8\times10^{-3}$ | $2.6\times10^{-3}$ | $8.5\times10^{-3}$ |
| 7.0 | $1.2\times10^{-4}$ | $5.5\times10^{-4}$ | $2.2\times10^{-3}$ | $2.9\times10^{-3}$ | $9.3\times10^{-3}$ |
| 9.0 | $1.4\times10^{-4}$ | $6.0\times10^{-4}$ | $2.4\times10^{-3}$ | $3.2\times10^{-3}$ | $1.0\times10^{-2}$ |
| 10.0 | $1.4\times10^{-4}$ | $6.1\times10^{-4}$ | $2.5\times10^{-3}$ | $3.3\times10^{-3}$ | $1.0\times10^{-2}$ |
| 12.0 | $1.6\times10^{-4}$ | $6.6\times10^{-4}$ | $2.7\times10^{-3}$ | $3.6\times10^{-3}$ | $1.1\times10^{-2}$ |
| 13.5 | $1.7\times10^{-4}$ | $7.1\times10^{-4}$ | $2.9\times10^{-3}$ | $3.8\times10^{-3}$ | $1.1\times10^{-2}$ |

**Table 5.** As Table 3, but at a distance of 6-8 kpc from the Galactic centre.

| Time (Gyr) | Normalized Dust Mass | | | | |
|---|---|---|---|---|---|
| | Oxide | Carb. | Iron | Silicate | Total |
| 0.2 | $1.3\times10^{-5}$ | $9.0\times10^{-5}$ | $2.8\times10^{-4}$ | $5.0\times10^{-4}$ | $1.9\times10^{-3}$ |
| 0.4 | $2.8\times10^{-5}$ | $1.5\times10^{-4}$ | $4.9\times10^{-4}$ | $8.6\times10^{-4}$ | $3.2\times10^{-3}$ |
| 0.6 | $4.1\times10^{-5}$ | $2.1\times10^{-4}$ | $6.8\times10^{-4}$ | $1.1\times10^{-3}$ | $4.1\times10^{-3}$ |
| 0.8 | $5.4\times10^{-5}$ | $2.5\times10^{-4}$ | $8.5\times10^{-4}$ | $1.4\times10^{-3}$ | $4.9\times10^{-3}$ |
| 1.0 | $6.6\times10^{-5}$ | $3.0\times10^{-4}$ | $1.0\times10^{-3}$ | $1.6\times10^{-3}$ | $5.6\times10^{-3}$ |
| 1.3 | $4.2\times10^{-5}$ | $2.0\times10^{-4}$ | $7.0\times10^{-4}$ | $1.1\times10^{-3}$ | $3.8\times10^{-3}$ |
| 1.7 | $4.4\times10^{-5}$ | $2.1\times10^{-4}$ | $7.3\times10^{-4}$ | $1.1\times10^{-3}$ | $4.0\times10^{-3}$ |
| 2.0 | $5.0\times10^{-5}$ | $2.4\times10^{-4}$ | $8.4\times10^{-4}$ | $1.3\times10^{-3}$ | $4.4\times10^{-3}$ |
| 3.0 | $7.0\times10^{-5}$ | $3.3\times10^{-4}$ | $1.1\times10^{-3}$ | $1.7\times10^{-3}$ | $5.6\times10^{-3}$ |
| 5.0 | $9.7\times10^{-5}$ | $4.4\times10^{-4}$ | $1.6\times10^{-3}$ | $2.3\times10^{-3}$ | $6.9\times10^{-3}$ |
| 7.0 | $1.1\times10^{-4}$ | $5.0\times10^{-4}$ | $1.9\times10^{-3}$ | $2.6\times10^{-3}$ | $7.5\times10^{-3}$ |
| 9.0 | $1.2\times10^{-4}$ | $5.3\times10^{-4}$ | $2.1\times10^{-3}$ | $2.8\times10^{-3}$ | $7.8\times10^{-3}$ |
| 10.0 | $1.2\times10^{-4}$ | $5.5\times10^{-4}$ | $2.2\times10^{-3}$ | $2.9\times10^{-3}$ | $8.0\times10^{-3}$ |
| 12.0 | $1.3\times10^{-4}$ | $5.9\times10^{-4}$ | $2.4\times10^{-3}$ | $3.1\times10^{-3}$ | $8.6\times10^{-3}$ |
| 13.5 | $1.4\times10^{-4}$ | $6.2\times10^{-4}$ | $2.5\times10^{-3}$ | $3.3\times10^{-3}$ | $9.1\times10^{-3}$ |

**Table 6.** As Table 3, but at a distance of 8-10 kpc (solar neighbourhood) from the Galactic centre.

| Time (Gyr) | Normalized Dust Mass | | | | |
|---|---|---|---|---|---|
| | Oxide | Carb. | Iron | Silicate | Total |
| 0.2 | $7.0\times10^{-6}$ | $5.5\times10^{-5}$ | $1.7\times10^{-4}$ | $3.0\times10^{-4}$ | $1.0\times10^{-3}$ |
| 0.4 | $1.5\times10^{-5}$ | $1.0\times10^{-4}$ | $3.3\times10^{-4}$ | $5.6\times10^{-4}$ | $1.8\times10^{-3}$ |
| 0.6 | $2.3\times10^{-5}$ | $1.4\times10^{-4}$ | $4.7\times10^{-4}$ | $7.8\times10^{-4}$ | $2.5\times10^{-3}$ |
| 0.8 | $3.2\times10^{-5}$ | $1.8\times10^{-4}$ | $6.0\times10^{-4}$ | $9.9\times10^{-4}$ | $3.1\times10^{-3}$ |
| 1.0 | $4.1\times10^{-5}$ | $2.1\times10^{-4}$ | $7.3\times10^{-4}$ | $1.1\times10^{-3}$ | $3.6\times10^{-3}$ |
| 1.3 | $3.6\times10^{-5}$ | $1.9\times10^{-4}$ | $6.7\times10^{-4}$ | $1.0\times10^{-3}$ | $3.2\times10^{-3}$ |
| 1.7 | $3.8\times10^{-5}$ | $2.0\times10^{-4}$ | $7.0\times10^{-4}$ | $1.1\times10^{-3}$ | $3.3\times10^{-3}$ |
| 2.0 | $4.1\times10^{-5}$ | $2.2\times10^{-4}$ | $7.7\times10^{-4}$ | $1.1\times10^{-3}$ | $3.4\times10^{-3}$ |
| 3.0 | $5.6\times10^{-5}$ | $2.9\times10^{-4}$ | $1.0\times10^{-3}$ | $1.5\times10^{-3}$ | $4.3\times10^{-3}$ |
| 5.0 | $7.9\times10^{-5}$ | $3.9\times10^{-4}$ | $1.5\times10^{-3}$ | $2.0\times10^{-3}$ | $5.3\times10^{-3}$ |
| 7.0 | $9.5\times10^{-5}$ | $4.5\times10^{-4}$ | $1.7\times10^{-3}$ | $2.4\times10^{-3}$ | $5.9\times10^{-3}$ |
| 9.0 | $1.0\times10^{-4}$ | $4.9\times10^{-4}$ | $1.9\times10^{-3}$ | $2.6\times10^{-3}$ | $6.3\times10^{-3}$ |
| 10.0 | $1.1\times10^{-4}$ | $5.0\times10^{-4}$ | $2.0\times10^{-3}$ | $2.6\times10^{-3}$ | $6.4\times10^{-3}$ |
| 12.0 | $1.2\times10^{-4}$ | $5.3\times10^{-4}$ | $2.1\times10^{-3}$ | $2.8\times10^{-3}$ | $6.8\times10^{-3}$ |
| 13.5 | $1.2\times10^{-4}$ | $5.6\times10^{-4}$ | $2.2\times10^{-3}$ | $3.0\times10^{-3}$ | $7.2\times10^{-3}$ |

**Table 7.** As Table 3, but at a distance of 10-12 kpc from the Galactic centre.

| Time (Gyr) | Normalized Dust Mass | | | | |
|---|---|---|---|---|---|
| | Oxide | Carb. | Iron | Silicate | Total |
| 0.2 | $4.3\times10^{-6}$ | $4.0\times10^{-5}$ | $1.2\times10^{-4}$ | $2.2\times10^{-4}$ | $6.1\times10^{-4}$ |
| 0.4 | $1.0\times10^{-5}$ | $8.0\times10^{-5}$ | $2.6\times10^{-4}$ | $4.4\times10^{-4}$ | $1.2\times10^{-3}$ |
| 0.6 | $1.6\times10^{-5}$ | $1.1\times10^{-4}$ | $3.7\times10^{-4}$ | $6.2\times10^{-4}$ | $1.7\times10^{-3}$ |
| 0.8 | $2.2\times10^{-5}$ | $1.4\times10^{-4}$ | $4.9\times10^{-4}$ | $7.9\times10^{-4}$ | $2.1\times10^{-3}$ |
| 1.0 | $2.8\times10^{-5}$ | $1.7\times10^{-4}$ | $6.0\times10^{-4}$ | $9.6\times10^{-4}$ | $2.5\times10^{-3}$ |

| 1.3 | 2.8×10⁻⁵ | 1.7×10⁻⁴ | 6.0×10⁻⁴ | 9.4×10⁻⁴ | 2.4×10⁻³ |
| 1.7 | 3.1×10⁻⁵ | 1.8×10⁻⁴ | 6.4×10⁻⁴ | 1.0×10⁻³ | 2.6×10⁻³ |
| 2.0 | 3.4×10⁻⁵ | 2.0×10⁻⁴ | 7.1×10⁻⁴ | 1.0×10⁻³ | 2.8×10⁻³ |
| 3.0 | 4.6×10⁻⁵ | 2.7×10⁻⁴ | 9.8×10⁻⁴ | 1.4×10⁻³ | 3.4×10⁻³ |
| 5.0 | 6.6×10⁻⁵ | 3.6×10⁻⁴ | 1.4×10⁻³ | 1.9×10⁻³ | 4.4×10⁻³ |
| 7.0 | 8.0×10⁻⁵ | 4.2×10⁻⁴ | 1.6×10⁻³ | 2.2×10⁻³ | 4.8×10⁻³ |
| 9.0 | 9.0×10⁻⁵ | 4.6×10⁻⁴ | 1.8×10⁻³ | 2.4×10⁻³ | 5.1×10⁻³ |
| 10.0 | 9.4×10⁻⁵ | 4.7×10⁻⁴ | 1.8×10⁻³ | 2.5×10⁻³ | 5.2×10⁻³ |
| 12.0 | 1.0×10⁻⁴ | 5.0×10⁻⁴ | 2.0×10⁻³ | 2.6×10⁻³ | 5.5×10⁻³ |
| 13.5 | 1.0×10⁻⁴ | 5.2×10⁻⁴ | 2.1×10⁻³ | 2.8×10⁻³ | 5.8×10⁻³ |

**Table 8.** As Table 3, but at a distance of 12-14 kpc from the Galactic centre.

| Time (Gyr) | Normalized Dust Mass | | | | |
|---|---|---|---|---|---|
| | Oxide | Carb. | Iron | Silicate | Total |
| 0.2 | 2.3×10⁻⁶ | 2.5×10⁻⁵ | 7.6×10⁻⁵ | 1.3×10⁻⁴ | 3.2×10⁻⁴ |
| 0.4 | 6.6×10⁻⁶ | 5.8×10⁻⁵ | 1.8×10⁻⁴ | 3.1×10⁻⁴ | 7.5×10⁻⁴ |
| 0.6 | 1.0×10⁻⁵ | 8.4×10⁻⁵ | 2.7×10⁻⁴ | 4.5×10⁻⁴ | 1.1×10⁻³ |
| 0.8 | 1.4×10⁻⁵ | 1.1×10⁻⁴ | 3.6×10⁻⁴ | 5.9×10⁻⁴ | 1.4×10⁻³ |
| 1.0 | 1.9×10⁻⁵ | 1.3×10⁻⁴ | 4.5×10⁻⁴ | 7.2×10⁻⁴ | 1.7×10⁻³ |
| 1.3 | 2.0×10⁻⁵ | 1.4×10⁻⁴ | 4.8×10⁻⁴ | 7.5×10⁻⁴ | 1.7×10⁻³ |
| 1.7 | 2.2×10⁻⁵ | 1.5×10⁻⁴ | 5.4×10⁻⁴ | 8.2×10⁻⁴ | 1.9×10⁻³ |
| 2.0 | 2.5×10⁻⁵ | 1.7×10⁻⁴ | 5.9×10⁻⁴ | 9.0×10⁻⁴ | 2.0×10⁻³ |
| 3.0 | 3.5×10⁻⁵ | 2.3×10⁻⁴ | 8.4×10⁻⁴ | 1.2×10⁻³ | 2.5×10⁻³ |
| 5.0 | 5.3×10⁻⁵ | 3.2×10⁻⁴ | 1.2×10⁻³ | 1.7×10⁻³ | 3.3×10⁻³ |
| 7.0 | 6.6×10⁻⁵ | 3.8×10⁻⁴ | 1.5×10⁻³ | 2.0×10⁻³ | 3.7×10⁻³ |
| 9.0 | 7.5×10⁻⁵ | 4.2×10⁻⁴ | 1.6×10⁻³ | 2.2×10⁻³ | 4.1×10⁻³ |
| 10.0 | 8.0×10⁻⁵ | 4.3×10⁻⁴ | 1.7×10⁻³ | 2.2×10⁻³ | 4.1×10⁻³ |
| 12.0 | 8.7×10⁻⁵ | 4.6×10⁻⁴ | 1.8×10⁻³ | 2.4×10⁻³ | 4.4×10⁻³ |
| 13.5 | 9.3×10⁻⁵ | 4.8×10⁻⁴ | 1.9×10⁻³ | 2.5×10⁻³ | 4.6×10⁻³ |

**Table 9.** As Table 3, but at a distance of 14-16 kpc from the Galactic centre.

| Time (Gyr) | Normalized Dust Mass | | | | |
|---|---|---|---|---|---|
| | Oxide | Carb. | Iron | Silicate | Total |
| 0.2 | 1.3×10⁻⁶ | 1.6×10⁻⁵ | 4.7×10⁻⁵ | 8.3×10⁻⁵ | 1.6×10⁻⁴ |
| 0.4 | 4.0×10⁻⁶ | 4.0×10⁻⁵ | 1.2×10⁻⁴ | 2.1×10⁻⁴ | 4.3×10⁻⁴ |
| 0.6 | 6.9×10⁻⁶ | 6.2×10⁻⁵ | 2.0×10⁻⁴ | 3.3×10⁻⁴ | 6.9×10⁻⁴ |
| 0.8 | 9.8×10⁻⁶ | 8.1×10⁻⁵ | 2.7×10⁻⁴ | 4.3×10⁻⁴ | 9.0×10⁻⁴ |
| 1.0 | 1.2×10⁻⁵ | 1.0×10⁻⁴ | 3.4×10⁻⁴ | 5.3×10⁻⁴ | 1.1×10⁻³ |
| 1.3 | 1.4×10⁻⁵ | 1.1×10⁻⁴ | 3.8×10⁻⁴ | 5.8×10⁻⁴ | 1.1×10⁻³ |
| 1.7 | 1.6×10⁻⁵ | 1.2×10⁻⁴ | 4.3×10⁻⁴ | 6.5×10⁻⁴ | 1.3×10⁻³ |
| 2.0 | 1.8×10⁻⁵ | 1.3×10⁻⁴ | 4.8×10⁻⁴ | 7.2×10⁻⁴ | 1.4×10⁻³ |
| 3.0 | 2.5×10⁻⁵ | 1.9×10⁻⁴ | 6.9×10⁻⁴ | 9.8×10⁻⁴ | 1.8×10⁻³ |
| 5.0 | 4.0×10⁻⁵ | 2.8×10⁻⁴ | 1.0×10⁻³ | 1.4×10⁻³ | 2.4×10⁻³ |
| 7.0 | 5.3×10⁻⁵ | 3.4×10⁻⁴ | 1.3×10⁻³ | 1.7×10⁻³ | 2.8×10⁻³ |
| 9.0 | 6.2×10⁻⁵ | 3.8×10⁻⁴ | 1.5×10⁻³ | 1.9×10⁻³ | 3.1×10⁻³ |
| 10.0 | 6.6×10⁻⁵ | 4.0×10⁻⁴ | 1.5×10⁻³ | 2.0×10⁻³ | 3.2×10⁻³ |
| 12.0 | 7.3×10⁻⁵ | 4.3×10⁻⁴ | 1.7×10⁻³ | 2.2×10⁻³ | 3.4×10⁻³ |
| 13.5 | 7.8×10⁻⁵ | 4.5×10⁻⁴ | 1.8×10⁻³ | 2.3×10⁻³ | 3.6×10⁻³ |

**Table 10.** As Table 3, but at a distance of 16-18 kpc from the Galactic centre.

| Time (Gyr) | Normalized Dust Mass | | | | |
|---|---|---|---|---|---|
| | Oxide | Carb. | Iron | Silicate | Total |
| 0.2 | 7.7×10⁻⁷ | 1.0×10⁻⁵ | 2.9×10⁻⁵ | 5.1×10⁻⁵ | 8.6×10⁻⁵ |
| 0.4 | 2.2×10⁻⁶ | 2.5×10⁻⁵ | 8.0×10⁻⁵ | 1.3×10⁻⁴ | 2.3×10⁻⁴ |
| 0.6 | 4.2×10⁻⁶ | 4.3×10⁻⁵ | 1.4×10⁻⁴ | 2.2×10⁻⁴ | 4.0×10⁻⁴ |
| 0.8 | 6.2×10⁻⁶ | 5.8×10⁻⁵ | 1.9×10⁻⁴ | 3.1×10⁻⁴ | 5.5×10⁻⁴ |
| 1.0 | 8.2×10⁻⁶ | 7.3×10⁻⁵ | 2.5×10⁻⁴ | 3.8×10⁻⁴ | 6.8×10⁻⁴ |
| 1.3 | 9.2×10⁻⁶ | 8.3×10⁻⁵ | 2.8×10⁻⁴ | 4.3×10⁻⁴ | 7.5×10⁻⁴ |
| 1.7 | 1.0×10⁻⁵ | 9.5×10⁻⁵ | 3.3×10⁻⁴ | 4.9×10⁻⁴ | 8.5×10⁻⁴ |
| 2.0 | 1.2×10⁻⁵ | 1.0×10⁻⁴ | 3.7×10⁻⁴ | 5.4×10⁻⁴ | 9.4×10⁻⁴ |
| 3.0 | 1.7×10⁻⁵ | 1.4×10⁻⁴ | 5.3×10⁻⁴ | 7.5×10⁻⁴ | 1.2×10⁻³ |
| 5.0 | 2.8×10⁻⁵ | 2.2×10⁻⁴ | 8.4×10⁻⁴ | 1.1×10⁻³ | 1.7×10⁻³ |
| 7.0 | 3.8×10⁻⁵ | 2.8×10⁻⁴ | 1.0×10⁻³ | 1.4×10⁻³ | 2.1×10⁻³ |
| 9.0 | 4.6×10⁻⁵ | 3.2×10⁻⁴ | 1.2×10⁻³ | 1.6×10⁻³ | 2.5×10⁻³ |
| 10.0 | 5.0×10⁻⁵ | 3.4×10⁻⁴ | 1.3×10⁻³ | 1.7×10⁻³ | 2.6×10⁻³ |
| 12.0 | 5.7×10⁻⁵ | 3.7×10⁻⁴ | 1.4×10⁻³ | 1.9×10⁻³ | 2.9×10⁻³ |
| 13.5 | 6.1×10⁻⁵ | 3.9×10⁻⁴ | 1.5×10⁻³ | 2.0×10⁻³ | 3.1×10⁻³ |

**Table 11a.** Appearance (and disappearance∗) temperatures (in K) of stable condensates for the Milky Way Galaxy over time and space.

| Distance (kpc) | 16-18 | 2-4 | 16-18 | 2-4 |
|---|---|---|---|---|
| Time (Gyr) | 1 | 13.5 | 9 | 9 |
| Al₂O₃ | 1519.6 | 1701.3 | 1603.1 | 1692.8 |
| CaAl₄O₇ | 1501.5 | 1658.0 | 1576.9 | 1652.6 |
| Al₂O₃ (*) | 1499.0 | 1643.8 | 1571.5 | 1640.3 |
| CaTiO₃ | 1454.5 | 1576.0 | 1515.7 | 1572.0 |
| Melilite | 1387.8 | 1514.5 | 1451.0 | 1511.0 |
| Spinel | 1238.5 | 1386.6 | 1316.1 | 1379.6 |
| CaAl₄O₇ (*) | 1376.6 | 1386.1 | 1418.4 | 1379.3 |
| Plagioclase | 1248.8 | 1358.1 | 1312.7 | 1360.1 |
| Spinel (*) | 1182.8 | 1351.7 | 1312.5 | 1356.1 |
| CaMgSi₂O₆ | 1247.9 | - | 1301.1 | 1353.6 |
| CaMgSiO₄ | - | 1356.7 | - | - |
| Melilite (*) | 1246.5 | 1355.9 | 1299.9 | 1353.0 |
| Plagioclase | 1237.5 | - | - | - |
| Olivine | 1236.5 | 1354.5 | 1292.5 | 1349.2 |
| CaTiSiO₅ | 1248.0 | 1348.1 | 1297.7 | 1346.8 |
| CaTiO₃ (*) | 1248.0 | 1348.1 | 1297.7 | 1346.8 |
| Al₂TiO₅ | 1239.5 | - | - | - |
| CaTiSiO₅ (*) | 1239.4 | - | - | - |
| Clinopyroxene | 1208.0 | 1306.2 | 1257.3 | 1304.6 |
| Olivine (*) | 1196.3 | 1280.7 | 1244.1 | 1286.4 |
| Fe-Metal | 1208.2 | 1298.2 | 1260.7 | 1296.8 |
| Al₂SiO₅ | 1184.6 | - | - | - |
| SiO₂ | 1177.7 | 1264.1 | 1228.2 | 1270.8 |
| Cr₂O₃ | 1126.8 | 1240.5 | 1181.8 | 1238.4 |
| Co | 1087.3 | 1162.2 | 1126.5 | 1159.4 |
| Ni₅P₂ | 1074.3 | 1050.6 | 1095.8 | 1087.8 |
| KAlSi₃O₈ | - | - | - | 1011.3 |

**Table 11b.** The maximum value of the normalized mass of stable condensates in the simulated temperature range for the Milky Way Galaxy over time and space.

| Distance (kpc) | 16-18 | 2-4 | 16-18 | 2-4 |
|---|---|---|---|---|
| Time (Gyr) | 1 | 13.5 | 9 | 9 |
| Al₂O₃ | 6.87×10⁻³ | 3.18×10⁻¹ | 5.10×10⁻² | 2.67×10⁻¹ |
| CaAl₄O₇ | 1.78×10⁻² | 5.51×10⁻¹ | 1.09×10⁻¹ | 4.80×10⁻¹ |
| Ca₂Al₂SiO₇ | 3.77×10⁻² | 6.96×10⁻¹ | 2.30×10⁻¹ | 6.58×10⁻¹ |
| Ca₂MgSi₂O₇ | 2.81×10⁻² | 3.30×10⁻¹ | 8.12×10⁻¹ | 1.83×10⁻¹ |
| CaAl₂Si₂O₈ | 3.82×10⁻² | 1.18×10⁰ | 2.34×10⁻¹ | 1.03×10⁰ |
| NaAlSi₃O₈ | 7.56×10⁻⁹ | 3.98×10⁻¹ | 5.59×10⁻² | 7.95×10⁻¹ |
| Al₂TiO₅ | 1.81×10⁻³ | - | - | - |
| MgAl₂O₄ | 1.81×10⁻² | 3.89×10⁻¹ | 4.37×10⁻³ | 2.54×10⁻¹ |
| FeAl₂O₄ | 4.71×10⁻⁴ | 2.82×10⁻³ | 2.87×10⁻⁵ | 2.06×10⁻³ |
| Al₂SiO₅ | 2.06×10⁻² | - | - | - |
| CaTiO₃ | 1.35×10⁻³ | 2.40×10⁻² | 6.47×10⁻³ | 2.12×10⁻² |
| CaTiSiO₅ | 1.95×10⁻³ | 3.47×10⁻² | 9.33×10⁻³ | 3.06×10⁻² |
| CaMgSi₂O₆ | 1.04×10⁻¹ | - | 2.42×10⁻¹ | 6.31×10⁻¹ |
| CaMgSiO₄ | - | 5.67×10⁻¹ | - | - |
| Mg₂SiO₄ | 3.18×10⁻¹ | 4.67×10⁰ | 1.26×10⁰ | 3.80×10⁰ |
| Fe₂SiO₄ | 1.64×10⁻⁷ | 6.49×10⁻⁵ | 2.71×10⁻⁶ | 4.41×10⁻⁵ |
| MgSiO₃ | 5.89×10⁻¹ | 7.45×10⁰ | 2.19×10⁰ | 6.21×10⁰ |
| FeSiO₃ | 1.18×10⁻³ | 1.14×10⁻¹ | 1.01×10⁻² | 8.44×10⁻² |
| SiO₂ | 1.63×10⁻¹ | 1.03×10⁰ | 7.13×10⁻¹ | 1.45×10⁰ |
| Cr₂O₃ | 7.42×10⁻¹ | 1.05×10⁰ | 3.76×10⁻¹ | 1.09×10⁻¹ |
| Fe-metal | 2.42×10⁻¹ | 4.07×10⁰ | 1.30×10⁰ | 3.88×10⁰ |
| Ni₅P₂ | 5.82×10⁻³ | 4.12×10⁻² | 3.57×10⁻² | 7.14×10⁻² |
| KAlSi₃O₈ | - | - | - | 1.68×10⁻² |
| Total | 1.13×10⁰ | 1.42×10¹ | 4.72×10⁰ | 1.32×10¹ |

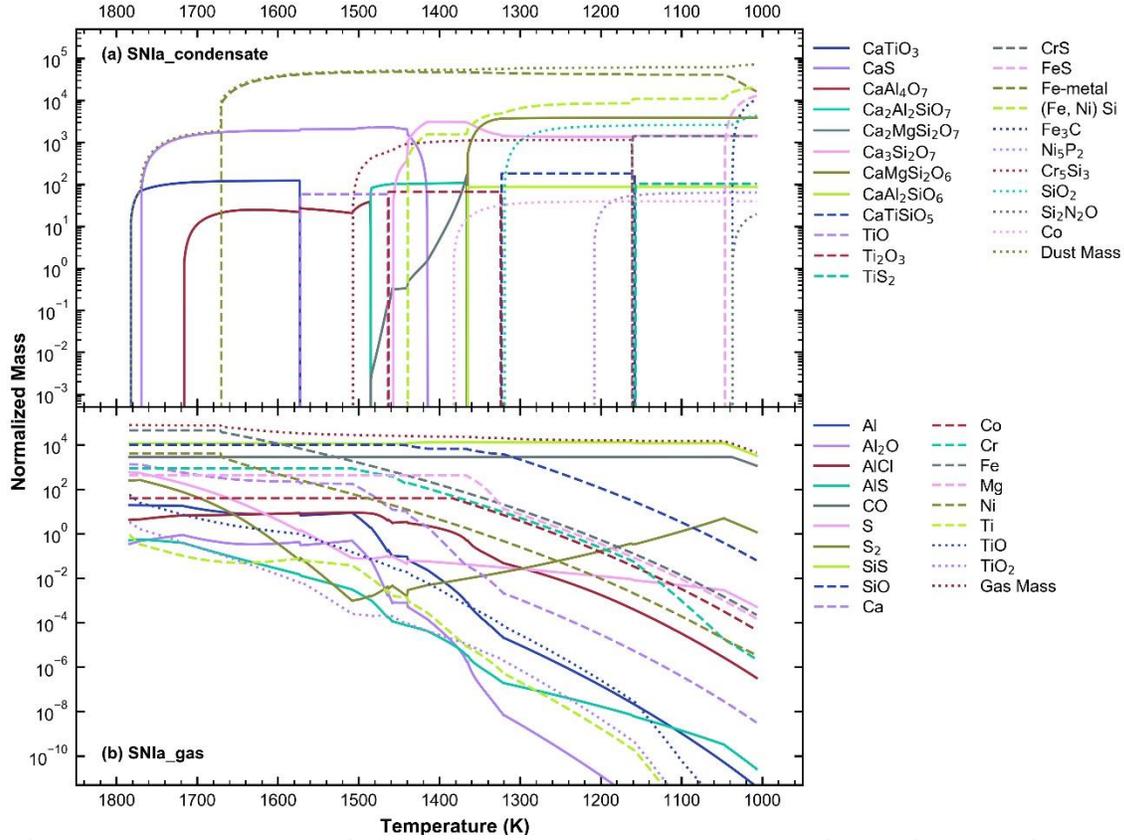

**Figure 4.** The normalized mass distribution of, (a) the condensed phases, (b) the gaseous species as a function of temperature for supernova SN Ia ejecta at an assumed pressure of $10^{-5}$ bar.

**Table 12.** Appearance temperatures (in K), disappearance temperatures (in K) and the peak mass value of the stable condensates for the supernova SN Ia ejecta at a pressure of $10^{-5}$ bar.

| Condensate | Appearance_Temp | Disappearance_Temp | Normalized_Mass |
|---|---|---|---|
| $CaTiO_3$ | 1783 | 1572.9 | $1.23 \times 10^2$ |
| CaS | 1770 | 1415.0 | $2.31 \times 10^3$ |
| $CaAl_4O_7$ | 1717 | 1484.9 | $3.86 \times 10^1$ |
| Fe-Metal | 1671 | - | $4.78 \times 10^4$ |
| TiO | 1573 | 1463.9 | $5.87 \times 10^1$ |
| $Cr_5Si_3$ | 1508 | 1160.1 | $1.14 \times 10^3$ |
| Melilite | 1486 | 1365.9 | $2.76 \times 10^2$ |
| $Ti_2O_3$ | 1464 | 1323.9 | $6.61 \times 10^1$ |
| $Ca_3Si_2O_7$ | 1458 | - | $3.10 \times 10^3$ |
| (Fe, Ni) Si | 1440 | - | $2.07 \times 10^4$ |
| Co | 1383 | - | $3.92 \times 10^1$ |
| $CaMgSi_2O_6$ | 1366 | - | $3.83 \times 10^3$ |
| $CaAl_2SiO_6$ | 1366 | - | $8.71 \times 10^1$ |
| $CaTiSiO_5$ | 1324 | 1157.9 | $1.80 \times 10^2$ |
| $SiO_2$ | 1320 | - | $4.36 \times 10^3$ |
| $Ni_5P_2$ | 1209 | - | $6.38 \times 10^1$ |
| CrS | 1162 | - | $1.40 \times 10^3$ |
| $TiS_2$ | 1158 | - | $1.03 \times 10^2$ |
| FeS | 1047 | - | $1.29 \times 10^4$ |
| $Si_2N_2O$ | 1038 | - | $1.95 \times 10^1$ |
| $Fe_3C$ | 1037.3 | - | $1.06 \times 10^4$ |
| Total | - | - | $7.18 \times 10^4$ |

The thermodynamical calculations performed for distinct epochs at different radial distances support the result deduced from mass-balance calculations. In order to understand the spatial evolution of distinct groups in the Galaxy, we performed thermodynamical calculations for the innermost and outermost annular ring at distinct epochs, which are presented in Fig. 3. It can be inferred from these panels that the condensation sequence remains almost the same in two distinct compositions, but the condensation temperatures decrease as we move away from the Galactic centre (Table 11a). Similarly, the condensed dust mass is more in the inner annular ring than the outer ring (Table 11b). It can also be inferred from thermodynamical calculations that the silicate dust mass is largest among all-groups, followed by iron dust mass. It should be noted that thermodynamical calculations include the time evolution of circumstellar winds depending upon its cooling rate and produce distinct compositions of dust at different times. For instance, 'initial' condensates are oxide-type in all the cases, but at temperatures lower than 1450 K (Tables 11a), the condensed dust is mainly of silicate-type. This redistribution modifies the order of condensed dust mass in different groups. Further, the other assumptions made in thermodynamical calculations, such as stellar yields affect the mass distribution of condensed dust. Further, the pressure value is chosen to be $10^{-5}$ bar for all the simulations. However, this assumption does not affect the result because the effect of pressure is primarily only on the condensation temperature. It does not influence the order of condensed dust mass. The calculations also assume the existence of thermodynamical equilibrium. Although non-equilibrium prevails in astrophysical environments, it mainly affects the abundance of 'later' condensates in a condensation sequence (Gupta & Sahijpal 2020). In order to estimate the general composition, an averaged-out ordering of an ensemble of dust grain mass can produce the solution with certainty. This order of condensed dust mass complies well with the order as predicted by mass-balance calculations.

### 4.3 Contribution from SN Ia in the dust evolution

Along with the dust composition and its abundance evolution in the Galaxy, we have also made an attempt to understand the stellar sources for the origin of dust grains. In order to accomplish this goal, we explored the condensation sequence in distinct stellar sources. To understand the condensation trends, we considered in detail the case of supernova type SN Ia. The supernovae SN Ia are considered to be associated with the explosion of a white dwarf star subsequent to the accretion from a binary companion. The panels (a) and (b) in Fig. 4 represent the normalized mass distribution of condensates and gaseous species, respectively, for the composition of SN Ia ejecta. At an initial stage, a completely vaporized and homogenized system assemblage was assumed at a pressure of $10^{-5}$ bar. The system was then allowed to cool following the thermodynamical constraints. Around 90 per cent of Al remained in the monatomic form before the onset of stability of any solid phase. The rest of the aluminium remained in AlCl, AlS, and $Al_2O$. Titanium stabilized in oxides such as TiO and $TiO_2$, and a small amount in the monatomic form. Carbon got trapped in a stable CO bond because it was less abundant than oxygen. Calcium, magnesium, and iron remained in their monatomic form.

With the decrease in temperature, $CaTiO_3$ is the first condensate at 1783 K (Fig. 4; Table 12). With the appearance of oxide dust, $CaTiO_3$, the mass of titanium oxides decreases from the gaseous phase and increases in the solid form. Because the abundance of calcium is an order of magnitude higher than titanium, the appearance of $CaTiO_3$ does not have any significant effect on the amount of calcium species. The amount of Ca slightly decreases from the gaseous phase after the stability of the next condensates, CaS. With the stability of the next oxide condensate, $CaAl_4O_7$ at 1717 K, the amount of Al increases in the solid form. Before the condensation of metallic iron, the mass remains primarily in the gaseous form. The dust mass increases significantly after the stability of Fe-metal. Fe-dust is produced maximum with this composition. Subsequent to condensation of metallic iron, silicates- and oxides-grains also get stabilized, but their normalized masses are much smaller as compared to the condensed iron-dust mass. The mass of metallic iron decreases after the stability of (Fe, Ni) Si at 1440 K. This decreases the amount of SiO from the gaseous form which was initially ~55 per cent of the total Si. Further, the decrease in silicon oxide from gaseous form leads to the condensation of silicates. This also increases the remaining Si in SiS gaseous form from 45 to 52 per cent. The sulphur chemistry comes into action with the stability of troilite at 1047 K. This breaks down the SiS bond and also consumes metallic iron. The entire sulphur inventory released from SiS is locked up in FeS. Further cooling reduces the stability of CO bond, and the excess of iron reacts with it to stabilize as $Fe_3C$. With the condensation of $Fe_3C$, the carbon decreases in the gaseous form and increases in the solid form. At this stage of condensation, all major reactive and condensable elements stabilize in the solid form. Only the volatile elements such as H, He remain in gaseous form. Hence, the supernovae SN Ia contribute significantly to iron-dust production.

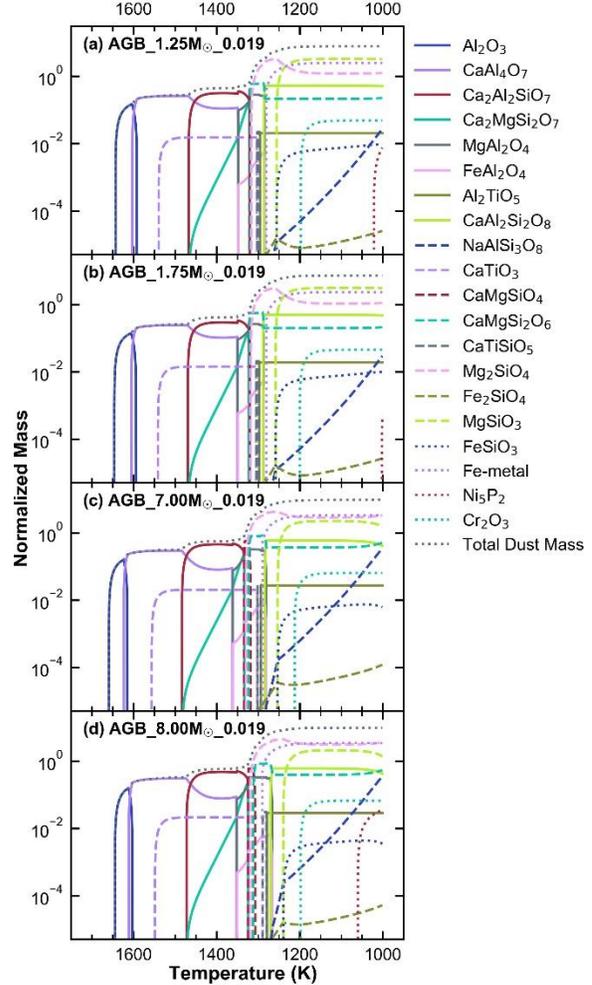

**Figure 5.** The normalized mass distribution of the condensed phases as a function of temperature in the winds of an AGB stellar model at a pressure of $10^{-5}$ bar for stellar masses of, (a) 1.25 $M_\odot$, (b) 1.75 $M_\odot$, (c) 7.00 $M_\odot$, (d) 8.00 $M_\odot$, at an assumed solar metallicity of 0.019.

**Table 13a.** Appearance (and disappearance∗) temperatures (in K) of stable condensates for the AGB stellar winds with distinct masses at a pressure of $10^{-5}$ bar at an assumed solar metallicity of 0.019. The dashes in table cells represent the absence of condensate in the corresponding model.

| Condensate | AGB Stellar Mass | | | |
|---|---|---|---|---|
| | 1.25M$_\odot$ | 1.75M$_\odot$ | 7.0M$_\odot$ | 8.0M$_\odot$ |
| Al$_2$O$_3$ | 1643.5 | 1646.1 | 1659.2 | 1644.6 |
| CaAl$_4$O$_7$ | 1603.4 | 1605.5 | 1623.7 | 1611.7 |
| Al$_2$O$_3$(*) | 1591.7 | 1593.8 | 1614.4 | 1602.8 |
| CaTiO$_3$ | 1539.8 | 1541.3 | 1556.3 | 1548.3 |
| Melilite | 1467.6 | 1469.5 | 1483.6 | 1473.0 |
| Spinel | 1348.2 | 1349.8 | 1361.5 | 1351.5 |
| CaAl$_4$O$_7$(*) | 1348.0 | 1349.6 | 1361.4 | 1351.4 |
| CaMgSiO$_4$ | 1321.5 | 1323.0 | 1334.4 | 1324.9 |
| Melilite (*) | 1319.9 | 1321.4 | 1332.6 | 1323.1 |
| CaMgSi$_2$O$_6$ | 1319.5 | 1321.2 | 1323.7 | 1312.7 |
| CaMgSiO$_4$(*) | 1319.5 | 1319.5 | 1318.0 | 1306.2 |
| Olivine | 1316.4 | 1318.0 | 1329.4 | 1319.6 |
| CaTiSiO$_5$ | 1302.9 | 1304.9 | 1301.1 | 1289.0 |
| CaTiO$_3$ (*) | 1302.8 | 1304.8 | 1301.0 | 1289.0 |
| Al$_2$TiO$_5$ | 1295.5 | 1297.7 | 1293.2 | 1280.0 |
| CaTiSiO$_5$(*) | 1295.4 | 1297.6 | 1293.1 | 1279.9 |
| Plagioclase | 1286.4 | 1288.8 | 1284.3 | 1270.3 |
| Spinel (*) | 1283.6 | 1286.0 | 1280.2 | 1265.8 |
| Fe-Metal | 1280.5 | 1280.3 | 1289.7 | 1290.6 |
| Clinopyroxene | 1255.9 | 1258.2 | 1253.1 | 1239.4 |
| Cr$_2$O$_3$ | 1197.9 | 1200.4 | 1212.8 | 1198.9 |
| Co | 1143.5 | 1143.4 | 1150.5 | 1151.3 |
| Ni$_5$P$_2$ | 1022.0 | 1002.5 | - | 1060.2 |

**Table 13b.** The maximum value of the normalized mass of stable condensates in the simulated temperature range for the AGB stellar winds with distinct masses at a pressure of $10^{-5}$ bar at an assumed solar metallicity of 0.019. The dashes in the table cells represent the absence of condensate in the corresponding model.

| Condensate | AGB Stellar Mass | | | |
|---|---|---|---|---|
| | 1.25M$_\odot$ | 1.75M$_\odot$ | 7.0M$_\odot$ | 8.0M$_\odot$ |
| Al$_2$O$_3$ | 1.47×10$^{-1}$ | 1.39×10$^{-1}$ | 1.56×10$^{-1}$ | 1.56×10$^{-1}$ |
| CaAl$_4$O$_7$ | 2.57×10$^{-1}$ | 2.41×10$^{-1}$ | 2.92×10$^{-1}$ | 3.01×10$^{-1}$ |
| Ca$_2$Al$_2$SiO$_7$ | 3.54×10$^{-1}$ | 3.33×10$^{-1}$ | 4.73×10$^{-1}$ | 4.94×10$^{-1}$ |
| Ca$_2$MgSi$_2$O$_7$ | 1.66×10$^{-1}$ | 1.60×10$^{-1}$ | 2.10×10$^{-1}$ | 2.25×10$^{-1}$ |
| MgAl$_2$O$_4$ | 2.80×10$^{-1}$ | 2.63×10$^{-1}$ | 3.18×10$^{-1}$ | 3.28×10$^{-1}$ |
| FeAl$_2$O$_4$ | 5.53×10$^{-3}$ | 5.21×10$^{-3}$ | 8.98×10$^{-3}$ | 6.60×10$^{-3}$ |
| Al$_2$TiO$_5$ | 2.03×10$^{-2}$ | 1.90×10$^{-2}$ | 2.70×10$^{-2}$ | 2.82×10$^{-2}$ |
| CaAl$_2$Si$_2$O$_8$ | 5.19×10$^{-1}$ | 4.87×10$^{-1}$ | 5.85×10$^{-1}$ | 6.01×10$^{-1}$ |
| NaAlSi$_3$O$_8$ | 2.72×10$^{-2}$ | 2.93×10$^{-2}$ | 3.41×10$^{-1}$ | 3.51×10$^{-1}$ |
| CaTiO$_3$ | 1.52×10$^{-2}$ | 1.42×10$^{-2}$ | 2.02×10$^{-2}$ | 2.11×10$^{-2}$ |
| CaMgSiO$_4$ | 4.09×10$^{-1}$ | 3.43×10$^{-1}$ | 5.67×10$^{-1}$ | 5.93×10$^{-1}$ |
| CaMgSi$_2$O$_6$ | 6.15×10$^{-1}$ | 5.77×10$^{-1}$ | 8.17×10$^{-1}$ | 8.54×10$^{-1}$ |
| CaTiSiO$_5$ | 2.19×10$^{-2}$ | 2.05×10$^{-2}$ | 2.91×10$^{-2}$ | 3.04×10$^{-2}$ |
| Mg$_2$SiO$_4$ | 3.28×10$^{0}$ | 3.04×10$^{0}$ | 4.22×10$^{0}$ | 4.41×10$^{0}$ |
| Fe$_2$SiO$_4$ | 2.54×10$^{-5}$ | 2.68×10$^{-5}$ | 1.20×10$^{-4}$ | 5.21×10$^{-5}$ |
| MgSiO$_3$ | 3.24×10$^{0}$ | 3.08×10$^{0}$ | 2.21×10$^{0}$ | 2.07×10$^{0}$ |
| FeSiO$_3$ | 9.67×10$^{-3}$ | 9.98×10$^{-3}$ | 7.42×10$^{-3}$ | 4.35×10$^{-3}$ |
| Fe-metal | 2.47×10$^{0}$ | 2.31×10$^{0}$ | 3.28×10$^{0}$ | 3.40×10$^{0}$ |
| Ni$_5$P$_2$ | 7.44×10$^{-3}$ | 4.99×10$^{-4}$ | - | 3.52×10$^{-2}$ |
| Cr$_2$O$_3$ | 4.77×10$^{-2}$ | 4.48×10$^{-2}$ | 6.33×10$^{-2}$ | 6.61×10$^{-2}$ |
| Total | 7.71×10$^{0}$ | 7.22×10$^{0}$ | 9.38×10$^{0}$ | 9.70×10$^{0}$ |

### 4.4 Contribution from AGB stars in the dust evolution

Low to intermediate-mass stars (< 8 M$_\odot$) evolve as red giants (RG) and asymptotic giant branch (AGB) stars. These stars were also simulated to explore the possibility of dust origin and to understand the grain condensation trends. The thermodynamical calculations were performed for AGB stellar models at two metallicities 0.019 (Z$_\odot$) and 0.0001 for 1.25, 1.75, 3, 4, 7 and 8 M$_\odot$ stars. Because the circumstellar environment and its composition depend on stellar mass, circumstellar envelopes in different mass AGB stars differ in chemical composition. The composition of the system assemblage is an essential input component in thermodynamical condensation calculations. Thus, different stars produce different condensation trends. In the first scenario of solar metallicity, C/O ratio is less than one for the AGB stellar models with mass other than 3 and 4 M$_\odot$. Therefore, these stars produce oxides- and silicates-dust grains at solar metallicity. The results of thermodynamical simulations of these four models are presented in Fig. 5 and Tables 13. In the second scenario of $10^{-4}$ metallicity, the C/O ratio is more than one for all the stellar masses. Therefore, carbonaceous grains are formed for all these models. This leads to the contribution of carbides grains in ISM from AGB stellar models (Fig. 6; Tables 14). In all models other than AGB stellar models, O-rich grains are formed owing to the relatively high abundance of oxygen than carbon. Thus, the low metallicity AGB stellar winds contribute the maximum mass of carbide grains in the ISM.

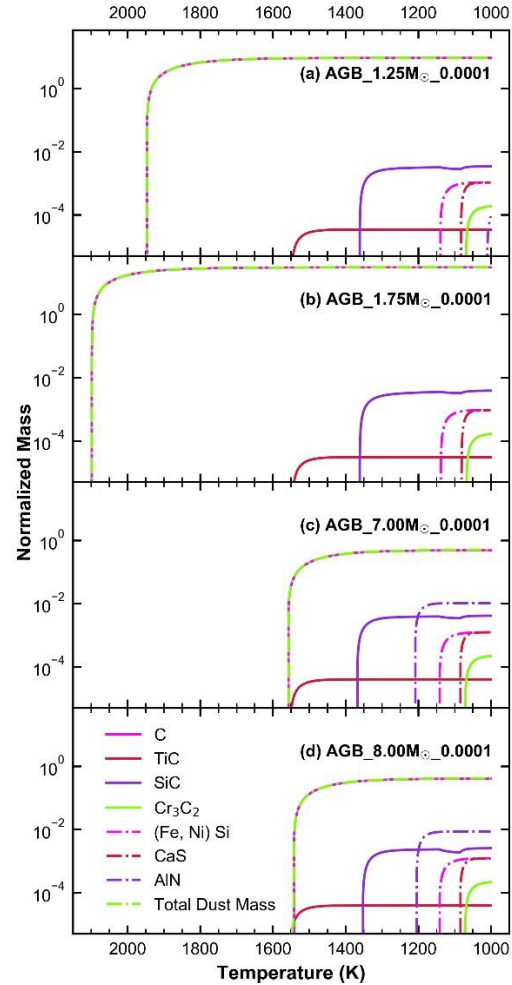

**Figure 6.** As Fig. 5, but for AGB stellar models at $10^{-4}$ metallicity. The major part of the dust mass is in the carbon (graphite) form.

**Table 14a.** As Table 13a, but for the AGB stellar winds at $10^{-4}$ metallicity.

| Condensate | AGB Stellar Mass | | | |
|---|---|---|---|---|
| | 1.25M☉ | 1.75M☉ | 7.0M☉ | 8.0M☉ |
| C | 1947.0 | 2098.1 | 1558 | 1543 |
| TiC | 1549.3 | 1547.1 | 1552 | 1551 |
| SiC | 1361.2 | 1361.2 | 1368 | 1353 |
| (Fe, Ni) Si | 1139.9 | 1138.2 | 1142 | 1142 |
| CaS | 1083.0 | 1081.1 | 1085 | 1085 |
| $Cr_3C_2$ | 1069.4 | 1067.9 | 1071 | 1072 |
| Co | 1024.0 | 1022.5 | 1026 | 1026 |
| AlN | 1011.0 | - | 1209 | 1205 |

**Table 14b.** As Table 13b, but for the AGB stellar winds at $10^{-4}$ metallicity.

| Condensate | AGB Stellar Mass | | | |
|---|---|---|---|---|
| | 1.25M☉ | 1.75M☉ | 7.0M☉ | 8.0M☉ |
| C | $9.20\times10^{0}$ | $3.06\times10^{1}$ | $4.81\times10^{-1}$ | $4.01\times10^{-1}$ |
| TiC | $3.44\times10^{-5}$ | $3.05\times10^{-5}$ | $3.98\times10^{-5}$ | $3.94\times10^{-5}$ |
| SiC | $3.44\times10^{-3}$ | $3.86\times10^{-3}$ | $4.11\times10^{-3}$ | $2.56\times10^{-3}$ |
| $Cr_3C_2$ | $1.85\times10^{-4}$ | $1.65\times10^{-4}$ | $2.16\times10^{-4}$ | $2.14\times10^{-4}$ |
| (Fe, Ni) Si | $1.07\times10^{-3}$ | $9.45\times10^{-4}$ | $1.23\times10^{-3}$ | $1.22\times10^{-3}$ |
| CaS | $1.05\times10^{-3}$ | $9.32\times10^{-4}$ | $1.22\times10^{-3}$ | $1.21\times10^{-3}$ |
| AlN | $9.07\times10^{-5}$ | - | $1.03\times10^{-2}$ | $8.59\times10^{-3}$ |
| Total | $9.21\times10^{0}$ | $3.06\times10^{1}$ | $4.98\times10^{-1}$ | $4.14\times10^{-1}$ |

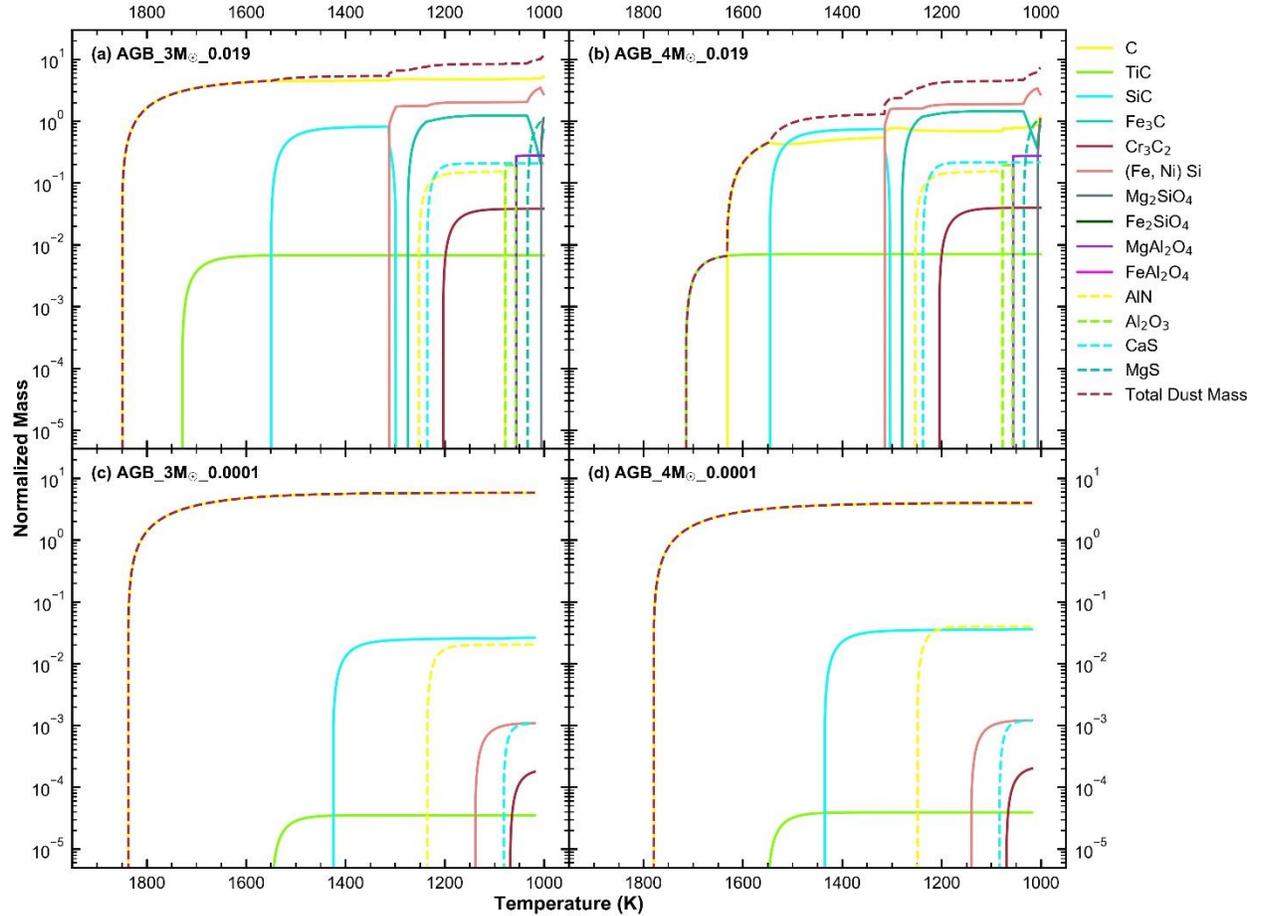

**Figure 7.** The normalized mass distribution of the condensed phases as a function of temperature in the winds of an AGB stellar model at an assumed pressure of $10^{-5}$ bar for stellar masses of, (a) 3 M☉ at solar metallicity (Z☉), (b) 4 M☉ at solar metallicity, (c) 3 M☉ at $10^{-4}$ metallicity, (d) 4 M☉ at $10^{-4}$ metallicity.

**Table 15a.** Appearance (and disappearance∗) temperatures (in K) of stable condensates for the AGB stellar winds with two distinct masses at a pressure of $10^{-5}$ bar at two distinct metallicities.

| Mass | 3M☉ | 4M☉ | 3M☉ | 4M☉ |
|---|---|---|---|---|
| Metal | (Z☉) | (Z☉) | $10^{-4}$ | $10^{-4}$ |
| C | 1850 | 1632 | 1838 | 1780 |
| TiC | 1729 | 1715 | 1548 | 1550 |
| SiC | 1550 | 1546 | 1425 | 1436 |
| (Fe, Ni) Si | 1313 | 1315 | 1139 | 1141 |
| SiC(*) | 1299 | 1304 | - | - |
| $Fe_3C$ | 1275 | 1280 | - | - |
| AlN | 1253 | 1254 | 1236 | 1249 |
| CaS | 1236 | 1238 | 1082 | 1084 |
| $Cr_3C_2$ | 1204 | 1205 | 1069 | 1070 |
| Co | 1144 | 1146 | 1023 | 1024 |
| $Al_2O_3$ | 1079 | 1078 | - | - |
| AlN(*) | 1078.9 | 1077.6 | - | - |
| Spinel | 1057 | 1057 | - | - |
| $Al_2O_3$(*) | 1056.9 | 1056.9 | - | - |
| MgS | 1034 | 1035 | - | - |
| Olivine | 1007 | 1007 | - | - |

**Table 15b.** The maximum value of the normalized mass of stable condensates in the simulated temperature range for the AGB stellar winds with two distinct masses at a pressure of $10^{-5}$ bar at two distinct metallicities.

| Mass | $3M_\odot$ | $4M_\odot$ | $3M_\odot$ | $4M_\odot$ |
| --- | --- | --- | --- | --- |
| Metal | ($Z_\odot$) | ($Z_\odot$) | $10^{-4}$ | $10^{-4}$ |
| C | $5.26\times10^0$ | $1.19\times10^0$ | $5.77\times10^0$ | $3.90\times10^0$ |
| TiC | $6.75\times10^{-3}$ | $7.05\times10^{-3}$ | $3.52\times10^{-5}$ | $3.90\times10^{-5}$ |
| SiC | $8.17\times10^{-1}$ | $7.45\times10^{-1}$ | $2.62\times10^{-2}$ | $3.60\times10^{-2}$ |
| $Fe_3C$ | $1.23\times10^0$ | $1.45\times10^0$ | - | - |
| $Cr_3C_2$ | $3.81\times10^{-2}$ | $3.96\times10^{-2}$ | $1.78\times10^{-4}$ | $2.01\times10^{-4}$ |
| (Fe, Ni) Si | $3.47\times10^0$ | $3.41\times10^0$ | $1.09\times10^{-3}$ | $1.21\times10^{-3}$ |
| $Mg_2SiO_4$ | $1.12\times10^0$ | $1.08\times10^0$ | - | - |
| $Fe_2SiO_4$ | $4.9\times10^{-18}$ | $4.4\times10^{-18}$ | - | - |
| $MgAl_2O_4$ | $2.77\times10^{-1}$ | $2.74\times10^{-1}$ | - | - |
| $FeAl_2O_4$ | $1.93\times10^{-7}$ | $1.88\times10^{-7}$ | - | - |
| AlN | $1.53\times10^{-1}$ | $1.55\times10^{-1}$ | $2.02\times10^{-2}$ | $3.99\times10^{-2}$ |
| $Al_2O_3$ | $1.95\times10^{-1}$ | $1.94\times10^{-1}$ | - | - |
| CaS | $2.07\times10^{-1}$ | $2.16\times10^{-1}$ | $1.07\times10^{-3}$ | $1.20\times10^{-3}$ |
| MgS | $9.70\times10^{-1}$ | $1.02\times10^0$ | - | - |
| Total | $1.13\times10^1$ | $7.36\times10^0$ | $5.82\times10^0$ | $3.98\times10^0$ |

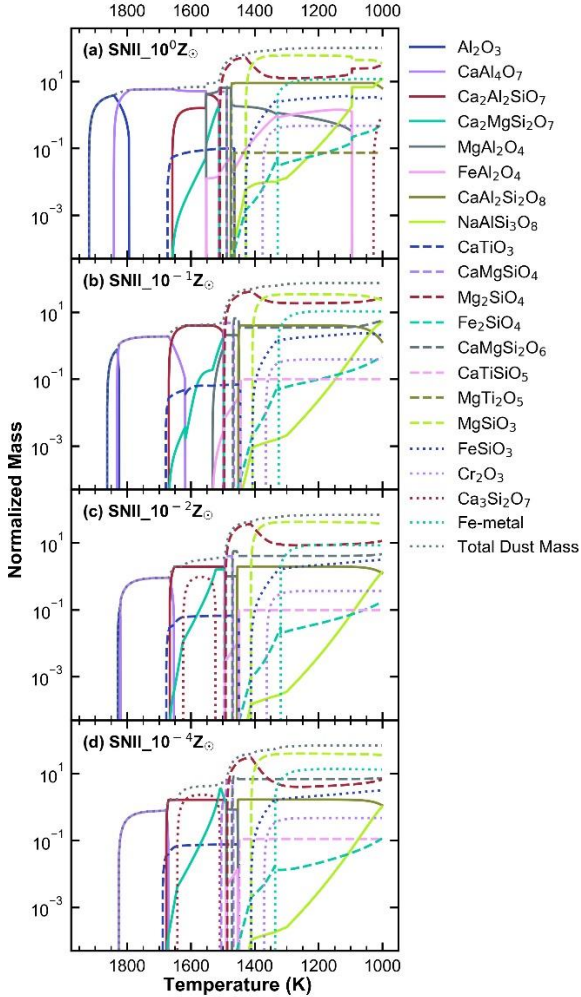

**Figure 8.** The normalized mass distribution of the condensed phases as a function of temperature at a pressure of $10^{-5}$ bar for supernovae (SN II, Ib/c) at, (a) solar metallicity ($Z_\odot$), (b) $0.1\times$ solar metallicity, (c) $0.01\times$ solar metallicity, (d) $10^{-4}\times$ solar metallicity.

**Table 16a.** Appearance (and disappearance∗) temperatures (in K) of stable condensates for supernovae (SN II+Ib/c) at solar ($Z_\odot$), $0.1\times$ solar, $0.01\times$ solar, and $10^{-4}\times$ solar metallicity at a pressure of $10^{-5}$ bar.

| | SN II, Ib/c with Metallicity | | | |
| --- | --- | --- | --- | --- |
| Condensate | $1\times$ | $0.1\times$ | $0.01\times$ | $0.0001\times$ |
| $Al_2O_3$ | 1919.5 | 1862.4 | 1829.5 | - |
| $CaAl_4O_7$ | 1841.1 | 1831.3 | 1821.6 | 1826.6 |
| $Al_2O_3$ (*) | 1793.4 | 1824.3 | 1820.4 | - |
| $CaTiO_3$ | 1674.0 | 1679.0 | 1677.6 | 1688.9 |
| Melilite | 1659.0 | 1669.5 | 1667.0 | 1677.4 |
| $CaAl_4O_7$ (*) | 1551.7 | 1617.3 | 1652.6 | 1670.4 |
| $Ca_3Si_2O_7$ | 1028.6 | - | 1624.9 | 1642.8 |
| $Ca_3Si_2O_7$ (*) | - | - | 1522.3 | 1509.0 |
| Spinel | 1552.9 | 1532.1 | 1494.4 | 1489.3 |
| $CaMgSiO_4$ | 1512.2 | 1498.4 | 1495.9 | 1505.3 |
| Melilite (*) | 1510.8 | 1496.8 | 1492.9 | 1487.9 |
| Olivine | 1509.8 | 1495.1 | 1491.6 | 1486.9 |
| $CaMgSi_2O_6$ | 1489.0 | 1468.5 | 1471.0 | 1471.5 |
| $CaMgSiO_4$ (*) | 1486.4 | 1464.4 | 1467.9 | 1466.8 |
| Plagioclase | 1475.3 | 1451.3 | 1455.2 | 1453.7 |
| $CaMgSi_2O_6$ (*) | 1472.2 | - | - | - |
| Spinel (*) | 1095.6 | 1448.9 | 1454.2 | 1452.8 |
| $MgTi_2O_5$ | 1463.9 | - | - | - |
| $CaTiSiO_5$ | - | 1444.7 | 1449.4 | 1448.4 |
| $CaTiO_3$ (*) | 1463.8 | 1444.7 | 1449.4 | 1448.4 |
| Clinopyroxene | 1428.4 | 1407.4 | 1412.2 | 1411.1 |
| $Cr_2O_3$ | 1375.1 | 1365.6 | 1362.3 | 1370.2 |
| Fe-Metal | 1328.9 | 1325.1 | 1318.1 | 1336.3 |
| Co | 1213.8 | 1180.9 | 1172.8 | 1167.9 |

**Table 16b.** The maximum value of the normalized mass of stable condensates in the simulated temperature range for supernovae (SNII + Ib/c) at solar ($Z_\odot$), $0.1\times$ solar, $0.01\times$ solar, and $10^{-4}\times$ solar metallicity at a pressure of $10^{-5}$ bar.

| | SN II, Ib/c with Metallicity | | | |
| --- | --- | --- | --- | --- |
| Condensate | $1\times$ | $0.1\times$ | $0.01\times$ | $0.0001\times$ |
| $Al_2O_3$ | $3.84\times10^0$ | $7.56\times10^{-1}$ | $1.19\times10^{-1}$ | $7.76\times10^{-1}$ |
| $CaAl_4O_7$ | $5.89\times10^0$ | $1.88\times10^0$ | $9.12\times10^{-1}$ | $7.76\times10^{-1}$ |
| $Ca_2Al_2SiO_7$ | $4.20\times10^0$ | $3.99\times10^0$ | $1.94\times10^0$ | $1.66\times10^0$ |
| $Ca_2MgSi_2O_7$ | $1.83\times10^0$ | $1.73\times10^0$ | $1.61\times10^0$ | $3.53\times10^0$ |
| $MgAl_2O_4$ | $6.44\times10^0$ | $2.07\times10^0$ | $1.00\times10^0$ | $8.55\times10^{-1}$ |
| $FeAl_2O_4$ | $1.43\times10^0$ | $2.70\times10^{-2}$ | $9.43\times10^{-3}$ | $1.36\times10^{-2}$ |
| $CaAl_2Si_2O_8$ | $9.03\times10^0$ | $4.05\times10^0$ | $1.97\times10^0$ | $1.68\times10^0$ |
| $NaAlSi_3O_8$ | $1.19\times10^1$ | $5.17\times10^0$ | $1.33\times10^0$ | $1.03\times10^0$ |
| $CaTiO_3$ | $9.87\times10^{-2}$ | $6.82\times10^{-2}$ | $6.70\times10^{-2}$ | $7.64\times10^{-2}$ |
| $CaMgSiO_4$ | $4.97\times10^0$ | $4.77\times10^0$ | $4.06\times10^0$ | $5.95\times10^0$ |
| $Mg_2SiO_4$ | $5.23\times10^1$ | $4.14\times10^1$ | $3.65\times10^1$ | $3.03\times10^1$ |
| $Fe_2SiO_4$ | $5.28\times10^{-1}$ | $4.59\times10^{-1}$ | $1.73\times10^{-1}$ | $1.15\times10^{-1}$ |
| $CaMgSi_2O_6$ | $6.87\times10^0$ | $6.60\times10^0$ | $5.62\times10^0$ | $8.23\times10^0$ |
| $CaTiSiO_5$ | - | $9.98\times10^{-2}$ | $9.83\times10^{-2}$ | $1.12\times10^{-1}$ |
| $MgTi_2O_5$ | $7.41\times10^{-2}$ | - | - | - |
| $MgSiO_3$ | $6.08\times10^1$ | $3.47\times10^1$ | $4.26\times10^1$ | $3.99\times10^1$ |
| $FeSiO_3$ | $3.69\times10^0$ | $2.34\times10^0$ | $3.12\times10^0$ | $3.17\times10^0$ |
| $Cr_2O_3$ | $4.75\times10^{-1}$ | $3.85\times10^{-1}$ | $3.67\times10^{-1}$ | $4.73\times10^{-1}$ |
| $Ca_3Si_2O_7$ | $9.36\times10^{-1}$ | - | $9.68\times10^{-1}$ | $2.35\times10^0$ |
| Fe-metal | $1.21\times10^1$ | $1.08\times10^1$ | $8.79\times10^0$ | $1.37\times10^1$ |
| Total | $1.01\times10^2$ | $7.48\times10^1$ | $6.89\times10^1$ | $6.93\times10^1$ |

### 4.5 Contribution from SN II and Ib/c in the dust evolution

The stars with mass greater than 11 $M_\odot$ explode as supernovae. The stars with the mass in the range 11-33 $M_\odot$ end their life as the supernovae type SN II, whereas the stellar masses in the range 34-100 $M_\odot$ evolve through WR phases prior to their explosion as supernovae SN Ib/c. These massive stars quickly recycle the galactic matter as compared to the low mass stars. Here, we explored 11–100 $M_\odot$ stars and obtained a weighted average yield using Salpeter IMF. The supernovae SNe II+Ib/c were simulated at four distinct metallicities. Although it is possible for us to perform thermodynamical condensation calculations for different phases of stellar evolution for individual stars (Gupta & Sahijpal 2020), we followed a generalized integrated approach based on IMF to make an assessment of dust from a stellar cluster that forms at almost a same instant and evolves to SN II+Ib/c over several tens of million years. This approach provides an overall assessment of SN II+Ib/c to a particular composition of the dust. This approach has a limitation. For instance, this work indicates that the AGB stars are the predominant source of C-rich grains. But it is already known that massive stars in their WC evolutionary phase enrich the ISM with carbide grains. The chemical composition obtained by weighted averaged over the complete range suppresses this kind of feature owing to the rareness of Wolf-Rayet phases. Still, we are not aiming to cover them all separately in this work. The future works in this field could include the detailed thermodynamical evolution of each stellar mass contributing to the Galactic dust inventories.

The appearance and disappearance temperatures for all the condensed phases in all the considered compositions of different metallicities are summarized in Table 16a. The peak value of normalized masses for the corresponding condensate is presented in Table 16b and the corresponding mass distribution in Fig. 8. The thermodynamical models indicate that the condensation temperatures of silicates- and oxides-dust are higher in the case of supernovae ejecta than their magnitudes in Galactic annular rings and AGB stellar models (Tables 11a, 13a, and 16a). The mass of silicates- and oxides-dust produced in these compositions is higher in comparison to models other than SN models. Further, the Fe-metal gets stabilized at lower temperatures. Also, the ratio of iron-dust mass to total dust mass is lower in these SNe II and Ib/c compositions in comparison to SN Ia and AGB models.

### 4.6 Influence of metallicity on dust evolution

The metallicity value directly affects the condensation of dust grains. Higher the metallicity value, the higher will be the abundance of condensable matter in the system assemblage. Therefore, in general, a positive correlation can be seen between the normalized dust mass and metallicity. This result is also in agreement with the observations of variation of dust-to-gas ratio with metallicity in nearby galaxies (Sandstrom et al. 2013). However, it is not the only factor affecting dust condensation. In the present study, we explored AGB stellar models corresponding to 3 and 4 $M_\odot$ at two distinct metallicities (Fig. 7; Tables 15). These two stellar masses were chosen because of the similarity in condensed carbonaceous dust at distinct metallicities. It can be seen in Fig. 7 that relatively simple chemistry takes place in the low metallicity region owing to the non-availability of refractory elements in the gaseous phase. Therefore, a lesser number of condensates are formed in the case of low metallicity AGB stars. In addition, we also considered the compositions of supernovae at four different metallicities and systematically studied the shift in the condensation temperatures along with the peak normalized masses of the condensates (Fig. 8; Tables 16). The general inference from the simulations is that more dust mass is condensed at high metallicity value.

## 5 CONCLUSION

In the present study, we have performed mass-balance and thermodynamical equilibrium condensation calculations in order to understand the abundance evolution of dust grains and their prominent stellar source in the Milky Way Galaxy. The mass distributions of distinct dust grain components have been computed for the distinct epochs over the entire Galaxy using mass-balance calculations. Thermodynamical condensation calculations have given the condensation sequences, condensation temperatures and normalized dust masses of the condensates for different considered compositions. We have performed numerical simulations for distinct stellar clusters and explored distinct regions of the Galaxy. Some of the major conclusions drawn from the present work are the following:

(1) The normalized dust mass decreases with the increase in distance from the Galactic centre. Not only the dust mass, but the condensation temperatures of the condensates also decrease in moving towards the outer annular rings of the Galaxy.
(2) Although the condensation sequences remain almost the same with the temporal evolution, the condensation temperatures as well as the normalized masses of the condensates increase. This gradual increase in the dust mass is because of the enrichment of ISM with heavier and refractory elements by successive stellar generations.
(3) The supernovae SN Ia are the most prominent sources for Fe-dust mass.
(4) Carbonaceous grains are primarily contributed by AGB stars. C, TiC, SiC, CaS, and AlN are the suite of refractory minerals for C-rich stellar environments. AGB stellar models at $10^{-4}$ metallicity, and AGB models with stellar mass 3–4 $M_\odot$ at solar metallicity produce carbides grains.
(5) Silicate- and oxide-type grains are condensed in AGB stellar models at four distinct stellar masses corresponding to solar metallicity, the Milky Way Galaxy across time and space and supernovae SNe II, Ib/c models at distinct metallicities. The maximum amount of oxides mass is produced by supernovae SNe II, Ib/c.

**ACKNOWLEDGEMENTS**

We are extremely grateful for the constructive criticism and helpful suggestions made by the reviewer which led to significant improvements in the article. AG acknowledges the Council of Scientific & Industrial Research (CSIR) for providing doctoral financial assistance (no. 09/135(0716)/2015-EMR-I), which was used to write the code. AG and SS acknowledge funding from the Indian Space Research Organisation (ISRO) through a PLANEX grant for providing and maintaining the theoretical laboratory facilities.

# Supplementary Data

**Table S1.** The normalized masses of the considered dust grain compositions at different epochs for the eight annular rings of the Galaxy.

| Annular ring 1 (2-4 kpc) | | | | | | | | | | | | | | | |
|---|---|---|---|---|---|---|---|---|---|---|---|---|---|---|---|
| Time (Gyr) | 0.2 | 0.4 | 0.6 | 0.8 | 1 | 1.3 | 1.7 | 2 | 3 | 5 | 7 | 9 | 10 | 12 | 13.5 |
| Gas density* | 2.21 | 2.87 | 3.04 | 3 | 2.9 | 27.5 | 38.9 | 41.9 | 40.8 | 28.2 | 19.2 | 14.5 | 14.3 | 13.5 | 12.3 |
| $CaTiO_3$ | $1.64 \times 10^{-6}$ | $2.71 \times 10^{-6}$ | $3.53 \times 10^{-6}$ | $4.20 \times 10^{-6}$ | $4.76 \times 10^{-6}$ | $1.79 \times 10^{-6}$ | $2.87 \times 10^{-6}$ | $3.57 \times 10^{-6}$ | $5.34 \times 10^{-6}$ | $8.74 \times 10^{-6}$ | $1.10 \times 10^{-5}$ | $1.25 \times 10^{-5}$ | $1.23 \times 10^{-5}$ | $1.27 \times 10^{-5}$ | $1.38 \times 10^{-5}$ |
| $Ti_2O_3$ | $9.64 \times 10^{-9}$ | $1.59 \times 10^{-8}$ | $2.07 \times 10^{-8}$ | $2.47 \times 10^{-8}$ | $2.80 \times 10^{-8}$ | $1.05 \times 10^{-8}$ | $1.69 \times 10^{-8}$ | $2.10 \times 10^{-8}$ | $3.14 \times 10^{-8}$ | $5.13 \times 10^{-8}$ | $6.46 \times 10^{-8}$ | $7.37 \times 10^{-8}$ | $7.22 \times 10^{-8}$ | $7.48 \times 10^{-8}$ | $8.11 \times 10^{-8}$ |
| TiC | $7.23 \times 10^{-8}$ | $1.20 \times 10^{-7}$ | $1.56 \times 10^{-7}$ | $1.85 \times 10^{-7}$ | $2.10 \times 10^{-7}$ | $7.86 \times 10^{-8}$ | $1.27 \times 10^{-7}$ | $1.57 \times 10^{-7}$ | $2.35 \times 10^{-7}$ | $3.85 \times 10^{-7}$ | $4.84 \times 10^{-7}$ | $5.53 \times 10^{-7}$ | $5.41 \times 10^{-7}$ | $5.61 \times 10^{-7}$ | $6.08 \times 10^{-7}$ |
| $Al_2O_3$ | $1.14 \times 10^{-5}$ | $2.24 \times 10^{-5}$ | $3.09 \times 10^{-5}$ | $3.77 \times 10^{-5}$ | $4.32 \times 10^{-5}$ | $1.35 \times 10^{-5}$ | $2.40 \times 10^{-5}$ | $3.13 \times 10^{-5}$ | $4.90 \times 10^{-5}$ | $7.26 \times 10^{-5}$ | $8.77 \times 10^{-5}$ | $9.89 \times 10^{-5}$ | $9.64 \times 10^{-5}$ | $1.01 \times 10^{-4}$ | $1.10 \times 10^{-4}$ |
| $CaAl_4O_7$ | $8.71 \times 10^{-6}$ | $1.71 \times 10^{-5}$ | $2.37 \times 10^{-5}$ | $2.88 \times 10^{-5}$ | $3.30 \times 10^{-5}$ | $1.03 \times 10^{-5}$ | $1.84 \times 10^{-5}$ | $2.39 \times 10^{-5}$ | $3.75 \times 10^{-5}$ | $5.56 \times 10^{-5}$ | $6.71 \times 10^{-5}$ | $7.56 \times 10^{-5}$ | $7.38 \times 10^{-5}$ | $7.69 \times 10^{-5}$ | $8.41 \times 10^{-5}$ |
| $Ca_2SiO_4$ | $1.08 \times 10^{-5}$ | $1.63 \times 10^{-5}$ | $2.00 \times 10^{-5}$ | $2.27 \times 10^{-5}$ | $2.49 \times 10^{-5}$ | $1.11 \times 10^{-5}$ | $1.68 \times 10^{-5}$ | $1.99 \times 10^{-5}$ | $2.71 \times 10^{-5}$ | $4.00 \times 10^{-5}$ | $4.79 \times 10^{-5}$ | $5.22 \times 10^{-5}$ | $4.99 \times 10^{-5}$ | $4.98 \times 10^{-5}$ | $5.30 \times 10^{-5}$ |
| $Ca_2Al_2SiO_7$ | $2.07 \times 10^{-5}$ | $3.12 \times 10^{-5}$ | $3.81 \times 10^{-5}$ | $4.34 \times 10^{-5}$ | $4.76 \times 10^{-5}$ | $2.12 \times 10^{-5}$ | $3.20 \times 10^{-5}$ | $3.80 \times 10^{-5}$ | $5.17 \times 10^{-5}$ | $7.64 \times 10^{-5}$ | $9.15 \times 10^{-5}$ | $9.98 \times 10^{-5}$ | $9.54 \times 10^{-5}$ | $9.52 \times 10^{-5}$ | $1.01 \times 10^{-4}$ |
| CaS | $5.09 \times 10^{-6}$ | $7.67 \times 10^{-6}$ | $9.37 \times 10^{-6}$ | $1.07 \times 10^{-5}$ | $1.17 \times 10^{-5}$ | $5.21 \times 10^{-6}$ | $7.86 \times 10^{-6}$ | $9.34 \times 10^{-6}$ | $1.27 \times 10^{-5}$ | $1.88 \times 10^{-5}$ | $2.25 \times 10^{-5}$ | $2.45 \times 10^{-5}$ | $2.34 \times 10^{-5}$ | $2.34 \times 10^{-5}$ | $2.49 \times 10^{-5}$ |
| $Mg_2SiO_4$ | $2.08 \times 10^{-4}$ | $3.24 \times 10^{-4}$ | $3.96 \times 10^{-4}$ | $4.48 \times 10^{-4}$ | $4.86 \times 10^{-4}$ | $2.17 \times 10^{-4}$ | $3.36 \times 10^{-4}$ | $3.99 \times 10^{-4}$ | $5.32 \times 10^{-4}$ | $6.94 \times 10^{-4}$ | $7.90 \times 10^{-4}$ | $8.80 \times 10^{-4}$ | $8.69 \times 10^{-4}$ | $9.29 \times 10^{-4}$ | $1.02 \times 10^{-3}$ |
| $MgAl_2O_4$ | $9.83 \times 10^{-7}$ | $1.53 \times 10^{-6}$ | $1.87 \times 10^{-6}$ | $2.11 \times 10^{-6}$ | $2.29 \times 10^{-6}$ | $1.03 \times 10^{-6}$ | $1.58 \times 10^{-6}$ | $1.89 \times 10^{-6}$ | $2.51 \times 10^{-6}$ | $3.27 \times 10^{-6}$ | $3.73 \times 10^{-6}$ | $4.15 \times 10^{-6}$ | $4.10 \times 10^{-6}$ | $4.38 \times 10^{-6}$ | $4.79 \times 10^{-6}$ |
| MgS | $7.78 \times 10^{-5}$ | $1.21 \times 10^{-4}$ | $1.48 \times 10^{-4}$ | $1.67 \times 10^{-4}$ | $1.82 \times 10^{-4}$ | $8.12 \times 10^{-5}$ | $1.25 \times 10^{-4}$ | $1.49 \times 10^{-4}$ | $1.99 \times 10^{-4}$ | $2.59 \times 10^{-4}$ | $2.95 \times 10^{-4}$ | $3.29 \times 10^{-4}$ | $3.25 \times 10^{-4}$ | $3.47 \times 10^{-4}$ | $3.79 \times 10^{-4}$ |
| $Ca_2MgSi_2O_7$ | $1.70 \times 10^{-5}$ | $2.63 \times 10^{-5}$ | $3.25 \times 10^{-5}$ | $3.71 \times 10^{-5}$ | $4.08 \times 10^{-5}$ | $1.73 \times 10^{-5}$ | $2.66 \times 10^{-5}$ | $3.19 \times 10^{-5}$ | $4.35 \times 10^{-5}$ | $6.14 \times 10^{-5}$ | $7.17 \times 10^{-5}$ | $7.71 \times 10^{-5}$ | $7.33 \times 10^{-5}$ | $7.28 \times 10^{-5}$ | $7.73 \times 10^{-5}$ |
| $MgSiO_3$ | $2.48 \times 10^{-4}$ | $3.83 \times 10^{-4}$ | $4.73 \times 10^{-4}$ | $5.41 \times 10^{-4}$ | $5.95 \times 10^{-4}$ | $2.52 \times 10^{-4}$ | $3.88 \times 10^{-4}$ | $4.65 \times 10^{-4}$ | $6.34 \times 10^{-4}$ | $8.95 \times 10^{-4}$ | $1.05 \times 10^{-3}$ | $1.12 \times 10^{-3}$ | $1.07 \times 10^{-3}$ | $1.06 \times 10^{-3}$ | $1.13 \times 10^{-3}$ |
| SiC | $7.92 \times 10^{-5}$ | $1.22 \times 10^{-4}$ | $1.51 \times 10^{-4}$ | $1.73 \times 10^{-4}$ | $1.90 \times 10^{-4}$ | $8.06 \times 10^{-5}$ | $1.24 \times 10^{-4}$ | $1.49 \times 10^{-4}$ | $2.03 \times 10^{-4}$ | $2.86 \times 10^{-4}$ | $3.34 \times 10^{-4}$ | $3.59 \times 10^{-4}$ | $3.42 \times 10^{-4}$ | $3.39 \times 10^{-4}$ | $3.60 \times 10^{-4}$ |
| $FeSiO_3$ | $2.47 \times 10^{-4}$ | $4.08 \times 10^{-4}$ | $5.36 \times 10^{-4}$ | $6.43 \times 10^{-4}$ | $7.32 \times 10^{-4}$ | $2.73 \times 10^{-4}$ | $4.46 \times 10^{-4}$ | $5.58 \times 10^{-4}$ | $8.50 \times 10^{-4}$ | $1.54 \times 10^{-3}$ | $1.98 \times 10^{-3}$ | $2.21 \times 10^{-3}$ | $2.13 \times 10^{-3}$ | $2.13 \times 10^{-3}$ | $2.26 \times 10^{-3}$ |
| Fe-metal | $7.84 \times 10^{-5}$ | $1.30 \times 10^{-4}$ | $1.70 \times 10^{-4}$ | $2.04 \times 10^{-4}$ | $2.32 \times 10^{-4}$ | $8.66 \times 10^{-5}$ | $1.42 \times 10^{-4}$ | $1.77 \times 10^{-4}$ | $2.70 \times 10^{-4}$ | $4.89 \times 10^{-4}$ | $6.28 \times 10^{-4}$ | $7.01 \times 10^{-4}$ | $6.75 \times 10^{-4}$ | $6.75 \times 10^{-4}$ | $7.18 \times 10^{-4}$ |
| $FeAl_2O_4$ | $2.44 \times 10^{-6}$ | $4.03 \times 10^{-6}$ | $5.30 \times 10^{-6}$ | $6.35 \times 10^{-6}$ | $7.23 \times 10^{-6}$ | $2.70 \times 10^{-6}$ | $4.41 \times 10^{-6}$ | $5.52 \times 10^{-6}$ | $8.40 \times 10^{-6}$ | $1.52 \times 10^{-5}$ | $1.95 \times 10^{-5}$ | $2.18 \times 10^{-5}$ | $2.10 \times 10^{-5}$ | $2.10 \times 10^{-5}$ | $2.23 \times 10^{-5}$ |
| $Fe_3C$ | $4.99 \times 10^{-5}$ | $8.25 \times 10^{-5}$ | $1.08 \times 10^{-4}$ | $1.30 \times 10^{-4}$ | $1.48 \times 10^{-4}$ | $5.51 \times 10^{-5}$ | $9.02 \times 10^{-5}$ | $1.13 \times 10^{-4}$ | $1.72 \times 10^{-4}$ | $3.11 \times 10^{-4}$ | $4.00 \times 10^{-4}$ | $4.46 \times 10^{-4}$ | $4.30 \times 10^{-4}$ | $4.30 \times 10^{-4}$ | $4.57 \times 10^{-4}$ |
| FeS | $4.01 \times 10^{-5}$ | $6.22 \times 10^{-5}$ | $7.71 \times 10^{-5}$ | $8.86 \times 10^{-5}$ | $9.81 \times 10^{-5}$ | $4.11 \times 10^{-5}$ | $6.31 \times 10^{-5}$ | $7.59 \times 10^{-5}$ | $1.06 \times 10^{-4}$ | $1.57 \times 10^{-4}$ | $1.89 \times 10^{-4}$ | $2.06 \times 10^{-4}$ | $1.97 \times 10^{-4}$ | $1.96 \times 10^{-4}$ | $2.08 \times 10^{-4}$ |
| $KAlSi_3O_8$ | $1.47 \times 10^{-6}$ | $2.85 \times 10^{-6}$ | $3.97 \times 10^{-6}$ | $4.90 \times 10^{-6}$ | $5.71 \times 10^{-6}$ | $1.72 \times 10^{-6}$ | $2.96 \times 10^{-6}$ | $3.87 \times 10^{-6}$ | $6.21 \times 10^{-6}$ | $9.63 \times 10^{-6}$ | $1.22 \times 10^{-5}$ | $1.44 \times 10^{-5}$ | $1.44 \times 10^{-5}$ | $1.55 \times 10^{-5}$ | $1.72 \times 10^{-5}$ |
| $NaAlSi_3O_8$ | $1.35 \times 10^{-7}$ | $3.04 \times 10^{-7}$ | $4.05 \times 10^{-7}$ | $4.83 \times 10^{-7}$ | $5.40 \times 10^{-7}$ | $1.74 \times 10^{-7}$ | $3.24 \times 10^{-7}$ | $4.19 \times 10^{-7}$ | $6.25 \times 10^{-7}$ | $8.72 \times 10^{-7}$ | $1.02 \times 10^{-6}$ | $1.10 \times 10^{-6}$ | $1.05 \times 10^{-6}$ | $1.06 \times 10^{-6}$ | $1.14 \times 10^{-6}$ |
| Graphite | $5.49 \times 10^{-6}$ | $8.49 \times 10^{-6}$ | $1.24 \times 10^{-5}$ | $1.57 \times 10^{-5}$ | $1.85 \times 10^{-5}$ | $6.32 \times 10^{-6}$ | $1.05 \times 10^{-5}$ | $1.37 \times 10^{-5}$ | $1.89 \times 10^{-5}$ | $2.19 \times 10^{-5}$ | $2.21 \times 10^{-5}$ | $2.40 \times 10^{-5}$ | $2.37 \times 10^{-5}$ | $2.66 \times 10^{-5}$ | $3.01 \times 10^{-5}$ |
| $H_2O$ | $2.62 \times 10^{-3}$ | $3.91 \times 10^{-3}$ | $4.63 \times 10^{-3}$ | $5.16 \times 10^{-3}$ | $5.57 \times 10^{-3}$ | $2.70 \times 10^{-3}$ | $4.00 \times 10^{-3}$ | $4.67 \times 10^{-3}$ | $5.93 \times 10^{-3}$ | $6.75 \times 10^{-3}$ | $7.17 \times 10^{-3}$ | $7.95 \times 10^{-3}$ | $7.68 \times 10^{-3}$ | $8.39 \times 10^{-3}$ | $9.35 \times 10^{-3}$ |

**Annular ring 2**
**(4-6 kpc)**

| Time (Gyr) | 0.2 | 0.4 | 0.6 | 0.8 | 1 | 1.3 | 1.7 | 2 | 3 | 5 | 7 | 9 | 10 | 12 | 13.5 |
|---|---|---|---|---|---|---|---|---|---|---|---|---|---|---|---|
| Gas density* | 2.45 | 3.29 | 3.55 | 3.54 | 3.4 | 11.9 | 17.1 | 19.1 | 21.4 | 20 | 17.3 | 14.8 | 14.2 | 12.6 | 11.4 |
| $CaTiO_3$ | $1.39 \times 10^{-6}$ | $2.38 \times 10^{-6}$ | $3.18 \times 10^{-6}$ | $3.86 \times 10^{-6}$ | $4.46 \times 10^{-6}$ | $2.17 \times 10^{-6}$ | $2.71 \times 10^{-6}$ | $3.23 \times 10^{-6}$ | $4.64 \times 10^{-6}$ | $6.67 \times 10^{-6}$ | $7.84 \times 10^{-6}$ | $8.78 \times 10^{-6}$ | $9.09 \times 10^{-6}$ | $1.01 \times 10^{-5}$ | $1.10 \times 10^{-5}$ |
| $Ti_2O_3$ | $8.19 \times 10^{-9}$ | $1.40 \times 10^{-8}$ | $1.87 \times 10^{-8}$ | $2.27 \times 10^{-8}$ | $2.62 \times 10^{-8}$ | $1.27 \times 10^{-8}$ | $1.59 \times 10^{-8}$ | $1.90 \times 10^{-8}$ | $2.73 \times 10^{-8}$ | $3.92 \times 10^{-8}$ | $4.60 \times 10^{-8}$ | $5.16 \times 10^{-8}$ | $5.34 \times 10^{-8}$ | $5.92 \times 10^{-8}$ | $6.46 \times 10^{-8}$ |
| TiC | $6.14 \times 10^{-8}$ | $1.05 \times 10^{-7}$ | $1.40 \times 10^{-7}$ | $1.70 \times 10^{-7}$ | $1.97 \times 10^{-7}$ | $9.56 \times 10^{-8}$ | $1.19 \times 10^{-7}$ | $1.42 \times 10^{-7}$ | $2.04 \times 10^{-7}$ | $2.94 \times 10^{-7}$ | $3.45 \times 10^{-7}$ | $3.87 \times 10^{-7}$ | $4.00 \times 10^{-7}$ | $4.44 \times 10^{-7}$ | $4.84 \times 10^{-7}$ |
| $Al_2O_3$ | $9.33 \times 10^{-6}$ | $1.90 \times 10^{-5}$ | $2.72 \times 10^{-5}$ | $3.44 \times 10^{-5}$ | $4.07 \times 10^{-5}$ | $1.75 \times 10^{-5}$ | $2.22 \times 10^{-5}$ | $2.71 \times 10^{-5}$ | $4.07 \times 10^{-5}$ | $5.62 \times 10^{-5}$ | $6.45 \times 10^{-5}$ | $7.19 \times 10^{-5}$ | $7.41 \times 10^{-5}$ | $8.22 \times 10^{-5}$ | $8.96 \times 10^{-5}$ |
| $CaAl_4O_7$ | $7.13 \times 10^{-6}$ | $1.45 \times 10^{-5}$ | $2.08 \times 10^{-5}$ | $2.63 \times 10^{-5}$ | $3.12 \times 10^{-5}$ | $1.34 \times 10^{-5}$ | $1.70 \times 10^{-5}$ | $2.07 \times 10^{-5}$ | $3.11 \times 10^{-5}$ | $4.30 \times 10^{-5}$ | $4.94 \times 10^{-5}$ | $5.50 \times 10^{-5}$ | $5.67 \times 10^{-5}$ | $6.29 \times 10^{-5}$ | $6.85 \times 10^{-5}$ |
| $Ca_2SiO_4$ | $9.36 \times 10^{-6}$ | $1.46 \times 10^{-5}$ | $1.84 \times 10^{-5}$ | $2.13 \times 10^{-5}$ | $2.38 \times 10^{-5}$ | $1.25 \times 10^{-5}$ | $1.55 \times 10^{-5}$ | $1.81 \times 10^{-5}$ | $2.40 \times 10^{-5}$ | $3.19 \times 10^{-5}$ | $3.60 \times 10^{-5}$ | $3.91 \times 10^{-5}$ | $3.96 \times 10^{-5}$ | $4.25 \times 10^{-5}$ | $4.52 \times 10^{-5}$ |
| $Ca_2Al_2SiO_7$ | $1.79 \times 10^{-5}$ | $2.79 \times 10^{-5}$ | $3.51 \times 10^{-5}$ | $4.07 \times 10^{-5}$ | $4.55 \times 10^{-5}$ | $2.38 \times 10^{-5}$ | $2.97 \times 10^{-5}$ | $3.45 \times 10^{-5}$ | $4.59 \times 10^{-5}$ | $6.09 \times 10^{-5}$ | $6.87 \times 10^{-5}$ | $7.47 \times 10^{-5}$ | $7.57 \times 10^{-5}$ | $8.11 \times 10^{-5}$ | $8.63 \times 10^{-5}$ |
| CaS | $4.39 \times 10^{-6}$ | $6.85 \times 10^{-6}$ | $8.61 \times 10^{-6}$ | $1.00 \times 10^{-5}$ | $1.12 \times 10^{-5}$ | $5.84 \times 10^{-6}$ | $7.29 \times 10^{-6}$ | $8.47 \times 10^{-6}$ | $1.13 \times 10^{-5}$ | $1.50 \times 10^{-5}$ | $1.69 \times 10^{-5}$ | $1.83 \times 10^{-5}$ | $1.86 \times 10^{-5}$ | $1.99 \times 10^{-5}$ | $2.12 \times 10^{-5}$ |
| $Mg_2SiO_4$ | $1.82 \times 10^{-4}$ | $2.92 \times 10^{-4}$ | $3.68 \times 10^{-4}$ | $4.27 \times 10^{-4}$ | $4.76 \times 10^{-4}$ | $2.43 \times 10^{-4}$ | $3.08 \times 10^{-4}$ | $3.58 \times 10^{-4}$ | $4.69 \times 10^{-4}$ | $5.76 \times 10^{-4}$ | $6.31 \times 10^{-4}$ | $6.86 \times 10^{-4}$ | $7.05 \times 10^{-4}$ | $7.78 \times 10^{-4}$ | $8.40 \times 10^{-4}$ |
| $MgAl_2O_4$ | $8.59 \times 10^{-7}$ | $1.38 \times 10^{-6}$ | $1.74 \times 10^{-6}$ | $2.02 \times 10^{-6}$ | $2.24 \times 10^{-6}$ | $1.15 \times 10^{-6}$ | $1.45 \times 10^{-6}$ | $1.69 \times 10^{-6}$ | $2.21 \times 10^{-6}$ | $2.72 \times 10^{-6}$ | $2.98 \times 10^{-6}$ | $3.24 \times 10^{-6}$ | $3.33 \times 10^{-6}$ | $3.67 \times 10^{-6}$ | $3.96 \times 10^{-6}$ |
| MgS | $6.80 \times 10^{-5}$ | $1.09 \times 10^{-4}$ | $1.38 \times 10^{-4}$ | $1.60 \times 10^{-4}$ | $1.78 \times 10^{-4}$ | $9.09 \times 10^{-5}$ | $1.15 \times 10^{-4}$ | $1.34 \times 10^{-4}$ | $1.75 \times 10^{-4}$ | $2.15 \times 10^{-4}$ | $2.36 \times 10^{-4}$ | $2.56 \times 10^{-4}$ | $2.63 \times 10^{-4}$ | $2.90 \times 10^{-4}$ | $3.14 \times 10^{-4}$ |
| $Ca_2MgSi_2O_7$ | $1.44 \times 10^{-5}$ | $2.32 \times 10^{-5}$ | $2.95 \times 10^{-5}$ | $3.46 \times 10^{-5}$ | $3.87 \times 10^{-5}$ | $1.96 \times 10^{-5}$ | $2.46 \times 10^{-5}$ | $2.87 \times 10^{-5}$ | $3.83 \times 10^{-5}$ | $4.96 \times 10^{-5}$ | $5.51 \times 10^{-5}$ | $5.94 \times 10^{-5}$ | $5.98 \times 10^{-5}$ | $6.37 \times 10^{-5}$ | $6.74 \times 10^{-5}$ |
| $MgSiO_3$ | $2.10 \times 10^{-4}$ | $3.38 \times 10^{-4}$ | $4.31 \times 10^{-4}$ | $5.04 \times 10^{-4}$ | $5.65 \times 10^{-4}$ | $2.86 \times 10^{-4}$ | $3.58 \times 10^{-4}$ | $4.18 \times 10^{-4}$ | $5.58 \times 10^{-4}$ | $7.23 \times 10^{-4}$ | $8.03 \times 10^{-4}$ | $8.66 \times 10^{-4}$ | $8.72 \times 10^{-4}$ | $9.29 \times 10^{-4}$ | $9.83 \times 10^{-4}$ |
| SiC | $6.72 \times 10^{-5}$ | $1.08 \times 10^{-4}$ | $1.38 \times 10^{-4}$ | $1.61 \times 10^{-4}$ | $1.81 \times 10^{-4}$ | $9.14 \times 10^{-5}$ | $1.14 \times 10^{-4}$ | $1.34 \times 10^{-4}$ | $1.78 \times 10^{-4}$ | $2.31 \times 10^{-4}$ | $2.57 \times 10^{-4}$ | $2.77 \times 10^{-4}$ | $2.79 \times 10^{-4}$ | $2.97 \times 10^{-4}$ | $3.14 \times 10^{-4}$ |
| $FeSiO_3$ | $2.12 \times 10^{-4}$ | $3.63 \times 10^{-4}$ | $4.88 \times 10^{-4}$ | $5.95 \times 10^{-4}$ | $6.90 \times 10^{-4}$ | $3.38 \times 10^{-4}$ | $4.27 \times 10^{-4}$ | $5.15 \times 10^{-4}$ | $7.53 \times 10^{-4}$ | $1.15 \times 10^{-3}$ | $1.37 \times 10^{-3}$ | $1.52 \times 10^{-3}$ | $1.56 \times 10^{-3}$ | $1.70 \times 10^{-3}$ | $1.83 \times 10^{-3}$ |
| Fe-metal | $6.74 \times 10^{-5}$ | $1.15 \times 10^{-4}$ | $1.55 \times 10^{-4}$ | $1.89 \times 10^{-4}$ | $2.19 \times 10^{-4}$ | $1.07 \times 10^{-4}$ | $1.36 \times 10^{-4}$ | $1.64 \times 10^{-4}$ | $2.39 \times 10^{-4}$ | $3.65 \times 10^{-4}$ | $4.34 \times 10^{-4}$ | $4.83 \times 10^{-4}$ | $4.96 \times 10^{-4}$ | $5.39 \times 10^{-4}$ | $5.81 \times 10^{-4}$ |
| $FeAl_2O_4$ | $2.10 \times 10^{-7}$ | $3.59 \times 10^{-7}$ | $4.83 \times 10^{-7}$ | $5.88 \times 10^{-7}$ | $6.82 \times 10^{-7}$ | $3.34 \times 10^{-7}$ | $4.22 \times 10^{-7}$ | $5.09 \times 10^{-7}$ | $7.45 \times 10^{-7}$ | $1.13 \times 10^{-6}$ | $1.35 \times 10^{-6}$ | $1.50 \times 10^{-6}$ | $1.54 \times 10^{-6}$ | $1.68 \times 10^{-6}$ | $1.81 \times 10^{-6}$ |
| $Fe_3C$ | $4.33 \times 10^{-5}$ | $7.41 \times 10^{-5}$ | $9.96 \times 10^{-5}$ | $1.21 \times 10^{-4}$ | $1.41 \times 10^{-4}$ | $6.90 \times 10^{-5}$ | $8.71 \times 10^{-5}$ | $1.05 \times 10^{-4}$ | $1.54 \times 10^{-4}$ | $2.34 \times 10^{-4}$ | $2.79 \times 10^{-4}$ | $3.10 \times 10^{-4}$ | $3.19 \times 10^{-4}$ | $3.46 \times 10^{-4}$ | $3.73 \times 10^{-4}$ |
| FeS | $3.41 \times 10^{-5}$ | $5.48 \times 10^{-5}$ | $7.00 \times 10^{-5}$ | $8.23 \times 10^{-5}$ | $9.27 \times 10^{-5}$ | $4.69 \times 10^{-5}$ | $5.84 \times 10^{-5}$ | $6.83 \times 10^{-5}$ | $9.28 \times 10^{-5}$ | $1.24 \times 10^{-4}$ | $1.41 \times 10^{-4}$ | $1.54 \times 10^{-4}$ | $1.56 \times 10^{-4}$ | $1.67 \times 10^{-4}$ | $1.78 \times 10^{-4}$ |
| $KAlSi_3O_8$ | $1.18 \times 10^{-6}$ | $2.36 \times 10^{-6}$ | $3.41 \times 10^{-6}$ | $4.34 \times 10^{-6}$ | $5.19 \times 10^{-6}$ | $2.26 \times 10^{-6}$ | $2.77 \times 10^{-6}$ | $3.37 \times 10^{-6}$ | $5.13 \times 10^{-6}$ | $7.31 \times 10^{-6}$ | $8.64 \times 10^{-6}$ | $9.88 \times 10^{-6}$ | $1.04 \times 10^{-5}$ | $1.18 \times 10^{-5}$ | $1.32 \times 10^{-5}$ |
| $NaAlSi_3O_8$ | $1.16 \times 10^{-7}$ | $2.64 \times 10^{-7}$ | $3.69 \times 10^{-7}$ | $4.57 \times 10^{-7}$ | $5.32 \times 10^{-7}$ | $2.27 \times 10^{-7}$ | $2.96 \times 10^{-7}$ | $3.63 \times 10^{-7}$ | $5.27 \times 10^{-7}$ | $6.90 \times 10^{-7}$ | $7.66 \times 10^{-7}$ | $8.32 \times 10^{-7}$ | $8.50 \times 10^{-7}$ | $9.24 \times 10^{-7}$ | $9.86 \times 10^{-7}$ |
| Graphite | $4.75 \times 10^{-6}$ | $7.59 \times 10^{-6}$ | $1.12 \times 10^{-5}$ | $1.45 \times 10^{-5}$ | $1.75 \times 10^{-5}$ | $8.27 \times 10^{-6}$ | $1.02 \times 10^{-5}$ | $1.23 \times 10^{-5}$ | $1.67 \times 10^{-5}$ | $1.90 \times 10^{-5}$ | $1.94 \times 10^{-5}$ | $2.04 \times 10^{-5}$ | $2.08 \times 10^{-5}$ | $2.28 \times 10^{-5}$ | $2.47 \times 10^{-5}$ |
| $H_2O$ | $1.88 \times 10^{-3}$ | $2.87 \times 10^{-3}$ | $3.52 \times 10^{-3}$ | $4.03 \times 10^{-3}$ | $4.38 \times 10^{-3}$ | $2.34 \times 10^{-3}$ | $2.94 \times 10^{-3}$ | $3.41 \times 10^{-3}$ | $4.24 \times 10^{-3}$ | $4.64 \times 10^{-3}$ | $4.85 \times 10^{-3}$ | $5.17 \times 10^{-3}$ | $5.24 \times 10^{-3}$ | $5.66 \times 10^{-3}$ | $6.05 \times 10^{-3}$ |

**Annular ring 3
(6-8 kpc)**

| Time (Gyr) | 0.2 | 0.4 | 0.6 | 0.8 | 1 | 1.3 | 1.7 | 2 | 3 | 5 | 7 | 9 | 10 | 12 | 13.5 |
|---|---|---|---|---|---|---|---|---|---|---|---|---|---|---|---|
| Gas density* | 2.76 | 3.9 | 4.34 | 4.41 | 4.3 | 8.53 | 11.5 | 12.7 | 14.4 | 14.4 | 13.7 | 12.7 | 12.4 | 11.5 | 10.7 |
| $CaTiO_3$ | $1.08\times10^{-6}$ | $1.91\times10^{-6}$ | $2.63\times10^{-6}$ | $3.28\times10^{-6}$ | $3.87\times10^{-6}$ | $2.65\times10^{-6}$ | $2.79\times10^{-6}$ | $3.13\times10^{-6}$ | $4.33\times10^{-6}$ | $6.00\times10^{-6}$ | $6.96\times10^{-6}$ | $7.65\times10^{-6}$ | $7.95\times10^{-6}$ | $8.71\times10^{-6}$ | $9.36\times10^{-6}$ |
| $Ti_2O_3$ | $6.37\times10^{-9}$ | $1.12\times10^{-8}$ | $1.55\times10^{-8}$ | $1.93\times10^{-8}$ | $2.27\times10^{-8}$ | $1.56\times10^{-8}$ | $1.64\times10^{-8}$ | $1.84\times10^{-8}$ | $2.54\times10^{-8}$ | $3.53\times10^{-8}$ | $4.09\times10^{-8}$ | $4.49\times10^{-8}$ | $4.67\times10^{-8}$ | $5.12\times10^{-8}$ | $5.50\times10^{-8}$ |
| TiC | $4.77\times10^{-8}$ | $8.43\times10^{-8}$ | $1.16\times10^{-7}$ | $1.44\times10^{-7}$ | $1.71\times10^{-7}$ | $1.17\times10^{-7}$ | $1.23\times10^{-7}$ | $1.38\times10^{-7}$ | $1.91\times10^{-7}$ | $2.64\times10^{-7}$ | $3.07\times10^{-7}$ | $3.37\times10^{-7}$ | $3.50\times10^{-7}$ | $3.84\times10^{-7}$ | $4.12\times10^{-7}$ |
| $Al_2O_3$ | $6.56\times10^{-6}$ | $1.40\times10^{-5}$ | $2.10\times10^{-5}$ | $2.76\times10^{-5}$ | $3.37\times10^{-5}$ | $2.16\times10^{-5}$ | $2.26\times10^{-5}$ | $2.54\times10^{-5}$ | $3.58\times10^{-5}$ | $4.96\times10^{-5}$ | $5.70\times10^{-5}$ | $6.26\times10^{-5}$ | $6.49\times10^{-5}$ | $7.12\times10^{-5}$ | $7.65\times10^{-5}$ |
| $CaAl_4O_7$ | $5.02\times10^{-6}$ | $1.07\times10^{-5}$ | $1.61\times10^{-5}$ | $2.11\times10^{-5}$ | $2.58\times10^{-5}$ | $1.66\times10^{-5}$ | $1.73\times10^{-5}$ | $1.95\times10^{-5}$ | $2.74\times10^{-5}$ | $3.79\times10^{-5}$ | $4.36\times10^{-5}$ | $4.79\times10^{-5}$ | $4.96\times10^{-5}$ | $5.45\times10^{-5}$ | $5.85\times10^{-5}$ |
| $Ca_2SiO_4$ | $7.40\times10^{-6}$ | $1.21\times10^{-5}$ | $1.57\times10^{-5}$ | $1.87\times10^{-5}$ | $2.13\times10^{-5}$ | $1.47\times10^{-5}$ | $1.56\times10^{-5}$ | $1.73\times10^{-5}$ | $2.26\times10^{-5}$ | $2.93\times10^{-5}$ | $3.27\times10^{-5}$ | $3.50\times10^{-5}$ | $3.57\times10^{-5}$ | $3.81\times10^{-5}$ | $4.01\times10^{-5}$ |
| $Ca_2Al_2SiO_7$ | $1.18\times10^{-5}$ | $1.93\times10^{-5}$ | $2.50\times10^{-5}$ | $2.98\times10^{-5}$ | $3.39\times10^{-5}$ | $2.34\times10^{-5}$ | $2.48\times10^{-5}$ | $2.75\times10^{-5}$ | $3.60\times10^{-5}$ | $4.66\times10^{-5}$ | $5.20\times10^{-5}$ | $5.57\times10^{-5}$ | $5.69\times10^{-5}$ | $6.06\times10^{-5}$ | $6.39\times10^{-5}$ |
| CaS | $3.72\times10^{-6}$ | $6.08\times10^{-6}$ | $7.89\times10^{-6}$ | $9.41\times10^{-6}$ | $1.07\times10^{-5}$ | $7.39\times10^{-6}$ | $7.84\times10^{-6}$ | $8.69\times10^{-6}$ | $1.14\times10^{-5}$ | $1.47\times10^{-5}$ | $1.64\times10^{-5}$ | $1.76\times10^{-5}$ | $1.80\times10^{-5}$ | $1.91\times10^{-5}$ | $2.02\times10^{-5}$ |
| $Mg_2SiO_4$ | $1.41\times10^{-4}$ | $2.39\times10^{-4}$ | $3.13\times10^{-4}$ | $3.74\times10^{-4}$ | $4.26\times10^{-4}$ | $2.86\times10^{-4}$ | $3.05\times10^{-4}$ | $3.36\times10^{-4}$ | $4.31\times10^{-4}$ | $5.31\times10^{-4}$ | $5.80\times10^{-4}$ | $6.21\times10^{-4}$ | $6.39\times10^{-4}$ | $6.93\times10^{-4}$ | $7.36\times10^{-4}$ |
| $MgAl_2O_4$ | $6.67\times10^{-7}$ | $1.13\times10^{-6}$ | $1.48\times10^{-6}$ | $1.76\times10^{-6}$ | $2.01\times10^{-6}$ | $1.35\times10^{-6}$ | $1.44\times10^{-6}$ | $1.59\times10^{-6}$ | $2.03\times10^{-6}$ | $2.51\times10^{-6}$ | $2.74\times10^{-6}$ | $2.93\times10^{-6}$ | $3.02\times10^{-6}$ | $3.27\times10^{-6}$ | $3.47\times10^{-6}$ |
| MgS | $5.28\times10^{-5}$ | $8.93\times10^{-5}$ | $1.17\times10^{-4}$ | $1.40\times10^{-4}$ | $1.59\times10^{-4}$ | $1.07\times10^{-4}$ | $1.14\times10^{-4}$ | $1.26\times10^{-4}$ | $1.61\times10^{-4}$ | $1.98\times10^{-4}$ | $2.17\times10^{-4}$ | $2.32\times10^{-4}$ | $2.39\times10^{-4}$ | $2.59\times10^{-4}$ | $2.75\times10^{-4}$ |
| $Ca_2MgSi_2O_7$ | $1.13\times10^{-5}$ | $1.89\times10^{-5}$ | $2.49\times10^{-5}$ | $3.00\times10^{-5}$ | $3.44\times10^{-5}$ | $2.33\times10^{-5}$ | $2.47\times10^{-5}$ | $2.73\times10^{-5}$ | $3.57\times10^{-5}$ | $4.56\times10^{-5}$ | $5.03\times10^{-5}$ | $5.36\times10^{-5}$ | $5.45\times10^{-5}$ | $5.78\times10^{-5}$ | $6.07\times10^{-5}$ |
| $MgSiO_3$ | $1.64\times10^{-4}$ | $2.75\times10^{-4}$ | $3.63\times10^{-4}$ | $4.37\times10^{-4}$ | $5.01\times10^{-4}$ | $3.40\times10^{-4}$ | $3.60\times10^{-4}$ | $3.99\times10^{-4}$ | $5.20\times10^{-4}$ | $6.65\times10^{-4}$ | $7.34\times10^{-4}$ | $7.82\times10^{-4}$ | $7.95\times10^{-4}$ | $8.43\times10^{-4}$ | $8.85\times10^{-4}$ |
| SiC | $5.25\times10^{-5}$ | $8.80\times10^{-5}$ | $1.16\times10^{-4}$ | $1.40\times10^{-4}$ | $1.60\times10^{-4}$ | $1.09\times10^{-4}$ | $1.15\times10^{-4}$ | $1.27\times10^{-4}$ | $1.66\times10^{-4}$ | $2.12\times10^{-4}$ | $2.34\times10^{-4}$ | $2.50\times10^{-4}$ | $2.54\times10^{-4}$ | $2.69\times10^{-4}$ | $2.83\times10^{-4}$ |
| $FeSiO_3$ | $1.67\times10^{-4}$ | $2.98\times10^{-4}$ | $4.13\times10^{-4}$ | $5.16\times10^{-4}$ | $6.09\times10^{-4}$ | $4.24\times10^{-4}$ | $4.47\times10^{-4}$ | $5.10\times10^{-4}$ | $7.25\times10^{-4}$ | $1.04\times10^{-3}$ | $1.21\times10^{-3}$ | $1.32\times10^{-3}$ | $1.37\times10^{-3}$ | $1.48\times10^{-3}$ | $1.58\times10^{-3}$ |
| Fe-metal | $5.30\times10^{-5}$ | $9.46\times10^{-5}$ | $1.31\times10^{-4}$ | $1.64\times10^{-4}$ | $1.93\times10^{-4}$ | $1.34\times10^{-4}$ | $1.42\times10^{-4}$ | $1.62\times10^{-4}$ | $2.30\times10^{-4}$ | $3.29\times10^{-4}$ | $3.84\times10^{-4}$ | $4.20\times10^{-4}$ | $4.34\times10^{-4}$ | $4.69\times10^{-4}$ | $5.00\times10^{-4}$ |
| $FeAl_2O_4$ | $1.65\times10^{-7}$ | $2.95\times10^{-7}$ | $4.08\times10^{-7}$ | $5.10\times10^{-7}$ | $6.02\times10^{-7}$ | $4.19\times10^{-7}$ | $4.42\times10^{-7}$ | $5.03\times10^{-7}$ | $7.16\times10^{-7}$ | $1.03\times10^{-6}$ | $1.20\times10^{-6}$ | $1.31\times10^{-6}$ | $1.35\times10^{-6}$ | $1.46\times10^{-6}$ | $1.56\times10^{-6}$ |
| $Fe_3C$ | $3.40\times10^{-5}$ | $6.08\times10^{-5}$ | $8.43\times10^{-5}$ | $1.05\times10^{-4}$ | $1.24\times10^{-4}$ | $8.64\times10^{-5}$ | $9.12\times10^{-5}$ | $1.04\times10^{-4}$ | $1.48\times10^{-4}$ | $2.12\times10^{-4}$ | $2.47\times10^{-4}$ | $2.70\times10^{-4}$ | $2.79\times10^{-4}$ | $3.01\times10^{-4}$ | $3.21\times10^{-4}$ |
| FeS | $2.66\times10^{-5}$ | $4.46\times10^{-5}$ | $5.88\times10^{-5}$ | $7.10\times10^{-5}$ | $8.17\times10^{-5}$ | $5.57\times10^{-5}$ | $5.88\times10^{-5}$ | $6.53\times10^{-5}$ | $8.64\times10^{-5}$ | $1.13\times10^{-4}$ | $1.27\times10^{-4}$ | $1.37\times10^{-4}$ | $1.40\times10^{-4}$ | $1.49\times10^{-4}$ | $1.57\times10^{-4}$ |
| $KAlSi_3O_8$ | $8.53\times10^{-7}$ | $1.74\times10^{-6}$ | $2.60\times10^{-6}$ | $3.43\times10^{-6}$ | $4.23\times10^{-6}$ | $2.77\times10^{-6}$ | $2.87\times10^{-6}$ | $3.20\times10^{-6}$ | $4.54\times10^{-6}$ | $6.43\times10^{-6}$ | $7.57\times10^{-6}$ | $8.49\times10^{-6}$ | $8.92\times10^{-6}$ | $9.99\times10^{-6}$ | $1.09\times10^{-5}$ |
| $NaAlSi_3O_8$ | $8.21\times10^{-8}$ | $2.00\times10^{-7}$ | $2.96\times10^{-7}$ | $3.79\times10^{-7}$ | $4.55\times10^{-7}$ | $2.88\times10^{-7}$ | $3.02\times10^{-7}$ | $3.40\times10^{-7}$ | $4.68\times10^{-7}$ | $6.18\times10^{-7}$ | $6.85\times10^{-7}$ | $7.36\times10^{-7}$ | $7.54\times10^{-7}$ | $8.12\times10^{-7}$ | $8.61\times10^{-7}$ |
| Graphite | $3.74\times10^{-6}$ | $6.26\times10^{-6}$ | $9.23\times10^{-6}$ | $1.24\times10^{-5}$ | $1.54\times10^{-5}$ | $1.08\times10^{-5}$ | $1.10\times10^{-5}$ | $1.22\times10^{-5}$ | $1.57\times10^{-5}$ | $1.82\times10^{-5}$ | $1.87\times10^{-5}$ | $1.93\times10^{-5}$ | $1.97\times10^{-5}$ | $2.10\times10^{-5}$ | $2.22\times10^{-5}$ |
| $H_2O$ | $1.18\times10^{-3}$ | $1.93\times10^{-3}$ | $2.41\times10^{-3}$ | $2.85\times10^{-3}$ | $3.16\times10^{-3}$ | $2.15\times10^{-3}$ | $2.30\times10^{-3}$ | $2.51\times10^{-3}$ | $3.01\times10^{-3}$ | $3.35\times10^{-3}$ | $3.47\times10^{-3}$ | $3.55\times10^{-3}$ | $3.61\times10^{-3}$ | $3.85\times10^{-3}$ | $4.05\times10^{-3}$ |

**Annular ring 4 (8-10 kpc)**

| Time (Gyr) | 0.2 | 0.4 | 0.6 | 0.8 | 1 | 1.3 | 1.7 | 2 | 3 | 5 | 7 | 9 | 10 | 12 | 13.5 |
|---|---|---|---|---|---|---|---|---|---|---|---|---|---|---|---|
| Gas density* | 3.17 | 4.88 | 5.74 | 6.1 | 6.2 | 8.7 | 10.7 | 11.6 | 12.8 | 13 | 12.8 | 12.4 | 12.2 | 11.7 | 11.1 |
| $CaTiO_3$ | $6.68 \times 10^{-7}$ | $1.24 \times 10^{-6}$ | $1.75 \times 10^{-6}$ | $2.25 \times 10^{-6}$ | $2.73 \times 10^{-6}$ | $2.47 \times 10^{-6}$ | $2.57 \times 10^{-6}$ | $2.79 \times 10^{-6}$ | $3.74 \times 10^{-6}$ | $5.22 \times 10^{-6}$ | $6.16 \times 10^{-6}$ | $6.80 \times 10^{-6}$ | $7.09 \times 10^{-6}$ | $7.72 \times 10^{-6}$ | $8.21 \times 10^{-6}$ |
| $Ti_2O_3$ | $3.92 \times 10^{-9}$ | $7.28 \times 10^{-9}$ | $1.03 \times 10^{-8}$ | $1.32 \times 10^{-8}$ | $1.60 \times 10^{-8}$ | $1.45 \times 10^{-8}$ | $1.51 \times 10^{-8}$ | $1.64 \times 10^{-8}$ | $2.20 \times 10^{-8}$ | $3.07 \times 10^{-8}$ | $3.62 \times 10^{-8}$ | $3.99 \times 10^{-8}$ | $4.16 \times 10^{-8}$ | $4.53 \times 10^{-8}$ | $4.83 \times 10^{-8}$ |
| TiC | $2.94 \times 10^{-8}$ | $5.46 \times 10^{-8}$ | $7.73 \times 10^{-8}$ | $9.91 \times 10^{-8}$ | $1.20 \times 10^{-7}$ | $1.09 \times 10^{-7}$ | $1.13 \times 10^{-7}$ | $1.23 \times 10^{-7}$ | $1.65 \times 10^{-7}$ | $2.30 \times 10^{-7}$ | $2.71 \times 10^{-7}$ | $2.99 \times 10^{-7}$ | $3.12 \times 10^{-7}$ | $3.40 \times 10^{-7}$ | $3.62 \times 10^{-7}$ |
| $Al_2O_3$ | $3.31 \times 10^{-6}$ | $7.51 \times 10^{-6}$ | $1.18 \times 10^{-5}$ | $1.62 \times 10^{-5}$ | $2.07 \times 10^{-5}$ | $1.83 \times 10^{-5}$ | $1.93 \times 10^{-5}$ | $2.10 \times 10^{-5}$ | $2.84 \times 10^{-5}$ | $4.05 \times 10^{-5}$ | $4.85 \times 10^{-5}$ | $5.42 \times 10^{-5}$ | $5.66 \times 10^{-5}$ | $6.19 \times 10^{-5}$ | $6.60 \times 10^{-5}$ |
| $CaAl_4O_7$ | $2.53 \times 10^{-6}$ | $5.74 \times 10^{-6}$ | $9.01 \times 10^{-6}$ | $1.24 \times 10^{-5}$ | $1.58 \times 10^{-5}$ | $1.40 \times 10^{-5}$ | $1.47 \times 10^{-5}$ | $1.61 \times 10^{-5}$ | $2.17 \times 10^{-5}$ | $3.10 \times 10^{-5}$ | $3.71 \times 10^{-5}$ | $4.14 \times 10^{-5}$ | $4.33 \times 10^{-5}$ | $4.73 \times 10^{-5}$ | $5.05 \times 10^{-5}$ |
| $Ca_2SiO_4$ | $4.64 \times 10^{-6}$ | $8.14 \times 10^{-6}$ | $1.10 \times 10^{-5}$ | $1.36 \times 10^{-5}$ | $1.60 \times 10^{-5}$ | $1.42 \times 10^{-5}$ | $1.46 \times 10^{-5}$ | $1.56 \times 10^{-5}$ | $2.00 \times 10^{-5}$ | $2.62 \times 10^{-5}$ | $2.97 \times 10^{-5}$ | $3.19 \times 10^{-5}$ | $3.28 \times 10^{-5}$ | $3.49 \times 10^{-5}$ | $3.65 \times 10^{-5}$ |
| $Ca_2Al_2SiO_7$ | $5.91 \times 10^{-6}$ | $1.04 \times 10^{-5}$ | $1.40 \times 10^{-5}$ | $1.73 \times 10^{-5}$ | $2.03 \times 10^{-5}$ | $1.80 \times 10^{-5}$ | $1.86 \times 10^{-5}$ | $1.99 \times 10^{-5}$ | $2.54 \times 10^{-5}$ | $3.34 \times 10^{-5}$ | $3.79 \times 10^{-5}$ | $4.07 \times 10^{-5}$ | $4.18 \times 10^{-5}$ | $4.44 \times 10^{-5}$ | $4.65 \times 10^{-5}$ |
| CaS | $2.49 \times 10^{-6}$ | $4.37 \times 10^{-6}$ | $5.90 \times 10^{-6}$ | $7.29 \times 10^{-6}$ | $8.56 \times 10^{-6}$ | $7.59 \times 10^{-6}$ | $7.82 \times 10^{-6}$ | $8.39 \times 10^{-6}$ | $1.07 \times 10^{-5}$ | $1.40 \times 10^{-5}$ | $1.59 \times 10^{-5}$ | $1.71 \times 10^{-5}$ | $1.76 \times 10^{-5}$ | $1.87 \times 10^{-5}$ | $1.96 \times 10^{-5}$ |
| $Mg_2SiO_4$ | $8.34 \times 10^{-5}$ | $1.54 \times 10^{-4}$ | $2.12 \times 10^{-4}$ | $2.64 \times 10^{-4}$ | $3.12 \times 10^{-4}$ | $2.71 \times 10^{-4}$ | $2.79 \times 10^{-4}$ | $2.98 \times 10^{-4}$ | $3.70 \times 10^{-4}$ | $4.70 \times 10^{-4}$ | $5.26 \times 10^{-4}$ | $5.66 \times 10^{-4}$ | $5.83 \times 10^{-4}$ | $6.25 \times 10^{-4}$ | $6.57 \times 10^{-4}$ |
| $MgAl_2O_4$ | $3.93 \times 10^{-7}$ | $7.28 \times 10^{-7}$ | $1.00 \times 10^{-6}$ | $1.25 \times 10^{-6}$ | $1.47 \times 10^{-6}$ | $1.28 \times 10^{-6}$ | $1.32 \times 10^{-6}$ | $1.40 \times 10^{-6}$ | $1.75 \times 10^{-6}$ | $2.22 \times 10^{-6}$ | $2.48 \times 10^{-6}$ | $2.67 \times 10^{-6}$ | $2.75 \times 10^{-6}$ | $2.95 \times 10^{-6}$ | $3.10 \times 10^{-6}$ |
| MgS | $3.11 \times 10^{-5}$ | $5.76 \times 10^{-5}$ | $7.93 \times 10^{-5}$ | $9.86 \times 10^{-5}$ | $1.16 \times 10^{-4}$ | $1.01 \times 10^{-4}$ | $1.04 \times 10^{-4}$ | $1.11 \times 10^{-4}$ | $1.38 \times 10^{-4}$ | $1.75 \times 10^{-4}$ | $1.96 \times 10^{-4}$ | $2.11 \times 10^{-4}$ | $2.18 \times 10^{-4}$ | $2.33 \times 10^{-4}$ | $2.45 \times 10^{-4}$ |
| $Ca_2MgSi_2O_7$ | $6.96 \times 10^{-6}$ | $1.24 \times 10^{-5}$ | $1.70 \times 10^{-5}$ | $2.12 \times 10^{-5}$ | $2.51 \times 10^{-5}$ | $2.22 \times 10^{-5}$ | $2.29 \times 10^{-5}$ | $2.45 \times 10^{-5}$ | $3.12 \times 10^{-5}$ | $4.07 \times 10^{-5}$ | $4.59 \times 10^{-5}$ | $4.91 \times 10^{-5}$ | $5.03 \times 10^{-5}$ | $5.33 \times 10^{-5}$ | $5.57 \times 10^{-5}$ |
| $MgSiO_3$ | $1.01 \times 10^{-4}$ | $1.81 \times 10^{-4}$ | $2.48 \times 10^{-4}$ | $3.09 \times 10^{-4}$ | $3.67 \times 10^{-4}$ | $3.23 \times 10^{-4}$ | $3.34 \times 10^{-4}$ | $3.58 \times 10^{-4}$ | $4.55 \times 10^{-4}$ | $5.93 \times 10^{-4}$ | $6.69 \times 10^{-4}$ | $7.17 \times 10^{-4}$ | $7.34 \times 10^{-4}$ | $7.77 \times 10^{-4}$ | $8.12 \times 10^{-4}$ |
| SiC | $3.24 \times 10^{-5}$ | $5.79 \times 10^{-5}$ | $7.93 \times 10^{-5}$ | $9.88 \times 10^{-5}$ | $1.17 \times 10^{-4}$ | $1.03 \times 10^{-4}$ | $1.07 \times 10^{-4}$ | $1.14 \times 10^{-4}$ | $1.45 \times 10^{-4}$ | $1.90 \times 10^{-4}$ | $2.14 \times 10^{-4}$ | $2.29 \times 10^{-4}$ | $2.34 \times 10^{-4}$ | $2.48 \times 10^{-4}$ | $2.59 \times 10^{-4}$ |
| $FeSiO_3$ | $1.02 \times 10^{-4}$ | $1.98 \times 10^{-4}$ | $2.84 \times 10^{-4}$ | $3.67 \times 10^{-4}$ | $4.46 \times 10^{-4}$ | $4.09 \times 10^{-4}$ | $4.27 \times 10^{-4}$ | $4.68 \times 10^{-4}$ | $6.49 \times 10^{-4}$ | $9.21 \times 10^{-4}$ | $1.09 \times 10^{-3}$ | $1.19 \times 10^{-3}$ | $1.23 \times 10^{-3}$ | $1.33 \times 10^{-3}$ | $1.40 \times 10^{-3}$ |
| Fe-metal | $3.25 \times 10^{-5}$ | $6.28 \times 10^{-5}$ | $9.02 \times 10^{-5}$ | $1.16 \times 10^{-4}$ | $1.41 \times 10^{-4}$ | $1.30 \times 10^{-4}$ | $1.36 \times 10^{-4}$ | $1.49 \times 10^{-4}$ | $2.06 \times 10^{-4}$ | $2.92 \times 10^{-4}$ | $3.45 \times 10^{-4}$ | $3.77 \times 10^{-4}$ | $3.91 \times 10^{-4}$ | $4.21 \times 10^{-4}$ | $4.45 \times 10^{-4}$ |
| $FeAl_2O_4$ | $1.01 \times 10^{-7}$ | $1.95 \times 10^{-7}$ | $2.81 \times 10^{-7}$ | $3.62 \times 10^{-7}$ | $4.40 \times 10^{-7}$ | $4.04 \times 10^{-7}$ | $4.22 \times 10^{-7}$ | $4.63 \times 10^{-7}$ | $6.41 \times 10^{-7}$ | $9.10 \times 10^{-7}$ | $1.07 \times 10^{-6}$ | $1.17 \times 10^{-6}$ | $1.22 \times 10^{-6}$ | $1.31 \times 10^{-6}$ | $1.39 \times 10^{-6}$ |
| $Fe_3C$ | $2.09 \times 10^{-5}$ | $4.03 \times 10^{-5}$ | $5.79 \times 10^{-5}$ | $7.48 \times 10^{-5}$ | $9.09 \times 10^{-5}$ | $8.35 \times 10^{-5}$ | $8.70 \times 10^{-5}$ | $9.55 \times 10^{-5}$ | $1.32 \times 10^{-4}$ | $1.88 \times 10^{-4}$ | $2.21 \times 10^{-4}$ | $2.42 \times 10^{-4}$ | $2.51 \times 10^{-4}$ | $2.71 \times 10^{-4}$ | $2.86 \times 10^{-4}$ |
| FeS | $1.65 \times 10^{-5}$ | $2.94 \times 10^{-5}$ | $4.02 \times 10^{-5}$ | $5.02 \times 10^{-5}$ | $5.95 \times 10^{-5}$ | $5.26 \times 10^{-5}$ | $5.44 \times 10^{-5}$ | $5.84 \times 10^{-5}$ | $7.50 \times 10^{-5}$ | $9.99 \times 10^{-5}$ | $1.14 \times 10^{-4}$ | $1.24 \times 10^{-4}$ | $1.27 \times 10^{-4}$ | $1.36 \times 10^{-4}$ | $1.43 \times 10^{-4}$ |
| $KAlSi_3O_8$ | $4.78 \times 10^{-7}$ | $9.72 \times 10^{-7}$ | $1.48 \times 10^{-6}$ | $2.02 \times 10^{-6}$ | $2.57 \times 10^{-6}$ | $2.31 \times 10^{-6}$ | $2.44 \times 10^{-6}$ | $2.65 \times 10^{-6}$ | $3.59 \times 10^{-6}$ | $5.23 \times 10^{-6}$ | $6.39 \times 10^{-6}$ | $7.27 \times 10^{-6}$ | $7.68 \times 10^{-6}$ | $8.56 \times 10^{-6}$ | $9.26 \times 10^{-6}$ |
| $NaAlSi_3O_8$ | $3.97 \times 10^{-8}$ | $1.12 \times 10^{-7}$ | $1.83 \times 10^{-7}$ | $2.40 \times 10^{-7}$ | $2.99 \times 10^{-7}$ | $2.61 \times 10^{-7}$ | $2.69 \times 10^{-7}$ | $2.91 \times 10^{-7}$ | $3.80 \times 10^{-7}$ | $5.18 \times 10^{-7}$ | $5.98 \times 10^{-7}$ | $6.53 \times 10^{-7}$ | $6.72 \times 10^{-7}$ | $7.18 \times 10^{-7}$ | $7.51 \times 10^{-7}$ |
| Graphite | $2.30 \times 10^{-6}$ | $4.17 \times 10^{-6}$ | $6.11 \times 10^{-6}$ | $8.42 \times 10^{-6}$ | $1.10 \times 10^{-5}$ | $1.07 \times 10^{-5}$ | $1.11 \times 10^{-5}$ | $1.18 \times 10^{-5}$ | $1.45 \times 10^{-5}$ | $1.74 \times 10^{-5}$ | $1.83 \times 10^{-5}$ | $1.89 \times 10^{-5}$ | $1.91 \times 10^{-5}$ | $2.00 \times 10^{-5}$ | $2.08 \times 10^{-5}$ |
| $H_2O$ | $5.56 \times 10^{-4}$ | $1.01 \times 10^{-3}$ | $1.38 \times 10^{-3}$ | $1.66 \times 10^{-3}$ | $1.90 \times 10^{-3}$ | $1.63 \times 10^{-3}$ | $1.67 \times 10^{-3}$ | $1.70 \times 10^{-3}$ | $1.97 \times 10^{-3}$ | $2.22 \times 10^{-3}$ | $2.33 \times 10^{-3}$ | $2.46 \times 10^{-3}$ | $2.41 \times 10^{-3}$ | $2.54 \times 10^{-3}$ | $2.66 \times 10^{-3}$ |

**Annular ring 5 (10-12 kpc)**

| Time (Gyr) | 0.2 | 0.4 | 0.6 | 0.8 | 1 | 1.3 | 1.7 | 2 | 3 | 5 | 7 | 9 | 10 | 12 | 13.5 |
|---|---|---|---|---|---|---|---|---|---|---|---|---|---|---|---|
| Gas density* | 2.66 | 4.24 | 5.13 | 5.57 | 5.7 | 7.34 | 8.59 | 9.15 | 9.89 | 9.91 | 9.8 | 9.7 | 9.65 | 9.36 | 9.06 |
| $CaTiO_3$ | $4.94 \times 10^{-7}$ | $9.71 \times 10^{-7}$ | $1.39 \times 10^{-6}$ | $1.79 \times 10^{-6}$ | $2.19 \times 10^{-6}$ | $2.17 \times 10^{-6}$ | $2.32 \times 10^{-6}$ | $2.53 \times 10^{-6}$ | $3.40 \times 10^{-6}$ | $4.80 \times 10^{-6}$ | $5.68 \times 10^{-6}$ | $6.26 \times 10^{-6}$ | $6.53 \times 10^{-6}$ | $7.08 \times 10^{-6}$ | $7.50 \times 10^{-6}$ |
| $Ti_2O_3$ | $2.90 \times 10^{-9}$ | $5.70 \times 10^{-9}$ | $8.15 \times 10^{-9}$ | $1.05 \times 10^{-8}$ | $1.29 \times 10^{-8}$ | $1.27 \times 10^{-8}$ | $1.37 \times 10^{-8}$ | $1.49 \times 10^{-8}$ | $2.00 \times 10^{-8}$ | $2.82 \times 10^{-8}$ | $3.34 \times 10^{-8}$ | $3.68 \times 10^{-8}$ | $3.83 \times 10^{-8}$ | $4.16 \times 10^{-8}$ | $4.41 \times 10^{-8}$ |
| TiC | $2.18 \times 10^{-8}$ | $4.28 \times 10^{-8}$ | $6.11 \times 10^{-8}$ | $7.89 \times 10^{-8}$ | $9.64 \times 10^{-8}$ | $9.55 \times 10^{-8}$ | $1.02 \times 10^{-7}$ | $1.12 \times 10^{-7}$ | $1.50 \times 10^{-7}$ | $2.11 \times 10^{-7}$ | $2.50 \times 10^{-7}$ | $2.76 \times 10^{-7}$ | $2.88 \times 10^{-7}$ | $3.12 \times 10^{-7}$ | $3.30 \times 10^{-7}$ |
| $Al_2O_3$ | $2.12 \times 10^{-6}$ | $5.27 \times 10^{-6}$ | $8.37 \times 10^{-6}$ | $1.16 \times 10^{-5}$ | $1.50 \times 10^{-5}$ | $1.48 \times 10^{-5}$ | $1.62 \times 10^{-5}$ | $1.78 \times 10^{-5}$ | $2.43 \times 10^{-5}$ | $3.55 \times 10^{-5}$ | $4.29 \times 10^{-5}$ | $4.80 \times 10^{-5}$ | $5.02 \times 10^{-5}$ | $5.48 \times 10^{-5}$ | $5.83 \times 10^{-5}$ |
| $CaAl_4O_7$ | $1.42 \times 10^{-6}$ | $3.53 \times 10^{-6}$ | $5.61 \times 10^{-6}$ | $7.78 \times 10^{-6}$ | $1.01 \times 10^{-5}$ | $9.88 \times 10^{-6}$ | $1.08 \times 10^{-5}$ | $1.19 \times 10^{-5}$ | $1.62 \times 10^{-5}$ | $2.37 \times 10^{-5}$ | $2.87 \times 10^{-5}$ | $3.21 \times 10^{-5}$ | $3.36 \times 10^{-5}$ | $3.67 \times 10^{-5}$ | $3.90 \times 10^{-5}$ |
| $Ca_2SiO_4$ | $3.46 \times 10^{-6}$ | $6.50 \times 10^{-6}$ | $8.93 \times 10^{-6}$ | $1.12 \times 10^{-5}$ | $1.33 \times 10^{-5}$ | $1.28 \times 10^{-5}$ | $1.35 \times 10^{-5}$ | $1.45 \times 10^{-5}$ | $1.86 \times 10^{-5}$ | $2.47 \times 10^{-5}$ | $2.82 \times 10^{-5}$ | $3.03 \times 10^{-5}$ | $3.11 \times 10^{-5}$ | $3.30 \times 10^{-5}$ | $3.44 \times 10^{-5}$ |
| $Ca_2Al_2SiO_7$ | $4.40 \times 10^{-6}$ | $8.28 \times 10^{-6}$ | $1.14 \times 10^{-5}$ | $1.42 \times 10^{-5}$ | $1.69 \times 10^{-5}$ | $1.63 \times 10^{-5}$ | $1.72 \times 10^{-5}$ | $1.85 \times 10^{-5}$ | $2.37 \times 10^{-5}$ | $3.15 \times 10^{-5}$ | $3.59 \times 10^{-5}$ | $3.85 \times 10^{-5}$ | $3.96 \times 10^{-5}$ | $4.20 \times 10^{-5}$ | $4.39 \times 10^{-5}$ |
| CaS | $1.85 \times 10^{-6}$ | $3.48 \times 10^{-6}$ | $4.79 \times 10^{-6}$ | $5.98 \times 10^{-6}$ | $7.11 \times 10^{-6}$ | $6.87 \times 10^{-6}$ | $7.26 \times 10^{-6}$ | $7.80 \times 10^{-6}$ | $9.98 \times 10^{-6}$ | $1.33 \times 10^{-5}$ | $1.51 \times 10^{-5}$ | $1.62 \times 10^{-5}$ | $1.67 \times 10^{-5}$ | $1.77 \times 10^{-5}$ | $1.85 \times 10^{-5}$ |
| $Mg_2SiO_4$ | $5.81 \times 10^{-5}$ | $1.18 \times 10^{-4}$ | $1.67 \times 10^{-4}$ | $2.10 \times 10^{-4}$ | $2.51 \times 10^{-4}$ | $2.39 \times 10^{-4}$ | $2.52 \times 10^{-4}$ | $2.70 \times 10^{-4}$ | $3.36 \times 10^{-4}$ | $4.33 \times 10^{-4}$ | $4.87 \times 10^{-4}$ | $5.22 \times 10^{-4}$ | $5.38 \times 10^{-4}$ | $5.73 \times 10^{-4}$ | $6.00 \times 10^{-4}$ |
| $MgAl_2O_4$ | $2.74 \times 10^{-7}$ | $5.58 \times 10^{-7}$ | $7.86 \times 10^{-7}$ | $9.91 \times 10^{-7}$ | $1.19 \times 10^{-6}$ | $1.13 \times 10^{-6}$ | $1.19 \times 10^{-6}$ | $1.27 \times 10^{-6}$ | $1.58 \times 10^{-6}$ | $2.04 \times 10^{-6}$ | $2.30 \times 10^{-6}$ | $2.46 \times 10^{-6}$ | $2.54 \times 10^{-6}$ | $2.71 \times 10^{-6}$ | $2.83 \times 10^{-6}$ |
| MgS | $2.17 \times 10^{-5}$ | $4.42 \times 10^{-5}$ | $6.22 \times 10^{-5}$ | $7.84 \times 10^{-5}$ | $9.38 \times 10^{-5}$ | $8.93 \times 10^{-5}$ | $9.42 \times 10^{-5}$ | $1.01 \times 10^{-4}$ | $1.25 \times 10^{-4}$ | $1.62 \times 10^{-4}$ | $1.82 \times 10^{-4}$ | $1.95 \times 10^{-4}$ | $2.01 \times 10^{-4}$ | $2.14 \times 10^{-4}$ | $2.24 \times 10^{-4}$ |
| $Ca_2MgSi_2O_7$ | $5.14 \times 10^{-6}$ | $9.79 \times 10^{-6}$ | $1.36 \times 10^{-5}$ | $1.71 \times 10^{-5}$ | $2.04 \times 10^{-5}$ | $1.97 \times 10^{-5}$ | $2.09 \times 10^{-5}$ | $2.25 \times 10^{-5}$ | $2.86 \times 10^{-5}$ | $3.79 \times 10^{-5}$ | $4.30 \times 10^{-5}$ | $4.61 \times 10^{-5}$ | $4.73 \times 10^{-5}$ | $4.99 \times 10^{-5}$ | $5.20 \times 10^{-5}$ |
| $MgSiO_3$ | $7.50 \times 10^{-5}$ | $1.43 \times 10^{-4}$ | $1.98 \times 10^{-4}$ | $2.49 \times 10^{-4}$ | $2.98 \times 10^{-4}$ | $2.88 \times 10^{-4}$ | $3.05 \times 10^{-4}$ | $3.28 \times 10^{-4}$ | $4.17 \times 10^{-4}$ | $5.52 \times 10^{-4}$ | $6.28 \times 10^{-4}$ | $6.72 \times 10^{-4}$ | $6.89 \times 10^{-4}$ | $7.28 \times 10^{-4}$ | $7.58 \times 10^{-4}$ |
| SiC | $2.40 \times 10^{-5}$ | $4.56 \times 10^{-5}$ | $6.32 \times 10^{-5}$ | $7.96 \times 10^{-5}$ | $9.53 \times 10^{-5}$ | $9.20 \times 10^{-5}$ | $9.74 \times 10^{-5}$ | $1.05 \times 10^{-4}$ | $1.33 \times 10^{-4}$ | $1.76 \times 10^{-4}$ | $2.01 \times 10^{-4}$ | $2.15 \times 10^{-4}$ | $2.20 \times 10^{-4}$ | $2.33 \times 10^{-4}$ | $2.42 \times 10^{-4}$ |
| $FeSiO_3$ | $7.39 \times 10^{-5}$ | $1.55 \times 10^{-4}$ | $2.27 \times 10^{-4}$ | $2.96 \times 10^{-4}$ | $3.64 \times 10^{-4}$ | $3.66 \times 10^{-4}$ | $3.94 \times 10^{-4}$ | $4.32 \times 10^{-4}$ | $6.01 \times 10^{-4}$ | $8.60 \times 10^{-4}$ | $1.02 \times 10^{-3}$ | $1.11 \times 10^{-3}$ | $1.15 \times 10^{-3}$ | $1.23 \times 10^{-3}$ | $1.30 \times 10^{-3}$ |
| Fe-metal | $2.35 \times 10^{-5}$ | $4.94 \times 10^{-5}$ | $7.21 \times 10^{-5}$ | $9.40 \times 10^{-5}$ | $1.15 \times 10^{-4}$ | $1.16 \times 10^{-4}$ | $1.25 \times 10^{-4}$ | $1.37 \times 10^{-4}$ | $1.91 \times 10^{-4}$ | $2.73 \times 10^{-4}$ | $3.22 \times 10^{-4}$ | $3.52 \times 10^{-4}$ | $3.65 \times 10^{-4}$ | $3.92 \times 10^{-4}$ | $4.12 \times 10^{-4}$ |
| $FeAl_2O_4$ | $7.31 \times 10^{-8}$ | $1.54 \times 10^{-7}$ | $2.24 \times 10^{-7}$ | $2.93 \times 10^{-7}$ | $3.59 \times 10^{-7}$ | $3.62 \times 10^{-7}$ | $3.89 \times 10^{-7}$ | $4.27 \times 10^{-7}$ | $5.93 \times 10^{-7}$ | $8.50 \times 10^{-7}$ | $1.00 \times 10^{-6}$ | $1.10 \times 10^{-6}$ | $1.14 \times 10^{-6}$ | $1.22 \times 10^{-6}$ | $1.28 \times 10^{-6}$ |
| $Fe_3C$ | $1.51 \times 10^{-5}$ | $3.17 \times 10^{-5}$ | $4.63 \times 10^{-5}$ | $6.04 \times 10^{-5}$ | $7.42 \times 10^{-5}$ | $7.47 \times 10^{-5}$ | $8.03 \times 10^{-5}$ | $8.82 \times 10^{-5}$ | $1.22 \times 10^{-4}$ | $1.75 \times 10^{-4}$ | $2.07 \times 10^{-4}$ | $2.26 \times 10^{-4}$ | $2.34 \times 10^{-4}$ | $2.52 \times 10^{-4}$ | $2.65 \times 10^{-4}$ |
| FeS | $1.22 \times 10^{-5}$ | $2.32 \times 10^{-5}$ | $3.22 \times 10^{-5}$ | $4.05 \times 10^{-5}$ | $4.85 \times 10^{-5}$ | $4.69 \times 10^{-5}$ | $4.97 \times 10^{-5}$ | $5.35 \times 10^{-5}$ | $6.88 \times 10^{-5}$ | $9.27 \times 10^{-5}$ | $1.07 \times 10^{-4}$ | $1.15 \times 10^{-4}$ | $1.19 \times 10^{-4}$ | $1.26 \times 10^{-4}$ | $1.32 \times 10^{-4}$ |
| $KAlSi_3O_8$ | $3.44 \times 10^{-7}$ | $7.16 \times 10^{-7}$ | $1.09 \times 10^{-6}$ | $1.48 \times 10^{-6}$ | $1.89 \times 10^{-6}$ | $1.88 \times 10^{-6}$ | $2.06 \times 10^{-6}$ | $2.26 \times 10^{-6}$ | $3.08 \times 10^{-6}$ | $4.58 \times 10^{-6}$ | $5.67 \times 10^{-6}$ | $6.48 \times 10^{-6}$ | $6.86 \times 10^{-6}$ | $7.63 \times 10^{-6}$ | $8.21 \times 10^{-6}$ |
| $NaAlSi_3O_8$ | $2.36 \times 10^{-8}$ | $7.96 \times 10^{-8}$ | $1.39 \times 10^{-7}$ | $1.83 \times 10^{-7}$ | $2.29 \times 10^{-7}$ | $2.21 \times 10^{-7}$ | $2.36 \times 10^{-7}$ | $2.56 \times 10^{-7}$ | $3.35 \times 10^{-7}$ | $4.64 \times 10^{-7}$ | $5.36 \times 10^{-7}$ | $5.80 \times 10^{-7}$ | $5.98 \times 10^{-7}$ | $6.36 \times 10^{-7}$ | $6.65 \times 10^{-7}$ |
| Graphite | $1.67 \times 10^{-6}$ | $3.28 \times 10^{-6}$ | $4.79 \times 10^{-6}$ | $6.61 \times 10^{-6}$ | $8.71 \times 10^{-6}$ | $9.47 \times 10^{-6}$ | $1.04 \times 10^{-5}$ | $1.12 \times 10^{-5}$ | $1.38 \times 10^{-5}$ | $1.69 \times 10^{-5}$ | $1.79 \times 10^{-5}$ | $1.85 \times 10^{-5}$ | $1.87 \times 10^{-5}$ | $1.94 \times 10^{-5}$ | $2.01 \times 10^{-5}$ |
| $H_2O$ | $2.94 \times 10^{-4}$ | $5.90 \times 10^{-4}$ | $7.98 \times 10^{-4}$ | $9.78 \times 10^{-4}$ | $1.14 \times 10^{-3}$ | $1.06 \times 10^{-3}$ | $1.11 \times 10^{-3}$ | $1.17 \times 10^{-3}$ | $1.27 \times 10^{-3}$ | $1.49 \times 10^{-3}$ | $1.50 \times 10^{-3}$ | $1.53 \times 10^{-3}$ | $1.47 \times 10^{-3}$ | $1.54 \times 10^{-3}$ | $1.60 \times 10^{-3}$ |

**Annular ring 6
(12-14 kpc)**

| Time (Gyr) | 0.2 | 0.4 | 0.6 | 0.8 | 1 | 1.3 | 1.7 | 2 | 3 | 5 | 7 | 9 | 10 | 12 | 13.5 |
|---|---|---|---|---|---|---|---|---|---|---|---|---|---|---|---|
| Gas density* | 2.3 | 3.8 | 4.73 | 5.27 | 5.5 | 6.67 | 7.54 | 7.91 | 8.33 | 8.16 | 8 | 7.94 | 7.9 | 7.73 | 7.55 |
| $CaTiO_3$ | $3.13 \times 10^{-7}$ | $7.00 \times 10^{-7}$ | $1.02 \times 10^{-6}$ | $1.32 \times 10^{-6}$ | $1.63 \times 10^{-6}$ | $1.72 \times 10^{-6}$ | $1.90 \times 10^{-6}$ | $2.08 \times 10^{-6}$ | $2.84 \times 10^{-6}$ | $4.15 \times 10^{-6}$ | $5.04 \times 10^{-6}$ | $5.63 \times 10^{-6}$ | $5.89 \times 10^{-6}$ | $6.40 \times 10^{-6}$ | $6.78 \times 10^{-6}$ |
| $Ti_2O_3$ | $1.84 \times 10^{-9}$ | $4.11 \times 10^{-9}$ | $5.99 \times 10^{-9}$ | $7.78 \times 10^{-9}$ | $9.55 \times 10^{-9}$ | $1.01 \times 10^{-8}$ | $1.12 \times 10^{-8}$ | $1.22 \times 10^{-8}$ | $1.67 \times 10^{-8}$ | $2.44 \times 10^{-8}$ | $2.96 \times 10^{-8}$ | $3.30 \times 10^{-8}$ | $3.46 \times 10^{-8}$ | $3.76 \times 10^{-8}$ | $3.98 \times 10^{-8}$ |
| TiC | $1.38 \times 10^{-8}$ | $3.08 \times 10^{-8}$ | $4.49 \times 10^{-8}$ | $5.83 \times 10^{-8}$ | $7.16 \times 10^{-8}$ | $7.57 \times 10^{-8}$ | $8.37 \times 10^{-8}$ | $9.18 \times 10^{-8}$ | $1.25 \times 10^{-7}$ | $1.83 \times 10^{-7}$ | $2.22 \times 10^{-7}$ | $2.48 \times 10^{-7}$ | $2.59 \times 10^{-7}$ | $2.82 \times 10^{-7}$ | $2.98 \times 10^{-7}$ |
| $Al_2O_3$ | $1.10 \times 10^{-6}$ | $3.27 \times 10^{-6}$ | $5.37 \times 10^{-6}$ | $7.53 \times 10^{-6}$ | $9.81 \times 10^{-6}$ | $1.03 \times 10^{-5}$ | $1.17 \times 10^{-5}$ | $1.31 \times 10^{-5}$ | $1.83 \times 10^{-5}$ | $2.80 \times 10^{-5}$ | $3.51 \times 10^{-5}$ | $4.01 \times 10^{-5}$ | $4.24 \times 10^{-5}$ | $4.66 \times 10^{-5}$ | $4.96 \times 10^{-5}$ |
| $CaAl_4O_7$ | $7.39 \times 10^{-7}$ | $2.19 \times 10^{-6}$ | $3.59 \times 10^{-6}$ | $5.04 \times 10^{-6}$ | $6.57 \times 10^{-6}$ | $6.92 \times 10^{-6}$ | $7.86 \times 10^{-6}$ | $8.76 \times 10^{-6}$ | $1.23 \times 10^{-5}$ | $1.87 \times 10^{-5}$ | $2.35 \times 10^{-5}$ | $2.69 \times 10^{-5}$ | $2.84 \times 10^{-5}$ | $3.12 \times 10^{-5}$ | $3.32 \times 10^{-5}$ |
| $Ca_2SiO_4$ | $2.21 \times 10^{-6}$ | $4.72 \times 10^{-6}$ | $6.68 \times 10^{-6}$ | $8.45 \times 10^{-6}$ | $1.01 \times 10^{-5}$ | $1.04 \times 10^{-5}$ | $1.13 \times 10^{-5}$ | $1.23 \times 10^{-5}$ | $1.59 \times 10^{-5}$ | $2.19 \times 10^{-5}$ | $2.56 \times 10^{-5}$ | $2.78 \times 10^{-5}$ | $2.87 \times 10^{-5}$ | $3.05 \times 10^{-5}$ | $3.19 \times 10^{-5}$ |
| $Ca_2Al_2SiO_7$ | $2.11 \times 10^{-6}$ | $4.51 \times 10^{-6}$ | $6.38 \times 10^{-6}$ | $8.07 \times 10^{-6}$ | $9.69 \times 10^{-6}$ | $9.98 \times 10^{-6}$ | $1.08 \times 10^{-5}$ | $1.17 \times 10^{-5}$ | $1.52 \times 10^{-5}$ | $2.09 \times 10^{-5}$ | $2.44 \times 10^{-5}$ | $2.65 \times 10^{-5}$ | $2.74 \times 10^{-5}$ | $2.92 \times 10^{-5}$ | $3.04 \times 10^{-5}$ |
| CaS | $1.26 \times 10^{-6}$ | $2.69 \times 10^{-6}$ | $3.80 \times 10^{-6}$ | $4.81 \times 10^{-6}$ | $5.78 \times 10^{-6}$ | $5.95 \times 10^{-6}$ | $6.46 \times 10^{-6}$ | $6.99 \times 10^{-6}$ | $9.08 \times 10^{-6}$ | $1.25 \times 10^{-5}$ | $1.46 \times 10^{-5}$ | $1.58 \times 10^{-5}$ | $1.64 \times 10^{-5}$ | $1.74 \times 10^{-5}$ | $1.82 \times 10^{-5}$ |
| $Mg_2SiO_4$ | $3.36 \times 10^{-5}$ | $8.11 \times 10^{-5}$ | $1.20 \times 10^{-4}$ | $1.54 \times 10^{-4}$ | $1.86 \times 10^{-4}$ | $1.89 \times 10^{-4}$ | $2.05 \times 10^{-4}$ | $2.21 \times 10^{-4}$ | $2.80 \times 10^{-4}$ | $3.73 \times 10^{-4}$ | $4.31 \times 10^{-4}$ | $4.67 \times 10^{-4}$ | $4.83 \times 10^{-4}$ | $5.15 \times 10^{-4}$ | $5.39 \times 10^{-4}$ |
| $MgAl_2O_4$ | $1.58 \times 10^{-7}$ | $3.83 \times 10^{-7}$ | $5.64 \times 10^{-7}$ | $7.26 \times 10^{-7}$ | $8.79 \times 10^{-7}$ | $8.92 \times 10^{-7}$ | $9.67 \times 10^{-7}$ | $1.04 \times 10^{-6}$ | $1.32 \times 10^{-6}$ | $1.76 \times 10^{-6}$ | $2.03 \times 10^{-6}$ | $2.20 \times 10^{-6}$ | $2.28 \times 10^{-6}$ | $2.43 \times 10^{-6}$ | $2.54 \times 10^{-6}$ |
| MgS | $1.25 \times 10^{-5}$ | $3.03 \times 10^{-5}$ | $4.47 \times 10^{-5}$ | $5.75 \times 10^{-5}$ | $6.96 \times 10^{-5}$ | $7.06 \times 10^{-5}$ | $7.65 \times 10^{-5}$ | $8.25 \times 10^{-5}$ | $1.05 \times 10^{-4}$ | $1.39 \times 10^{-4}$ | $1.61 \times 10^{-4}$ | $1.74 \times 10^{-4}$ | $1.81 \times 10^{-4}$ | $1.92 \times 10^{-4}$ | $2.01 \times 10^{-4}$ |
| $Ca_2MgSi_2O_7$ | $3.26 \times 10^{-6}$ | $7.09 \times 10^{-6}$ | $1.01 \times 10^{-5}$ | $1.28 \times 10^{-5}$ | $1.54 \times 10^{-5}$ | $1.59 \times 10^{-5}$ | $1.73 \times 10^{-5}$ | $1.88 \times 10^{-5}$ | $2.43 \times 10^{-5}$ | $3.34 \times 10^{-5}$ | $3.90 \times 10^{-5}$ | $4.23 \times 10^{-5}$ | $4.37 \times 10^{-5}$ | $4.63 \times 10^{-5}$ | $4.82 \times 10^{-5}$ |
| $MgSiO_3$ | $4.75 \times 10^{-5}$ | $1.03 \times 10^{-4}$ | $1.47 \times 10^{-4}$ | $1.86 \times 10^{-4}$ | $2.25 \times 10^{-4}$ | $2.32 \times 10^{-4}$ | $2.53 \times 10^{-4}$ | $2.74 \times 10^{-4}$ | $3.55 \times 10^{-4}$ | $4.87 \times 10^{-4}$ | $5.69 \times 10^{-4}$ | $6.17 \times 10^{-4}$ | $6.37 \times 10^{-4}$ | $6.75 \times 10^{-4}$ | $7.03 \times 10^{-4}$ |
| SiC | $1.52 \times 10^{-5}$ | $3.30 \times 10^{-5}$ | $4.69 \times 10^{-5}$ | $5.95 \times 10^{-5}$ | $7.18 \times 10^{-5}$ | $7.40 \times 10^{-5}$ | $8.08 \times 10^{-5}$ | $8.75 \times 10^{-5}$ | $1.13 \times 10^{-4}$ | $1.56 \times 10^{-4}$ | $1.82 \times 10^{-4}$ | $1.97 \times 10^{-4}$ | $2.04 \times 10^{-4}$ | $2.16 \times 10^{-4}$ | $2.25 \times 10^{-4}$ |
| $FeSiO_3$ | $4.50 \times 10^{-5}$ | $1.11 \times 10^{-4}$ | $1.68 \times 10^{-4}$ | $2.22 \times 10^{-4}$ | $2.75 \times 10^{-4}$ | $2.96 \times 10^{-4}$ | $3.28 \times 10^{-4}$ | $3.63 \times 10^{-4}$ | $5.14 \times 10^{-4}$ | $7.62 \times 10^{-4}$ | $9.22 \times 10^{-4}$ | $1.02 \times 10^{-3}$ | $1.06 \times 10^{-3}$ | $1.14 \times 10^{-3}$ | $1.20 \times 10^{-3}$ |
| Fe-metal | $1.43 \times 10^{-5}$ | $3.53 \times 10^{-5}$ | $5.32 \times 10^{-5}$ | $7.04 \times 10^{-5}$ | $8.72 \times 10^{-5}$ | $9.39 \times 10^{-5}$ | $1.04 \times 10^{-4}$ | $1.15 \times 10^{-4}$ | $1.63 \times 10^{-4}$ | $2.42 \times 10^{-4}$ | $2.93 \times 10^{-4}$ | $3.24 \times 10^{-4}$ | $3.37 \times 10^{-4}$ | $3.62 \times 10^{-4}$ | $3.81 \times 10^{-4}$ |
| $FeAl_2O_4$ | $4.45 \times 10^{-8}$ | $1.10 \times 10^{-7}$ | $1.66 \times 10^{-7}$ | $2.19 \times 10^{-7}$ | $2.71 \times 10^{-7}$ | $2.92 \times 10^{-7}$ | $3.25 \times 10^{-7}$ | $3.59 \times 10^{-7}$ | $5.07 \times 10^{-7}$ | $7.53 \times 10^{-7}$ | $9.11 \times 10^{-7}$ | $1.01 \times 10^{-6}$ | $1.05 \times 10^{-6}$ | $1.13 \times 10^{-6}$ | $1.18 \times 10^{-6}$ |
| $Fe_3C$ | $9.18 \times 10^{-6}$ | $2.27 \times 10^{-5}$ | $3.42 \times 10^{-5}$ | $4.52 \times 10^{-5}$ | $5.60 \times 10^{-5}$ | $6.03 \times 10^{-5}$ | $6.70 \times 10^{-5}$ | $7.41 \times 10^{-5}$ | $1.05 \times 10^{-4}$ | $1.55 \times 10^{-4}$ | $1.88 \times 10^{-4}$ | $2.08 \times 10^{-4}$ | $2.16 \times 10^{-4}$ | $2.33 \times 10^{-4}$ | $2.45 \times 10^{-4}$ |
| FeS | $7.80 \times 10^{-6}$ | $1.69 \times 10^{-5}$ | $2.39 \times 10^{-5}$ | $3.04 \times 10^{-5}$ | $3.66 \times 10^{-5}$ | $3.78 \times 10^{-5}$ | $4.12 \times 10^{-5}$ | $4.47 \times 10^{-5}$ | $5.84 \times 10^{-5}$ | $8.11 \times 10^{-5}$ | $9.58 \times 10^{-5}$ | $1.05 \times 10^{-4}$ | $1.09 \times 10^{-4}$ | $1.16 \times 10^{-4}$ | $1.21 \times 10^{-4}$ |
| $KAlSi_3O_8$ | $2.14 \times 10^{-7}$ | $4.87 \times 10^{-7}$ | $7.35 \times 10^{-7}$ | $9.94 \times 10^{-7}$ | $1.27 \times 10^{-6}$ | $1.35 \times 10^{-6}$ | $1.53 \times 10^{-6}$ | $1.70 \times 10^{-6}$ | $2.36 \times 10^{-6}$ | $3.65 \times 10^{-6}$ | $4.69 \times 10^{-6}$ | $5.47 \times 10^{-6}$ | $5.84 \times 10^{-6}$ | $6.55 \times 10^{-6}$ | $7.07 \times 10^{-6}$ |
| $NaAlSi_3O_8$ | $1.05 \times 10^{-8}$ | $4.84 \times 10^{-8}$ | $9.64 \times 10^{-8}$ | $1.34 \times 10^{-7}$ | $1.65 \times 10^{-7}$ | $1.68 \times 10^{-7}$ | $1.83 \times 10^{-7}$ | $1.99 \times 10^{-7}$ | $2.64 \times 10^{-7}$ | $3.74 \times 10^{-7}$ | $4.44 \times 10^{-7}$ | $4.89 \times 10^{-7}$ | $5.08 \times 10^{-7}$ | $5.43 \times 10^{-7}$ | $5.69 \times 10^{-7}$ |
| Graphite | $1.03 \times 10^{-6}$ | $2.34 \times 10^{-6}$ | $3.49 \times 10^{-6}$ | $4.76 \times 10^{-6}$ | $6.33 \times 10^{-6}$ | $7.40 \times 10^{-6}$ | $8.59 \times 10^{-6}$ | $9.54 \times 10^{-6}$ | $1.22 \times 10^{-5}$ | $1.58 \times 10^{-5}$ | $1.74 \times 10^{-5}$ | $1.82 \times 10^{-5}$ | $1.84 \times 10^{-5}$ | $1.90 \times 10^{-5}$ | $1.95 \times 10^{-5}$ |
| $H_2O$ | $1.22 \times 10^{-4}$ | $2.96 \times 10^{-4}$ | $4.19 \times 10^{-4}$ | $5.24 \times 10^{-4}$ | $6.21 \times 10^{-4}$ | $6.19 \times 10^{-4}$ | $6.61 \times 10^{-4}$ | $7.05 \times 10^{-4}$ | $7.43 \times 10^{-4}$ | $8.02 \times 10^{-4}$ | $7.36 \times 10^{-4}$ | $7.81 \times 10^{-4}$ | $7.12 \times 10^{-4}$ | $7.49 \times 10^{-4}$ | $7.78 \times 10^{-4}$ |

**Annular ring 7
(14-16 kpc)**

| Time (Gyr) | 0.2 | 0.4 | 0.6 | 0.8 | 1 | 1.3 | 1.7 | 2 | 3 | 5 | 7 | 9 | 10 | 12 | 13.5 |
|---|---|---|---|---|---|---|---|---|---|---|---|---|---|---|---|
| Gas density* | 2.03 | 3.43 | 4.36 | 4.96 | 5.3 | 6.17 | 6.82 | 7.09 | 7.33 | 7 | 6.73 | 6.61 | 6.55 | 6.4 | 6.26 |
| $CaTiO_3$ | $2.02\times10^{-7}$ | $4.92\times10^{-7}$ | $7.49\times10^{-7}$ | $9.83\times10^{-7}$ | $1.21\times10^{-6}$ | $1.33\times10^{-6}$ | $1.50\times10^{-6}$ | $1.66\times10^{-6}$ | $2.30\times10^{-6}$ | $3.48\times10^{-6}$ | $4.37\times10^{-6}$ | $4.99\times10^{-6}$ | $5.25\times10^{-6}$ | $5.76\times10^{-6}$ | $6.12\times10^{-6}$ |
| $Ti_2O_3$ | $1.19\times10^{-9}$ | $2.89\times10^{-9}$ | $4.40\times10^{-9}$ | $5.78\times10^{-9}$ | $7.12\times10^{-9}$ | $7.82\times10^{-9}$ | $8.83\times10^{-9}$ | $9.76\times10^{-9}$ | $1.35\times10^{-8}$ | $2.04\times10^{-8}$ | $2.56\times10^{-8}$ | $2.93\times10^{-8}$ | $3.09\times10^{-8}$ | $3.38\times10^{-8}$ | $3.59\times10^{-8}$ |
| TiC | $8.89\times10^{-9}$ | $2.16\times10^{-8}$ | $3.30\times10^{-8}$ | $4.33\times10^{-8}$ | $5.34\times10^{-8}$ | $5.86\times10^{-8}$ | $6.62\times10^{-8}$ | $7.32\times10^{-8}$ | $1.01\times10^{-7}$ | $1.53\times10^{-7}$ | $1.92\times10^{-7}$ | $2.20\times10^{-7}$ | $2.31\times10^{-7}$ | $2.54\times10^{-7}$ | $2.69\times10^{-7}$ |
| $Al_2O_3$ | $6.16\times10^{-7}$ | $1.93\times10^{-6}$ | $3.42\times10^{-6}$ | $4.88\times10^{-6}$ | $6.41\times10^{-6}$ | $7.03\times10^{-6}$ | $8.17\times10^{-6}$ | $9.18\times10^{-6}$ | $1.31\times10^{-5}$ | $2.10\times10^{-5}$ | $2.76\times10^{-5}$ | $3.25\times10^{-5}$ | $3.47\times10^{-5}$ | $3.87\times10^{-5}$ | $4.15\times10^{-5}$ |
| $CaAl_4O_7$ | $4.12\times10^{-7}$ | $1.29\times10^{-6}$ | $2.29\times10^{-6}$ | $3.27\times10^{-6}$ | $4.29\times10^{-6}$ | $4.71\times10^{-6}$ | $5.47\times10^{-6}$ | $6.15\times10^{-6}$ | $8.78\times10^{-6}$ | $1.41\times10^{-5}$ | $1.84\times10^{-5}$ | $2.18\times10^{-5}$ | $2.32\times10^{-5}$ | $2.59\times10^{-5}$ | $2.78\times10^{-5}$ |
| $Ca_2SiO_4$ | $1.45\times10^{-6}$ | $3.33\times10^{-6}$ | $4.94\times10^{-6}$ | $6.36\times10^{-6}$ | $7.69\times10^{-6}$ | $8.25\times10^{-6}$ | $9.16\times10^{-6}$ | $9.99\times10^{-6}$ | $1.32\times10^{-5}$ | $1.88\times10^{-5}$ | $2.27\times10^{-5}$ | $2.52\times10^{-5}$ | $2.62\times10^{-5}$ | $2.81\times10^{-5}$ | $2.94\times10^{-5}$ |
| $Ca_2Al_2SiO_7$ | $9.23\times10^{-7}$ | $2.12\times10^{-6}$ | $3.15\times10^{-6}$ | $4.05\times10^{-6}$ | $4.90\times10^{-6}$ | $5.25\times10^{-6}$ | $5.83\times10^{-6}$ | $6.36\times10^{-6}$ | $8.39\times10^{-6}$ | $1.20\times10^{-5}$ | $1.45\times10^{-5}$ | $1.60\times10^{-5}$ | $1.67\times10^{-5}$ | $1.79\times10^{-5}$ | $1.88\times10^{-5}$ |
| CaS | $8.75\times10^{-7}$ | $2.01\times10^{-6}$ | $2.98\times10^{-6}$ | $3.83\times10^{-6}$ | $4.64\times10^{-6}$ | $4.97\times10^{-6}$ | $5.52\times10^{-6}$ | $6.02\times10^{-6}$ | $7.95\times10^{-6}$ | $1.13\times10^{-5}$ | $1.37\times10^{-5}$ | $1.52\times10^{-5}$ | $1.58\times10^{-5}$ | $1.70\times10^{-5}$ | $1.78\times10^{-5}$ |
| $Mg_2SiO_4$ | $1.99\times10^{-5}$ | $5.28\times10^{-5}$ | $8.39\times10^{-5}$ | $1.11\times10^{-4}$ | $1.36\times10^{-4}$ | $1.44\times10^{-4}$ | $1.60\times10^{-4}$ | $1.73\times10^{-4}$ | $2.23\times10^{-4}$ | $3.10\times10^{-4}$ | $3.70\times10^{-4}$ | $4.10\times10^{-4}$ | $4.28\times10^{-4}$ | $4.59\times10^{-4}$ | $4.81\times10^{-4}$ |
| $MgAl_2O_4$ | $9.38\times10^{-8}$ | $2.49\times10^{-7}$ | $3.96\times10^{-7}$ | $5.23\times10^{-7}$ | $6.44\times10^{-7}$ | $6.81\times10^{-7}$ | $7.54\times10^{-7}$ | $8.18\times10^{-7}$ | $1.05\times10^{-6}$ | $1.46\times10^{-6}$ | $1.75\times10^{-6}$ | $1.94\times10^{-6}$ | $2.02\times10^{-6}$ | $2.16\times10^{-6}$ | $2.27\times10^{-6}$ |
| MgS | $7.42\times10^{-6}$ | $1.97\times10^{-5}$ | $3.13\times10^{-5}$ | $4.14\times10^{-5}$ | $5.10\times10^{-5}$ | $5.39\times10^{-5}$ | $5.97\times10^{-5}$ | $6.48\times10^{-5}$ | $8.34\times10^{-5}$ | $1.16\times10^{-4}$ | $1.38\times10^{-4}$ | $1.53\times10^{-4}$ | $1.60\times10^{-4}$ | $1.71\times10^{-4}$ | $1.80\times10^{-4}$ |
| $Ca_2MgSi_2O_7$ | $2.10\times10^{-6}$ | $4.97\times10^{-6}$ | $7.42\times10^{-6}$ | $9.56\times10^{-6}$ | $1.16\times10^{-5}$ | $1.24\times10^{-5}$ | $1.39\times10^{-5}$ | $1.51\times10^{-5}$ | $2.00\times10^{-5}$ | $2.85\times10^{-5}$ | $3.45\times10^{-5}$ | $3.83\times10^{-5}$ | $3.99\times10^{-5}$ | $4.26\times10^{-5}$ | $4.46\times10^{-5}$ |
| $MgSiO_3$ | $3.07\times10^{-5}$ | $7.25\times10^{-5}$ | $1.08\times10^{-4}$ | $1.39\times10^{-4}$ | $1.69\times10^{-4}$ | $1.81\times10^{-4}$ | $2.02\times10^{-4}$ | $2.21\times10^{-4}$ | $2.91\times10^{-4}$ | $4.16\times10^{-4}$ | $5.03\times10^{-4}$ | $5.59\times10^{-4}$ | $5.81\times10^{-4}$ | $6.22\times10^{-4}$ | $6.50\times10^{-4}$ |
| SiC | $9.80\times10^{-6}$ | $2.32\times10^{-5}$ | $3.46\times10^{-5}$ | $4.45\times10^{-5}$ | $5.40\times10^{-5}$ | $5.79\times10^{-5}$ | $6.46\times10^{-5}$ | $7.05\times10^{-5}$ | $9.30\times10^{-5}$ | $1.33\times10^{-4}$ | $1.61\times10^{-4}$ | $1.79\times10^{-4}$ | $1.86\times10^{-4}$ | $1.99\times10^{-4}$ | $2.08\times10^{-4}$ |
| $FeSiO_3$ | $2.80\times10^{-5}$ | $7.62\times10^{-5}$ | $1.23\times10^{-4}$ | $1.65\times10^{-4}$ | $2.07\times10^{-4}$ | $2.32\times10^{-4}$ | $2.64\times10^{-4}$ | $2.95\times10^{-4}$ | $4.23\times10^{-4}$ | $6.55\times10^{-4}$ | $8.21\times10^{-4}$ | $9.29\times10^{-4}$ | $9.74\times10^{-4}$ | $1.06\times10^{-3}$ | $1.11\times10^{-3}$ |
| Fe-metal | $8.90\times10^{-6}$ | $2.42\times10^{-5}$ | $3.90\times10^{-5}$ | $5.25\times10^{-5}$ | $6.57\times10^{-5}$ | $7.36\times10^{-5}$ | $8.38\times10^{-5}$ | $9.35\times10^{-5}$ | $1.34\times10^{-4}$ | $2.08\times10^{-4}$ | $2.61\times10^{-4}$ | $2.95\times10^{-4}$ | $3.09\times10^{-4}$ | $3.35\times10^{-4}$ | $3.53\times10^{-4}$ |
| $FeAl_2O_4$ | $2.77\times10^{-8}$ | $7.53\times10^{-8}$ | $1.21\times10^{-7}$ | $1.63\times10^{-7}$ | $2.04\times10^{-7}$ | $2.29\times10^{-7}$ | $2.61\times10^{-7}$ | $2.91\times10^{-7}$ | $4.18\times10^{-7}$ | $6.47\times10^{-7}$ | $8.12\times10^{-7}$ | $9.18\times10^{-7}$ | $9.62\times10^{-7}$ | $1.04\times10^{-6}$ | $1.10\times10^{-6}$ |
| $Fe_3C$ | $5.72\times10^{-6}$ | $1.55\times10^{-5}$ | $2.51\times10^{-5}$ | $3.37\times10^{-5}$ | $4.22\times10^{-5}$ | $4.73\times10^{-5}$ | $5.38\times10^{-5}$ | $6.01\times10^{-5}$ | $8.64\times10^{-5}$ | $1.34\times10^{-4}$ | $1.68\times10^{-4}$ | $1.90\times10^{-4}$ | $1.99\times10^{-4}$ | $2.15\times10^{-4}$ | $2.27\times10^{-4}$ |
| FeS | $5.12\times10^{-6}$ | $1.19\times10^{-5}$ | $1.77\times10^{-5}$ | $2.28\times10^{-5}$ | $2.76\times10^{-5}$ | $2.96\times10^{-5}$ | $3.30\times10^{-5}$ | $3.61\times10^{-5}$ | $4.79\times10^{-5}$ | $6.90\times10^{-5}$ | $8.40\times10^{-5}$ | $9.39\times10^{-5}$ | $9.81\times10^{-5}$ | $1.06\times10^{-4}$ | $1.11\times10^{-4}$ |
| $KAlSi_3O_8$ | $1.38\times10^{-7}$ | $3.30\times10^{-7}$ | $5.10\times10^{-7}$ | $6.86\times10^{-7}$ | $8.70\times10^{-7}$ | $9.57\times10^{-7}$ | $1.11\times10^{-6}$ | $1.24\times10^{-6}$ | $1.74\times10^{-6}$ | $2.80\times10^{-6}$ | $3.73\times10^{-6}$ | $4.49\times10^{-6}$ | $4.84\times10^{-6}$ | $5.51\times10^{-6}$ | $6.00\times10^{-6}$ |
| $NaAlSi_3O_8$ | $5.04\times10^{-9}$ | $2.77\times10^{-8}$ | $6.51\times10^{-8}$ | $9.78\times10^{-8}$ | $1.25\times10^{-7}$ | $1.31\times10^{-7}$ | $1.42\times10^{-7}$ | $1.55\times10^{-7}$ | $2.03\times10^{-7}$ | $2.95\times10^{-7}$ | $3.62\times10^{-7}$ | $4.07\times10^{-7}$ | $4.26\times10^{-7}$ | $4.61\times10^{-7}$ | $4.84\times10^{-7}$ |
| Graphite | $6.49\times10^{-7}$ | $1.61\times10^{-6}$ | $2.52\times10^{-6}$ | $3.44\times10^{-6}$ | $4.55\times10^{-6}$ | $5.56\times10^{-6}$ | $6.73\times10^{-6}$ | $7.64\times10^{-6}$ | $1.03\times10^{-5}$ | $1.42\times10^{-5}$ | $1.64\times10^{-5}$ | $1.77\times10^{-5}$ | $1.80\times10^{-5}$ | $1.87\times10^{-5}$ | $1.92\times10^{-5}$ |
| $H_2O$ | $4.23\times10^{-5}$ | $1.19\times10^{-4}$ | $2.00\times10^{-4}$ | $2.57\times10^{-4}$ | $3.09\times10^{-4}$ | $2.90\times10^{-4}$ | $3.17\times10^{-4}$ | $3.40\times10^{-4}$ | $3.29\times10^{-4}$ | $3.10\times10^{-4}$ | $2.15\times10^{-4}$ | $1.55\times10^{-4}$ | $1.60\times10^{-4}$ | $8.44\times10^{-5}$ | $8.78\times10^{-5}$ |

**Annular ring 8
(16-18 kpc)**

| Time (Gyr) | 0.2 | 0.4 | 0.6 | 0.8 | 1 | 1.3 | 1.7 | 2 | 3 | 5 | 7 | 9 | 10 | 12 | 13.5 |
|---|---|---|---|---|---|---|---|---|---|---|---|---|---|---|---|
| Gas density* | 1.81 | 3.11 | 4.01 | 4.63 | 5 | 5.78 | 6.37 | 6.62 | 6.87 | 6.62 | 6.36 | 6.23 | 6.17 | 6.04 | 5.92 |
| $CaTiO_3$ | $1.28\times10^{-7}$ | $3.21\times10^{-7}$ | $5.26\times10^{-7}$ | $7.11\times10^{-7}$ | $8.84\times10^{-7}$ | $9.95\times10^{-7}$ | $1.14\times10^{-6}$ | $1.26\times10^{-6}$ | $1.75\times10^{-6}$ | $2.71\times10^{-6}$ | $3.49\times10^{-6}$ | $4.10\times10^{-6}$ | $4.36\times10^{-6}$ | $4.86\times10^{-6}$ | $5.20\times10^{-6}$ |
| $Ti_2O_3$ | $7.5\times10^{-10}$ | $1.88\times10^{-9}$ | $3.09\times10^{-9}$ | $4.18\times10^{-9}$ | $5.20\times10^{-9}$ | $5.84\times10^{-9}$ | $6.68\times10^{-9}$ | $7.41\times10^{-9}$ | $1.03\times10^{-8}$ | $1.59\times10^{-8}$ | $2.05\times10^{-8}$ | $2.41\times10^{-8}$ | $2.56\times10^{-8}$ | $2.85\times10^{-8}$ | $3.05\times10^{-8}$ |
| TiC | $5.65\times10^{-9}$ | $1.41\times10^{-8}$ | $2.32\times10^{-8}$ | $3.13\times10^{-8}$ | $3.90\times10^{-8}$ | $4.38\times10^{-8}$ | $5.01\times10^{-8}$ | $5.55\times10^{-8}$ | $7.72\times10^{-8}$ | $1.19\times10^{-7}$ | $1.54\times10^{-7}$ | $1.80\times10^{-7}$ | $1.92\times10^{-7}$ | $2.14\times10^{-7}$ | $2.29\times10^{-7}$ |
| $Al_2O_3$ | $3.46\times10^{-7}$ | $1.05\times10^{-6}$ | $2.02\times10^{-6}$ | $3.04\times10^{-6}$ | $4.04\times10^{-6}$ | $4.56\times10^{-6}$ | $5.38\times10^{-6}$ | $6.07\times10^{-6}$ | $8.72\times10^{-6}$ | $1.44\times10^{-5}$ | $1.96\times10^{-5}$ | $2.40\times10^{-5}$ | $2.60\times10^{-5}$ | $2.97\times10^{-5}$ | $3.22\times10^{-5}$ |
| $CaAl_4O_7$ | $2.32\times10^{-7}$ | $7.03\times10^{-7}$ | $1.35\times10^{-6}$ | $2.03\times10^{-6}$ | $2.70\times10^{-6}$ | $3.05\times10^{-6}$ | $3.60\times10^{-6}$ | $4.06\times10^{-6}$ | $5.84\times10^{-6}$ | $9.64\times10^{-6}$ | $1.31\times10^{-5}$ | $1.61\times10^{-5}$ | $1.74\times10^{-5}$ | $1.99\times10^{-5}$ | $2.16\times10^{-5}$ |
| $Ca_2SiO_4$ | $9.52\times10^{-7}$ | $2.20\times10^{-6}$ | $3.49\times10^{-6}$ | $4.63\times10^{-6}$ | $5.67\times10^{-6}$ | $6.24\times10^{-6}$ | $7.03\times10^{-6}$ | $7.70\times10^{-6}$ | $1.02\times10^{-5}$ | $1.50\times10^{-5}$ | $1.86\times10^{-5}$ | $2.13\times10^{-5}$ | $2.24\times10^{-5}$ | $2.44\times10^{-5}$ | $2.58\times10^{-5}$ |
| $Ca_2Al_2SiO_7$ | $6.06\times10^{-7}$ | $1.40\times10^{-6}$ | $2.22\times10^{-6}$ | $2.95\times10^{-6}$ | $3.61\times10^{-6}$ | $3.97\times10^{-6}$ | $4.48\times10^{-6}$ | $4.91\times10^{-6}$ | $6.53\times10^{-6}$ | $9.54\times10^{-6}$ | $1.19\times10^{-5}$ | $1.36\times10^{-5}$ | $1.43\times10^{-5}$ | $1.55\times10^{-5}$ | $1.64\times10^{-5}$ |
| CaS | $5.74\times10^{-7}$ | $1.32\times10^{-6}$ | $2.10\times10^{-6}$ | $2.79\times10^{-6}$ | $3.42\times10^{-6}$ | $3.76\times10^{-6}$ | $4.24\times10^{-6}$ | $4.65\times10^{-6}$ | $6.18\times10^{-6}$ | $9.04\times10^{-6}$ | $1.12\times10^{-5}$ | $1.28\times10^{-5}$ | $1.35\times10^{-5}$ | $1.47\times10^{-5}$ | $1.56\times10^{-5}$ |
| $Mg_2SiO_4$ | $1.15\times10^{-5}$ | $3.16\times10^{-5}$ | $5.49\times10^{-5}$ | $7.63\times10^{-5}$ | $9.58\times10^{-5}$ | $1.05\times10^{-4}$ | $1.18\times10^{-4}$ | $1.29\times10^{-4}$ | $1.67\times10^{-4}$ | $2.39\times10^{-4}$ | $2.95\times10^{-4}$ | $3.36\times10^{-4}$ | $3.55\times10^{-4}$ | $3.87\times10^{-4}$ | $4.08\times10^{-4}$ |
| $MgAl_2O_4$ | $5.43\times10^{-8}$ | $1.49\times10^{-7}$ | $2.59\times10^{-7}$ | $3.60\times10^{-7}$ | $4.52\times10^{-7}$ | $4.95\times10^{-7}$ | $5.56\times10^{-7}$ | $6.07\times10^{-7}$ | $7.90\times10^{-7}$ | $1.13\times10^{-6}$ | $1.39\times10^{-6}$ | $1.59\times10^{-6}$ | $1.67\times10^{-6}$ | $1.83\times10^{-6}$ | $1.93\times10^{-6}$ |
| MgS | $4.30\times10^{-6}$ | $1.18\times10^{-5}$ | $2.05\times10^{-5}$ | $2.85\times10^{-5}$ | $3.58\times10^{-5}$ | $3.91\times10^{-5}$ | $4.41\times10^{-5}$ | $4.81\times10^{-5}$ | $6.25\times10^{-5}$ | $8.92\times10^{-5}$ | $1.10\times10^{-4}$ | $1.26\times10^{-4}$ | $1.32\times10^{-4}$ | $1.45\times10^{-4}$ | $1.53\times10^{-4}$ |
| $Ca_2MgSi_2O_7$ | $1.33\times10^{-6}$ | $3.24\times10^{-6}$ | $5.19\times10^{-6}$ | $6.91\times10^{-6}$ | $8.48\times10^{-6}$ | $9.34\times10^{-6}$ | $1.06\times10^{-5}$ | $1.16\times10^{-5}$ | $1.54\times10^{-5}$ | $2.25\times10^{-5}$ | $2.82\times10^{-5}$ | $3.22\times10^{-5}$ | $3.40\times10^{-5}$ | $3.70\times10^{-5}$ | $3.90\times10^{-5}$ |
| $MgSiO_3$ | $1.95\times10^{-5}$ | $4.72\times10^{-5}$ | $7.58\times10^{-5}$ | $1.01\times10^{-4}$ | $1.24\times10^{-4}$ | $1.36\times10^{-4}$ | $1.54\times10^{-4}$ | $1.69\times10^{-4}$ | $2.24\times10^{-4}$ | $3.29\times10^{-4}$ | $4.11\times10^{-4}$ | $4.70\times10^{-4}$ | $4.95\times10^{-4}$ | $5.40\times10^{-4}$ | $5.69\times10^{-4}$ |
| SiC | $6.22\times10^{-6}$ | $1.51\times10^{-5}$ | $2.42\times10^{-5}$ | $3.22\times10^{-5}$ | $3.95\times10^{-5}$ | $4.35\times10^{-5}$ | $4.92\times10^{-5}$ | $5.39\times10^{-5}$ | $7.16\times10^{-5}$ | $1.05\times10^{-4}$ | $1.31\times10^{-4}$ | $1.50\times10^{-4}$ | $1.58\times10^{-4}$ | $1.72\times10^{-4}$ | $1.82\times10^{-4}$ |
| $FeSiO_3$ | $1.71\times10^{-5}$ | $4.79\times10^{-5}$ | $8.45\times10^{-5}$ | $1.19\times10^{-4}$ | $1.51\times10^{-4}$ | $1.74\times10^{-4}$ | $2.02\times10^{-4}$ | $2.26\times10^{-4}$ | $3.27\times10^{-4}$ | $5.21\times10^{-4}$ | $6.74\times10^{-4}$ | $7.86\times10^{-4}$ | $8.33\times10^{-4}$ | $9.18\times10^{-4}$ | $9.76\times10^{-4}$ |
| Fe-metal | $5.44\times10^{-6}$ | $1.52\times10^{-5}$ | $2.68\times10^{-5}$ | $3.78\times10^{-5}$ | $4.80\times10^{-5}$ | $5.53\times10^{-5}$ | $6.41\times10^{-5}$ | $7.18\times10^{-5}$ | $1.04\times10^{-4}$ | $1.65\times10^{-4}$ | $2.14\times10^{-4}$ | $2.50\times10^{-4}$ | $2.65\times10^{-4}$ | $2.92\times10^{-4}$ | $3.10\times10^{-4}$ |
| $FeAl_2O_4$ | $1.69\times10^{-8}$ | $4.74\times10^{-8}$ | $8.35\times10^{-8}$ | $1.18\times10^{-7}$ | $1.49\times10^{-7}$ | $1.72\times10^{-7}$ | $1.99\times10^{-7}$ | $2.23\times10^{-7}$ | $3.24\times10^{-7}$ | $5.14\times10^{-7}$ | $6.66\times10^{-7}$ | $7.77\times10^{-7}$ | $8.23\times10^{-7}$ | $9.07\times10^{-7}$ | $9.64\times10^{-7}$ |
| $Fe_3C$ | $3.49\times10^{-6}$ | $9.77\times10^{-6}$ | $1.72\times10^{-5}$ | $2.43\times10^{-5}$ | $3.08\times10^{-5}$ | $3.55\times10^{-5}$ | $4.11\times10^{-5}$ | $4.61\times10^{-5}$ | $6.68\times10^{-5}$ | $1.06\times10^{-4}$ | $1.37\times10^{-4}$ | $1.60\times10^{-4}$ | $1.70\times10^{-4}$ | $1.87\times10^{-4}$ | $1.99\times10^{-4}$ |
| FeS | $3.34\times10^{-6}$ | $7.84\times10^{-6}$ | $1.25\times10^{-5}$ | $1.65\times10^{-5}$ | $2.03\times10^{-5}$ | $2.24\times10^{-5}$ | $2.53\times10^{-5}$ | $2.77\times10^{-5}$ | $3.71\times10^{-5}$ | $5.46\times10^{-5}$ | $6.84\times10^{-5}$ | $7.86\times10^{-5}$ | $8.29\times10^{-5}$ | $9.09\times10^{-5}$ | $9.64\times10^{-5}$ |
| $KAlSi_3O_8$ | $8.83\times10^{-8}$ | $2.12\times10^{-7}$ | $3.44\times10^{-7}$ | $4.69\times10^{-7}$ | $5.92\times10^{-7}$ | $6.64\times10^{-7}$ | $7.72\times10^{-7}$ | $8.63\times10^{-7}$ | $1.21\times10^{-6}$ | $1.97\times10^{-6}$ | $2.70\times10^{-6}$ | $3.34\times10^{-6}$ | $3.65\times10^{-6}$ | $4.24\times10^{-6}$ | $4.67\times10^{-6}$ |
| $NaAlSi_3O_8$ | $2.37\times10^{-9}$ | $1.43\times10^{-8}$ | $4.13\times10^{-8}$ | $6.87\times10^{-8}$ | $9.27\times10^{-8}$ | $1.05\times10^{-7}$ | $1.14\times10^{-7}$ | $1.22\times10^{-7}$ | $1.55\times10^{-7}$ | $2.23\times10^{-7}$ | $2.79\times10^{-7}$ | $3.22\times10^{-7}$ | $3.41\times10^{-7}$ | $3.75\times10^{-7}$ | $3.97\times10^{-7}$ |
| Graphite | $4.02\times10^{-7}$ | $1.02\times10^{-6}$ | $1.72\times10^{-6}$ | $2.41\times10^{-6}$ | $3.20\times10^{-6}$ | $3.98\times10^{-6}$ | $4.94\times10^{-6}$ | $5.71\times10^{-6}$ | $8.04\times10^{-6}$ | $1.17\times10^{-5}$ | $1.43\times10^{-5}$ | $1.61\times10^{-5}$ | $1.66\times10^{-5}$ | $1.75\times10^{-5}$ | $1.80\times10^{-5}$ |
| $H_2O$ | $1.12\times10^{-5}$ | $3.66\times10^{-5}$ | $6.44\times10^{-5}$ | $8.83\times10^{-5}$ | $1.08\times10^{-4}$ | $1.09\times10^{-4}$ | $1.19\times10^{-4}$ | $1.26\times10^{-4}$ | $9.02\times10^{-5}$ | $1.96\times10^{-5}$ | $2.33\times10^{-5}$ | $2.59\times10^{-5}$ | $2.70\times10^{-5}$ | $2.89\times10^{-5}$ | $3.03\times10^{-5}$ |

\* Gas surface mass density ($M_\odot$ pc$^{-2}$). The absolute dust mass density (for any species) can be obtained by multiplying the gas mass density with the normalized mass density.